%% file: bg19b.tex
\documentclass[aps,prb,twocolumn,showpacs,floatfix,longbibliography]{revtex4-1}
\usepackage{graphicx,amsfonts,amssymb,amsmath}
\usepackage{textcomp}
\usepackage{esint}
\usepackage[colorlinks,linktocpage,bookmarks=false,citecolor=blue,linkcolor=red,urlcolor=blue]{hyperref}
\usepackage[T1]{fontenc}
\usepackage[dvipsnames]{xcolor}
\usepackage[english]{babel}
\usepackage{afterpage}
\usepackage{bbold}

\allowdisplaybreaks

\def\phiset{\{\phi_a\}}
\def\varphiset{\{\varphi_a\}}
\def\Jset{\{J_a\}}

\def\varphipset{\{\varphi'_a\}}
\def\Jpset{\{J'_a\}}
\def\phifset{\{\phi_f\}}
\def\Jfset{\{J_f\}} 
\def\Jfpset{\{J'_f\}} 

\begin{document}

\graphicspath{{./figures_submit/}}

\title{Bose-glass phase of a one-dimensional disordered Bose fluid: Metastable states, quantum tunneling and droplets}

\author{Nicolas Dupuis}
\author{Romain Daviet}
\affiliation{Sorbonne Universit\'e, CNRS, Laboratoire de Physique Th\'eorique de la Mati\`ere Condens\'ee, LPTMC, F-75005 Paris, France}

\date{May 13, 2020} 

\begin{abstract}
	We study a one-dimensional disordered Bose fluid using bosonization, the replica method and a nonperturbative functional renormalization-group approach. We find that the Bose-glass phase is described by a fully attractive strong-disorder fixed point characterized by a singular disorder correlator whose functional dependence assumes a cuspy form that is related to the existence of metastable states. At nonzero momentum scale $k$, quantum tunneling between the ground state and low-lying metastable states leads to a rounding of the cusp singularity into a quantum boundary layer (QBL). The width of the QBL depends on an effective Luttinger parameter $K_k\sim k^\theta$ that vanishes with an exponent $\theta=z-1$ related to the dynamical critical exponent $z$. The QBL encodes the existence of rare ``superfluid'' regions, controls the low-energy dynamics and yields a (dissipative) conductivity vanishing as $\omega^2$ in the low-frequency limit. These results reveal the glassy properties (pinning, ``shocks'' or static avalanches) of the Bose-glass phase and can be understood within the ``droplet'' picture put forward for the description of glassy (classical) systems.
\end{abstract}

\maketitle 

\input{./definition.tex}

\tableofcontents

\section{Introduction}

In quantum many-body systems, disorder may lead to a localization transition. In the absence of interaction, this is due to the Anderson localization of single-particle wavefunctions,\cite{Anderson58a,Anderson_50years} a phenomenon that is now well understood and has been observed in various experiments ranging from microwaves to cold atoms.\cite{Aspect09} The situation is considerably more complex when disorder competes with collective effects due to interactions between particles.   

In interacting disordered boson systems, one expects a competition between superfluidity and localization. This question was first addressed in one dimension by Giamarchi and Schulz (GS) who showed, using a perturbative renormalization-group (RG) approach, that the system undergoes a transition between a superfluid and a localized phase.\cite{Giamarchi87,Giamarchi88} Scaling arguments have led to the conclusion that the localized phase, dubbed Bose glass, also exists in higher dimensions and is generically characterized by a nonzero compressibility, the absence of a gap in the excitation spectrum and an infinite superfluid susceptibility.\cite{Fisher89}  Experimentally, the superfluid--Bose-glass transition has regained a considerable interest thanks to the observation of a localization transition in cold atomic gases\cite{Billy08,Roati08,Pasienski10} as well as in magnetic insulators.\cite{Hong10,Yamada11,Zheludev13} The Bose-glass phase is also relevant for the physics of one-dimensional Fermi fluids,\cite{Giamarchi_book} 
charge-density waves in metals\cite{Fukuyama78} and superinductors.\cite{Houzet19}

\begin{figure}
\centerline{\includegraphics[width=4.5cm]{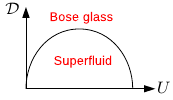}}
\caption{Schematic phase diagram of a disordered one-dimensional Bose fluid as a function of boson repulsion $U$ and disorder strength $\calD$.} 
\label{fig_phasedia}
\end{figure}

Most previous studies of one-dimensional disordered bosons have focused on the determination of the phase diagram and the nature of the superfluid--Bose-glass transition (one exception being the Gaussian Variational Method of Ref.~\onlinecite{Giamarchi96} that we briefly comment in Sec.~\ref{subsubsec_self_energy}). The original GS approach, which is valid in the limit of weak disorder, is based on the replica formalism, bosonization and a one-loop RG calculation.\cite{Giamarchi87,Giamarchi88} It suggests the phase diagram shown in Fig.~\ref{fig_phasedia} (possibly with an additional transition line that would lead to the existence of two distinct Bose-glass phases) and shows that for weak disorder (i.e. large $U/\calD$ in Fig.~\ref{fig_phasedia}), the transition is in the Berezinskii-Kosterlitz-Thouless (BKT) universality class with a universal critical Luttinger parameter $K_c=3/2$. This scenario has been confirmed by a two-loop calculation.\cite{Ristivojevic12,Ristivojevic14} To study the strong-disorder limit (small $U/\calD$ in Fig.~\ref{fig_phasedia}), different approaches (including numerical simulations) have been used.\cite{Altman04,Altman08,Altman10,Pielawa13,Pollet13,Pollet14,Yao16,Doggen17} In this regime, the physics of rare weak links plays a crucial role. Nevertheless, the transition is still believed to be of BKT type but with a nonuniversal critical Luttinger parameter $K_c>3/2$.

In this paper, we consider the weak-disorder limit where the bosonization approach\cite{Giamarchi87,Giamarchi88} is a valid starting point. By using a nonperturbative functional renormalization-group (FRG) approach, we can go beyond the perturbative RG of GS and follow the RG flow in the strong-disorder regime. This allows us to determine the physical properties of the Bose-glass phase. Perturbative implementations of the FRG in disordered (classical) systems have a long history.\cite{Fisher85,Narayan92,Nattermann92,Balents93,Chauve01,Ledoussal04,Tarjus04,Ledoussal10} The nonperturbative version\cite{Berges02,Delamotte12,Kopietz_book} has been used more recently to study the random-field Ising model\cite{Tarjus08,Tissier08,Tissier12,Tissier12a,Tarjus19} and random elastic manifold models.\cite{Balog19a} Here we report the first application to a quantum system.\cite{Dupuis19} 

An essential feature of the FRG approach is that the renormalized disorder correlator may assume a cuspy functional form whose origin lies in the existence of many different microscopic, locally stable, configurations.\cite{Balents96} This metastability leads in turn to a host of effects specific to disordered systems: non-ergodicity, pinning and ``shocks'' (static avalanches), depinning transition and avalanches, chaotic behavior, slow dynamics and aging, etc. In this paper, we describe in detail the FRG approach to one-dimensional disordered bosons and show that it gives a fairly complete picture of the Bose-glass phase while emphasizing the importance of metastable states. 

The outline of the paper is as follows. In Sec.~\ref{sec_frg} we present the bosonization approach and the replica formalism. We introduce the scale-dependent effective action $\Gamma_k$, the main quantity of interest in the FRG approach, and its exact flow equation. We then discuss the truncation of $\Gamma_k$ used to obtain an approximate solution of the flow equation. In Sec.~\ref{sec_transition} we consider the superfluid--Bose-glass transition and recover the standard BKT flow equations in agreement with GS. In Sec.~\ref{sec_bg} we show that the Bose-glass phase is described by a strong-disorder fixed point where the renormalized Luttinger parameter vanishes and the disorder correlator assumes a cuspy functional form. At finite momentum scale $k$, the cusp singularity is rounded into a QBL whose size depends on an effective Luttinger parameter $K_k\sim k^\theta$ that vanishes with an exponent $\theta=1-z$ where $z$ is the dynamical critical exponent. This rounding is explained by the quantum tunneling between the ground state and low-lying metastable states, which is expected to give rise to (rare) ``superfluid'' regions with significant density fluctuations and therefore reduced fluctuations (i.e. a nonzero rigidity) of the phase of the boson field. Furthermore, we find that the QBL is responsible for the low-frequency equilibrium dynamics and in particular the vanishing of the conductivity as $\w^2$. Finally, we show that our results can be understood within the ``droplet picture'' of glassy systems.\cite{Fisher88b}

We set $\hbar=k_B=1$ throughout the paper.

\section{FRG approach} 
\label{sec_frg} 

\subsection{Model and replica formalism}
\label{subsec_model} 

We consider a one-dimensional Bose fluid described by the Hamiltonian $\hat H_0+\hat H_{\rm dis}$. In the absence of disorder, at low energies $\hat H_0$ can be approximated by the Tomonaga-Luttinger Hamiltonian\cite{Giamarchi_book,Haldane81,Cazalilla11}
\begin{equation}
\hat H_0 = \int dx \frac{v}{2\pi} \left\{ \frac{1}{K} (\dx \hat\varphi)^2 + K (\dx \hat\theta)^2 \right\} , 
\label{H0}
\end{equation}
where $\hat\theta$ is the phase of the boson operator $\hat\psi(x)=e^{i\hat\theta(x)}\hat\rho(x)^{1/2}$ and $\hat\varphi$ is related to the density operator {\it via} 
\beq 
\hat\rho(x) =\rho_0 - \frac{1}{\pi}\dx\hat\varphi(x) 
+ 2\rho_2 \cos(2\pi\rho_0 x-2\hat\varphi(x)) , 
\eeq 
where $\rho_0$ is the average density and $\rho_2$ a nonuniversal parameter that depends on microscopic details. $\hat\varphi$ and $\hat\theta$ satisfy the commutation relations $[\hat\theta(x),\partial_y\hat\varphi(y)]=i\pi\delta(x-y)$. $v$ denotes the sound-mode velocity and the dimensionless parameter $K$, which encodes the strength of boson-boson interactions, is the Luttinger parameter. The constant $\rho_2$ depends on microscopic details of the system. The ground state of $\hat H_0$ is a Luttinger liquid, i.e. a superfluid state with superfluid stiffness $\rho_s=vK/\pi$ and compressibility $\kappa=d\rho_0/d\mu=K/\pi v$.\cite{Giamarchi_book}

The disorder contributes to the Hamiltonian a term\cite{Giamarchi87,Giamarchi88}  
\begin{equation}
\hat H_{\rm dis} = \int dx \left\{ - \frac{1}{\pi} \eta \dx \hat\varphi + \rho_2 [ \xi^* e^{2i\hat\varphi} + \hc ] \right\} , 
\label{Hdis} 
\end{equation}
where $\eta(x)$ (real) and $\xi(x)$ (complex) denote random potentials with Fourier components near 0 and $\pm 2\pi\rho_0$, respectively. $\eta$ can be eliminated by a shift of $\hat\varphi$,
\beq 
\hat\varphi(x) \to \hat\varphi(x) + \alpha(x) , \quad \dx \alpha = \frac{K}{v} \eta ,
\eeq
and is not considered in the following. 

In the functional-integral formalism, after integrating out the field $\theta$, one obtains the Euclidean (imaginary-time) action
\begin{align}
S[\varphi;\xi] ={}& \int_{x,\tau} \biggl\{ \frac{v}{2\pi K} \left[ (\dx\varphi)^2 + v^{-2} (\dtau \varphi)^2 \right] \nonumber \\ & 
+ \rho_2 [ \xi^* e^{2i\varphi} + \cc ] \biggr\} ,
\label{action0}
\end{align}
where we use the notation $\int_{x,\tau}=\inttau\int dx$ and $\varphi(x,\tau)$ is a bosonic field with $\tau\in[0,\beta]$. The model is regularized by a UV cutoff $\Lambda$ acting both on momenta and frequencies. We shall only consider the zero-temperature limit $\beta=1/T\to\infty$ but $\beta$ will be kept finite at intermediate stages of calculations. Equation~(\ref{action0}) shows that the Luttinger parameter controls the quantum fluctuations of the boson density ($K\to 0$ corresponds to the classical limit).

The partition function 
\beq
\calZ[J;\xi] \equiv e^{W[J;\xi]} = \int\calD[\varphi]\, e^{-S[\varphi;\xi]+\int_{x,\tau} J\varphi}
\label{ZJxi}
\eeq 
is a functional of both the external source $J$ and the random potential $\xi$. The thermodynamics depends on the average free energy $\overline{W[J;\xi]}$ (the overbar denotes the average over disorder) but full information on the system requires also knowledge of higher moments of $W[J;\xi]$. The latter can be obtained by considering $n$ copies (or replicas) of the system. Quite differently from the standard but controversial use of the replica trick, in which the analytic continuation $n\to 0$ opens the possibility of a spontaneous breaking of the replica symmetry (a signature of glassy properties),\cite{Mezard87} in the FRG approach one usually explicitly breaks the replica symmetry by introducing $n$ external sources acting on each replica independently.\cite{Tarjus08} Thus we consider
\beq 
\calZ[\Jset] = \overline{\prod_{a=1}^n \calZ[J_a;\xi]} .
\eeq 
Assuming a zero-mean Gaussian random potential $\xi$ with variance 
\beq 
\overline{\xi^*(x)\xi(x')}=D_b\delta(x-x') , 
\eeq 
we can explicitly perform the disorder average. This leads to 
\beq
\calZ[\Jset] = \int \calD[\varphiset] \, 
e^{-S[\varphiset]+ \sum_a \int_{x,\tau} J_a\varphi_a } 
\label{partition}
\eeq
with the replicated action\cite{Giamarchi87,Giamarchi88}  
\begin{multline}
S[\varphiset] = \sum_a \int_{x,\tau} \frac{v}{2\pi K} \left\{ (\dx\varphi_a)^2 + \frac{(\dtau\varphi_a)^2}{v^2} \right\} \\
-\calD \sum_{a,b} \int_{x,\tau,\tau'} \cos[2\varphi_a(x,\tau)-2\varphi_b(x,\tau')] , 
\label{action} 
\end{multline}
where $\calD=\rho_2^2 D_b$ and $a,b=1\cdots n$ are replica indices. Note that the disorder part of the action in nonlocal in imaginary time. If we interpret $y=v\tau$ as a space coordinate, the action~(\ref{action}) also describes (two-dimensional) elastic manifolds in a (three-dimensional) disordered medium,\cite{Fisher86,Chauve00,Ledoussal02a,Ledoussal04,Ledoussal06a,Balents04} yet with a periodic structure and a perfectly correlated disorder in the $y$ direction.\cite{Balents93,Giamarchi96,Fedorenko08} The Luttinger parameter, which controls quantum fluctuations in the Bose fluid, defines the temperature of the classical model.\cite{Giamarchi96} 

Introducing the functional $W[\Jset]=\ln \calZ[\Jset]$ defined by 
\beq
\exp\left(W[\Jset]\right) = \overline{\exp\Bigl(\sum_a W[J_a;\xi]\Bigr)} , 
\eeq 
one easily obtains  
\begin{align}
W[\Jset] ={}& \sum_a W_1[J_a]+\half \sum_{a,b} W_2[J_a,J_b] \nonumber \\ & 
+ \frac{1}{3!} \sum_{a,b,c} W_3[J_a,J_b,J_c] + \cdots 
\label{WJ}
\end{align}
where 
\begin{equation}
\begin{split} 
W_1[J_a] &= \overline{W[J_a;\xi]} , \\ 
W_2[J_a,J_b] &= \overline{W[J_a;\xi]W[J_b;\xi]} - \overline{W[J_a;\xi]}\; \overline{W[J_b;\xi]},
\end{split} 
\label{cumdef}
\end{equation}
etc., are the cumulants of the random functional $W[J;\xi]$. Equation~(\ref{WJ}) expresses $W[J]$ as an expansion in increasing number of free replica sums.

\subsection{Scale-dependent effective action} 
\label{subsec_scaledepgamma} 

The strategy of the nonperturbative RG approach is to build a family of models indexed by a momentum scale $k$ such that fluctuations are smoothly taken into account as $k$ is lowered from the UV scale $\Lamb$ down to 0.\cite{Berges02,Delamotte12,Kopietz_book} This is achieved by adding to the action~(\ref{action}) the infrared regulator term 
\begin{equation}
\Delta S_k[\varphiset] = \half \sum_{a,q,\w} \varphi_a(-q,-i\w) R_{k}(q,i\w) \varphi_a(q,i\w) ,
\label{DeltaSk} 
\end{equation}
where $\w\equiv\wn=2\pi n/\beta$ ($n$ integer) is a Matsubara frequency. The cutoff function $R_k(q,i\w)$ is chosen so that fluctuation modes satisfying $|q|,|\w|/v_k\ll k$ are suppressed while those with $|q|\gg k$ or $|\w|/v_k\gg k$ are left unaffected ($v_k$ denotes the $k$-dependent sound-mode velocity); its precise form will be given below (the possibility to choose a nondiagonal function $R_{k,ab}$ is discussed in Appendix~\ref{app_sts}). 

The partition function
\beq
\calZ_k[\Jset] = \int \calD[\varphiset] \, 
e^{-S[\varphiset] - \Delta S_k[\varphiset]+ \sum_a \int_{x,\tau} J_a\varphi_a } 
\eeq
thus becomes $k$ dependent. The expectation value of the field reads 
\beq
\phi_{a,k}[x,\tau;\Jfset] = \frac{\delta\ln \calZ_k[\Jfset]}{\delta J_a(x,\tau)} =\mean{\varphi_a(x,\tau)} 
\eeq 
(to avoid confusion in the indices we denote by $\Jfset$ the $n$ external sources). 
The scale-dependent effective action 
\beq
\Gamma_k[\phiset] = - \ln \calZ_k[\Jset] + \sum_a \int_{x,\tau} J_a \phi_a - \Delta S_k[\phiset]
\eeq
is defined as a modified Legendre transform which includes the subtraction of $\Delta S_k[\phiset]$. Assuming that for $k=\Lamb$ the fluctuations are completely frozen by the $\Delta S_{\Lamb}$ term, $\Gamma_{\Lamb}[\phiset]=S[\phiset]$ as in mean-field theory. On the other hand the effective action of the original model~(\ref{action}) is given by $\Gamma_{k=0}$ provided that $R_{k=0}$ vanishes. The nonperturbative FRG approach aims at determining $\Gamma_{k=0}$ from $\Gamma_{\Lamb}$ using Wetterich's equation\cite{Wetterich93,Ellwanger94,Morris94} 
\beq
\dt \Gamma_k[\phiset] = \half \Tr \left\{ \dt R_k \bigl(\Gamma_k^{(2)}[\phiset] + R_k \bigr)^{-1} \right\} ,
\label{eqwet}
\eeq
where $\Gamma_k^{(2)}$ is the second-order functional derivative of $\Gamma_k$ and $t=\ln(k/\Lamb)$ a (negative) RG ``time''. The trace in~(\ref{eqwet}) involves a sum over momenta and frequencies as well as the replica index. 

A difficulty with the replica formalism is that one must invert the $n\times n$ matrix $\Gamma_k^{(2)}[\phiset] + R_k$ for arbitrary values of the fields $\phi_a$ (which are {\it a priori} all different since the sources $J_a$ are). To circumvent this difficulty we expand the effective action
\begin{align}
\Gamma_k[\phiset] ={}& \sum_a \Gamma_{1,k}[\phi_a] - \half \sum_{a,b} \Gamma_{2,k}[\phi_a,\phi_b] 
\nonumber \\ & 
+ \frac{1}{3!} \sum_{a,b,c} \Gamma_{3,k}[\phi_a,\phi_b,\phi_c] + \cdots 
\label{freerep} 
\end{align}
in increasing number of free replica sums.\cite{Tarjus08} The minus sign in~(\ref{freerep}) is chosen for later convenience. In Appendix~\ref{app_inversion} we show how the free replica sum expansion allows one to systemically obtain the inverse of the matrix $\Gamma_k^{(2)}+R_k$ (or any other matrix).

Since $\Gamma[\phiset]$ is the Legendre transform of $W[\Jset]$, we can relate the $\Gamma_i$'s to the cumulants $W_i$ using 
\begin{align}
\Gamma[\phiset] ={}& \sum_a \int_{x,\tau} J_a\phi_a - \sum_a W_1[J_a] - \half \sum_{a,b} W_2[J_a,J_b] 
\nonumber \\ & - \frac{1}{3!} \sum_{a,b,c} W_3[J_a,J_b,J_c] - \cdots 
\label{Gamma2}
\end{align} 
where the source $J_a$ is a functional of $\phifset$ defined by 
\begin{align}
J_a[x,\tau;\phifset] &= \frac{\delta\Gamma[\phifset]}{\delta\phi_a(x,\tau)} \nonumber \\
&= \frac{\delta\Gamma_1[\phi_a]}{\delta\phi_a(x,\tau)}  
- \sum_b \frac{\delta\Gamma_2[\phi_a,\phi_b]}{\delta\phi_a(x,\tau)}  + \cdots 
\label{J2}
\end{align}  
For simplicity we consider the case $k=0$ (it is however easy to generalize the discussion to the case $k>0$ by subtracting the regulator term $\Delta S_k[\phiset]$ in the rhs of~(\ref{Gamma2})). 
From~(\ref{Gamma2}) and (\ref{J2}), by considering the one-replica term, one obtains 
\beq
\Gamma_1[\phi_a] = - W_1[J_a] + \int_{x,\tau} J_a[x,\tau;\phi_a] \phi_a(x,\tau) , 
\eeq
where
\beq
J_a[x,\tau;\phi_a] = \frac{\delta\Gamma_1[\phi_a]}{\delta\phi_a(x,\tau)} 
\label{Gam1b}
\eeq 
is a functional of the field $\phi_a$. We conclude that $\Gamma_1[\phi_a]$ is the Legendre transform of the first cumulant $W_1[J_a]= \overline{W[J_a;\xi]}$ and thus determines the thermodynamics of the system. Considering now the two-replica term, one easily finds\cite{Tarjus08} 
\beq
\Gamma_2[\phi_a,\phi_b] = W_2[J_a[\phi_a],J_b[\phi_b]] ,
\label{Gam2}
\eeq 
where $J_a[\phi_a]$ is the source defined in~(\ref{Gam1b}). Thus $\Gamma_2[\phi_a,\phi_b]$ is directly the second cumulant of $W[J;\xi]$ (with the proper choice of sources). We shall see that it encodes the existence of metastable states and ``shocks'' (static avalanches). To higher orders, the relation between the $\Gamma_i$'s and the $W_i$'s is more complicated,\cite{Tarjus08} but will be of no use in the following. By an abuse of language the $\Gamma_i$'s will be referred to as the cumulants of the renormalized disorder although this is correct {\it stricto sensu} only for $i\leq 2$.

Another quantity of interest is the disorder-averaged two-point correlation function. From the definition of $W_1[J_a]$ and the fact that its Legendre transform is $\Gamma_1[\phi_a]$, one obtains the connected propagator 
\begin{align} 
G_c(Q)
={}& W_1^{(2)}(Q,-Q) 
= \Gamma_1^{(2)-1}(Q,-Q,\phi) \nonumber \\
={}& \overline{ \mean{\varphi(Q)\varphi(-Q)} } 
- \overline{ \mean{\varphi(Q)} \mean{\varphi(-Q)} }  ,
\label{Gc}
\end{align} 
where 
\beq 
W_1^{(2)}(Q,-Q) = \frac{\delta^2 W_1[J_a]}{\delta J_a(-Q)\delta J_a(Q)} \biggl|_{J_a=0} 
\eeq 
and $Q=(q,i\w)$. The two-point vertex $\Gamma_1^{(2)}$ is computed in a constant (i.e. uniform and time-independent) field configuration $\phi$ corresponding to a vanishing source, i.e. $J_a[\phi]=0$ with $J_a[\phi]$ defined by~(\ref{Gam1b}). Similarly, from the definition of $W_2[J_a,J_b]$ one obtains the disconnected propagator
\begin{align}
G_d(Q)
={}& W_2^{(11)}(Q;-Q) \nonumber \\ 
={}& \overline{ \mean{\varphi(Q)} \mean{\varphi(-Q)} } 
- \overline{ \mean{\varphi(Q)}} \; \overline{ \mean{\varphi(-Q)} } ,
\label{Gd}
\end{align}
with the notation 
\beq 
W_2^{(11)}(Q;-Q) = \frac{\delta^2 W_2[J_a,J_b]}{\delta J_a(-Q)\delta J_b(Q)}\biggl|_{J_a=J_b=0} .
\eeq 
The relation~(\ref{Gam2}) between $W_2$ and $\Gamma_2$ then yields 
\beq
G_d(Q)
= G_c(Q) \Gamma_2^{(11)}(Q,\phi;-Q,\phi) G_c(Q) .
\eeq
A more general discussion of Green functions can be found in Appendix~\ref{app_GF}.

\subsection{RG equations for the disorder cumulants} 

\begin{figure}
\centerline{\includegraphics[width=8cm]{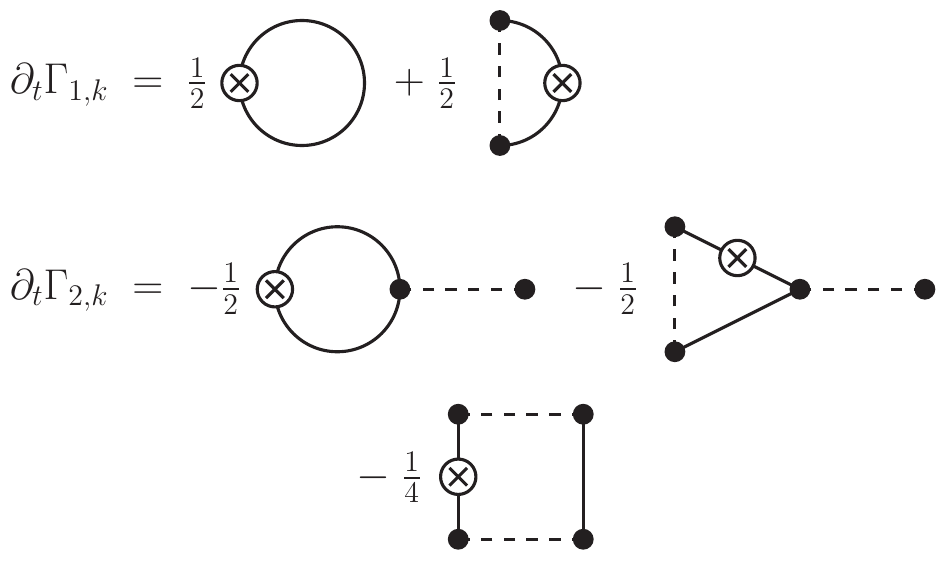}}
\caption{Diagrammatic representation of the RG equations~(\ref{rgeq1}) and (\ref{rgeq2}) satisfied by $\Gamma_{1,k}[\phi_a]$ and $\Gamma_{2,k}[\phi_a,\phi_b]$. The solid line stands for the propagator $P_k$, the cross for $\dt R_k$, and $2$ black dots attached to $n$ and $m$ propagators, respectively, and connected by dashed lines for $\Gamma^{(n,m)}_{2,k}$. (Symmetric diagrams, obtained by moving the cross to a different propagator line or exchanging the replica indices $a$ and $b$, are not shown.)}
\label{fig_rgeq} 
\end{figure}

RG equations for the cumulants $\Gamma_{i,k}[\phi_{a_1},\cdots,\phi_{a_i}]$ are obtained by inserting the free replica sum expansion~(\ref{freerep}) into the exact flow equation~(\ref{eqwet}). The propagator $G_k[\phiset]=(\Gamma^{(2)}_k[\phiset]+R_k)^{-1}$ can be written as a free replica sum expansion and related to $\Gamma_k^{(2)}[\phiset]$. Ignoring the cumulants $\Gamma_{i,k}$ with $i\geq 3$, one finds (see Appendix~\ref{app_inversion}),\cite{Tarjus08} 
\begin{align}
\dt \Gamma_{1,k}[\phi_a] ={}& \half \Tr \llbrace \dt R_k P_k[\phi_a] \rrbrace \nonumber \\ & 
- \half \tdt \Tr \llbrace P_k[\phi_a] \Gamma_{2,k}^{(11)}[\phi_a,\phi_a] \rrbrace 
\label{rgeq1}
\end{align}
and 
\begin{multline}
\dt \Gamma_{2,k}[\phi_a,\phi_b] = \half \tdt \Tr\Bigl\{ 
P_k[\phi_a] \Bigl[ \Gamma_{2,k}^{(20)}[\phi_a,\phi_b] \\
+ \Gamma_{2,k}^{(20)}[\phi_a,\phi_b] P_k[\phi_a] \Gamma_{2,k}^{(11)}[\phi_a,\phi_a] \\
+ \half \Gamma_{2,k}^{(11)}[\phi_a,\phi_b] P_k[\phi_b] \Gamma_{2,k}^{(11)}[\phi_b,\phi_a] 
+ \mbox{perm}(a,b) \Bigr] \Bigr\} , 
\label{rgeq2}
\end{multline}
where 
\beq 
P_k[\phi_a] = \bigl( \Gamma_{1,k}^{(2)}[\phi_a] + R_k \bigr)^{-1}
\eeq 
and ${\rm perm}(a,b)$ denotes all terms obtained by exchanging the replica indices $a$ and $b$. We have introduced the operator $\tdt=(\dt R_k)\partial_{R_k}$ acting only on the time dependence of $R_k$ and used $\tdt P_k[\phi_a]=-P_k[\phi_a](\dt R_k)P_k[\phi_a]$. Equations~(\ref{rgeq1}) and (\ref{rgeq2}) are represented diagrammatically in Fig.~\ref{fig_rgeq}. 

\subsection{Truncation scheme} 

The exact flow equation~(\ref{eqwet}) cannot be solved exactly and one has to resort to approximations. In the following we consider an ansatz for the effective action which includes only $\Gamma_{1,k}$ and $\Gamma_{2,k}$,
\begin{equation}
\begin{split} 
&\Gamma_{1,k}[\phi_a] = \int_{x,\tau} \llbrace \frac{Z_{x}}{2} (\dx\phi_a)^2 + \half \phi_a \Delta_k(-\dtau) \phi_a \rrbrace , \\ 
&\Gamma_{2,k}[\phi_a,\phi_b] = \int_{x,\tau\,\tau'} V_k(\phi_a(x,\tau)-\phi_b(x,\tau')) .
\end{split}
\label{ansatz}
\end{equation}
A similar truncation of the effective action has been used with success in the study of the random-field Ising model\cite{Tarjus08,Tissier08} as well as in the perturbative FRG approach to classical disordered systems where it becomes controlled, within an epsilon expansion, near the upper critical dimension. In most disordered systems studied so far, it is the minimum truncation that captures the physics of metastable states.\cite{Tarjus19} 

The initial condition $\Gamma_{\Lamb}[\{\phi_a\}]=S[\{\phi_a\}]$ implies $\Delta_{\Lamb}(i\w)=\w^2/\pi vK$ and $V_{\Lamb}(u)=2\calD \cos(2u)$. The $\pi$-periodic function $V_k(u)$ directly gives the renormalized second cumulant of the disorder. The form of $\Gamma_{1,k}$ and $\Gamma_{2,k}$ are strongly constrained by the statistical tilt symmetry\cite{Schulz88} (STS) discussed in Appendix~\ref{app_sts}. In particular $Z_{x}=v/\pi K$ remains equal to its initial value and no higher-order space derivatives are allowed. As for the part involving time derivatives, we assume a quadratic form with an unknown ``self-energy'' $\Delta_k(i\w)$ satisfying $\Delta_k(i\w=0)=0$ as required by the STS.\cite{not88} The infrared regulator ensures that the self-energy $\Delta_k(i\w)$ is a regular function of $\w$ near $\w=0$ and can therefore be written as $\Delta_k(i\w)=Z_x\w^2/v_k^2+\calO(\w^4)$. In addition to the running velocity $v_k$ one may define a $k$-dependent Luttinger parameter by $Z_x=v_k/\pi K_k$. The popular derivative expansion, which amounts to approximating $\Delta_k(i\w)=\w^2/v_k^2$ by a quadratic function, is questionable in the present case (as will be discussed in more detail in Sec.~\ref{subsubsec_self_energy}). The STS also ensures that the two-replica potential $V_k(\phi_a,\phi_b)$ is a function of $\phi_a-\phi_b$ only. Note that in random elastic manifold (classical) models with short-range correlated disorder, the STS implies that $\Gamma_{1,k}$ is not renormalized.\cite{Balog19a} In the quantum model, because the disorder part of the replicated action~(\ref{action}) is nonlocal in imaginary time, the dynamic part of $\Gamma_{1,k}$ is not constrained. 

As for the cutoff function we take 
\beq 
R_k(q,i\w) = ( Z_x q^2 + \Delta_k(i\w)) r\left( \frac{Z_x q^2 + \Delta_k(i\w)}{Z_xk^2} \right) ,
\label{Rk}  
\eeq
where $r(x)=\alpha/(e^x-1)$ with $\alpha$ a free parameter of order unity. Thus the regulator term $\Delta S_k[\varphi]$ suppresses fluctuations such that $q^2\ll k^2$ and $\Delta_k(i\w)\ll Z_xk^2$ but leaves unaffected those with $q^2\gg k^2$ or $\Delta_k(i\w)\gg Z_xk^2$, i.e., with $|q|\gg k$ or $|\w|\gg v_kk$. 

\subsubsection{Propagators}

From Eqs.~(\ref{Gc},\ref{Gd}) and the ansatz~(\ref{ansatz}) one obtains the propagators 
\begin{align}
G_{c,k}(Q) &= \frac{1}{Z_xq^2 +\Delta_k(i\w) + R_k(Q)} , \label{Gc1}\\ 
G_{d,k}(Q) &= - \beta\delta_{\w,0} \frac{V''_k(0)}{[Z_xq^2+ R_k(q,0)]^2}  \label{Gd1}
\end{align}
(see Appendix~\ref{app_vertices}).

\subsubsection{RG equations} 

By inserting the ansatz~(\ref{ansatz}) into the flow equation~(\ref{eqwet}) we obtain coupled RG equations for $\Delta_k(i\w)$ and $V_k(u)$. In practice it is convenient to define dimensionless variables $\tilde q=q/k$, $\tilde\w=\w/v_kk$, and introduce the dimensionless functions
\begin{align}
\delta_k(u) &= - \frac{K^2}{v^2} \frac{V_k''(u)}{k^3} , \label{dimvara} \\ 
\tilde \Delta_k(i\tw) &= \frac{\Delta(i\w)}{Z_x k^2} .\label{dimvarb} 
\end{align}
The flow equations then read 
\begin{align} 
\dt\delta_k(u) ={}& -3 \delta_k(u) - K_k l_1 \delta''_k(u) \nonumber \\ & 
+ \pi \bar l_2 [ \delta_k''(u) (\delta_k(u)-\delta_k(0)) + \delta'_k(u)^2 ] , \label{rgeq3a} \\ 
\dt \tilde\Delta_k(i\tw) ={}& - 2 \tilde\Delta_k(i\tw) + z_k \tw \partial_{\tw} \tilde\Delta_k(i\tw)  \nonumber \\ &
- \pi \delta''_k(0) [ \bar l_1(i\tw) - \bar l_1(0) ] , \label{rgeq3b} \\ 
\dt K_k ={}& \theta_k K_k , \qquad
\dt (K_k/v_k) = 0 , \label{rgeq3c} 
\end{align}
where $z_k=1+\theta_k$ is the running dynamical critical exponent and  
\begin{equation}
\theta_k = \frac{\pi}{2} \delta''_k(0) \bar m_\tau . 
\label{thetak}
\end{equation}
The derivation of Eqs.~(\ref{rgeq3a}-\ref{thetak}) is detailed in Appendix~\ref{app_rgeq} and the ``threshold'' functions $l_1,\bar l_2,\bar l_1(i\tw),\bar m_\tau$  are defined in Appendix~\ref{app_threshold}. Except for $\bar l_2$, the threshold functions depend on $k$. 

To numerically solve Eqs.~(\ref{rgeq3a}-\ref{thetak}), we expand $\delta_k(u)$ in circular harmonics,
\beq
\delta_k(u) = \sum_{p=1}^{p_{\rm max}}\delta_{p,k} \cos(2pu) ,
\label{deltaharm} 
\eeq 
with typically $p_{\rm max}$ in the range $[300,400]$. Note that the zeroth order harmonics $\delta_{0,k}$ always vanishes. To compute the dimensionless self-energy $\tDelta_k(i\tw)$ we use a $\tw$ grid of 100 points with $\tw\in[0,4]$. For $\Delta_k(i\w)$, the $\w$ grid contains 5000 points with $\w\in[0,\Lambda]$. The flow equations are integrated using fourth-order Runge-Kutta method with adaptative step size.

\section{Superfluid--Bose-glass transition} 
\label{sec_transition}

In the absence of disorder ($\calD_{\Lamb}=0$) the system is a one-dimensional superfluid with a sound-mode velocity $v$ and a Luttinger parameter $K$. To study the stability of this phase and the transition to the Bose-glass phase it is sufficient to approximate $\Delta_k(i\w)=Z_x\w^2/v_k^2$ and $V_k(u)=2\calD_k\cos(2u)$, i.e. $\delta_k(u)=\delta_{1,k}\cos(2u)$ with $\delta_{1,k}=8K^2\calD_k/v^2k^3$. This gives
\beq
\begin{split}
\dt \delta_{1,k} &= (-3+4K_kl_1)\delta_{1,k} + 4\pi \bar l_2 \delta_{1,k}^2 , \\
\theta_k &= -2\pi \bar m_\tau \delta_{1,k} .
\end{split}
\label{flow1} 
\eeq  
For $\theta_k=0$, $K_k=K$, $v_k=v$ and $\Delta_k(i\w)=\w^2/\pi vK$ (i.e. $\tilde\Delta_k(i\tw)=\tw^2$), one has $l_1=1/2$ regardless of the cutoff function $R_k$ (see Appendix~\ref{app_threshold}). We conclude that the superfluid phase is destabilized by an infinitesimal disorder when $K<3/2$. This conclusion may be drawn more directly from the scaling dimension $[\calD]=3-2K$ of the disorder strength in the superfluid phase.\cite{not140}

In the vicinity of $(K_k=3/2,\delta_{1,k}=0)$, to leading order Eqs.~(\ref{flow1}) yield  
\beq
\begin{split}
\dt \delta_{1,k} &= 2\delta_{1,k}(K_k-3/2) , \\ 
\dt K_k &= -3\pi \bar m_\tau \delta_{1,k},
\end{split}
\label{flow2} 
\eeq
where $\bar m_\tau\equiv \bar m_\tau|_{\theta_k=0}<0$. These equations are similar to those obtained in Refs.~\onlinecite{Giamarchi87,Giamarchi88,Ristivojevic12,Ristivojevic14}. They imply that the flow trajectories are parabolic, 
\beq
\delta_{1,k} = - \frac{1}{3\pi \bar m_\tau} \left(K_k-\frac{3}{2} \right)^2 + C ,
\label{flow3} 
\eeq 
where the constant $C$, which depends on the initial values $K_{\Lamb}=K$ and $\delta_{1,\Lamb}$, vanishes for the critical trajectory passing through the critical point $(K_k=3/2,\delta_{1,k}=0)$. Taking ($K_k,\sqrt{\delta_{1,k}}$) as variables, Eqs.~(\ref{flow2}) and (\ref{flow3}) reproduce the flow equations of the BKT transition.\cite{[{See, e.g., }]Chaikin_book}

For all trajectories that do not end up in the superfluid phase, the disorder strength $\delta_{1,k}$ rapidly diverges. The perturbative RG becomes uncontrolled once $\delta_{1,k}\sim 1$ and is of little use to understand the physical properties of the Bose-glass phase.

\section{Bose-glass phase}
\label{sec_bg}

\begin{figure}
	\centerline{\includegraphics[width=7.5cm,angle=0]{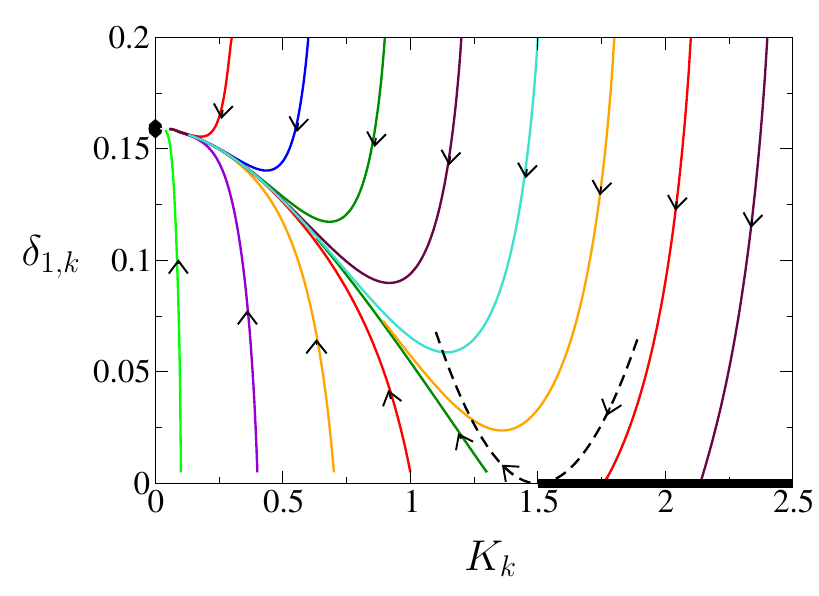}}
	\caption{Flow diagram projected onto the plane $(K_k,\delta_{1,k})$, where $K_k$ is the running Luttinger parameter and $\delta_{1,k}$ is the first harmonic of the dimensionless potential $\delta_k(u)$ defined by~(\ref{dimvara}) and (\ref{deltaharm}). The thick solid line $(K_k\geq 3/2,\delta_{1,k}=0)$ shows the attractive line of fixed points corresponding to the superfluid phase and the black dot $(K^*=0,\delta^*_1\simeq 0.159)$ the attractive fixed point corresponding to the Bose-glass phase. The dashed line shows the critical trajectory~(\ref{flow3}) (with $C=0$) obtained from perturbative RG.}
	\label{fig_flow_diagram} 
\end{figure}

The nonperturbative FRG approach allows us to follow the flow into the strong-disorder regime. All trajectories that do not end up in the superfluid phase ($\lim_{k\to 0}K_k>3/2$ and $\lim_{k\to 0}\delta_k(u)=0$) are attracted by a fixed point characterized by a vanishing Luttinger parameter $K^*=0$ and a two-replica potential $\delta^*(u)$ (Fig.~\ref{fig_flow_diagram}). The vanishing of $K_k\sim k^\theta$ is controlled by an exponent $\theta=\lim_{k\to 0}\theta_k=z-1$ which is related to the dynamical critical exponent $z$ of the Bose-glass phase. 

The vanishing of the Luttinger parameter $K_k$ has important consequences. First, it implies that
the renormalized superfluid stiffness
\beq 
\rho_{s,k}= \frac{v_k K_k}{\pi}= \frac{v}{\pi K} K_k^2 \sim k^{2\theta} 
\label{rhosk}
\eeq 
and the charge stiffness (or Drude weight, i.e. the weight of the zero-frequency delta peak in the conductivity, see Sec.~\ref{subsubsec_conductivity})
\beq
D_k = v_k K_k = \frac{v}{K} K_k^2 \sim k^{2\theta} 
\label{Dk} 
\eeq
vanish for $k\to 0$, whereas the compressibility,
\beq
\kappa = \frac{K_k}{\pi v_k} = \frac{K}{\pi v} = \frac{1}{\pi^2 Z_x} ,
\eeq
is unaffected by disorder. The fact that $\rho_{s,k}$ and $D_k$ are nonzero at any finite scale $k>0$ can be related to the infinite superfluid susceptibility.\cite{Fisher89} 
Second, it shows that quantum fluctuations are suppressed at low energies. We thus expect the phase field $\varphi(x,\tau)$ to have weak temporal (quantum) fluctuations and to adjust its
value in space so as to minimize the energy due to the random potential, a hallmark of pinning (this point is further discussed in Sec.~\ref{subsec_droplet}).

\begin{figure}
\centerline{\includegraphics[width=6cm]{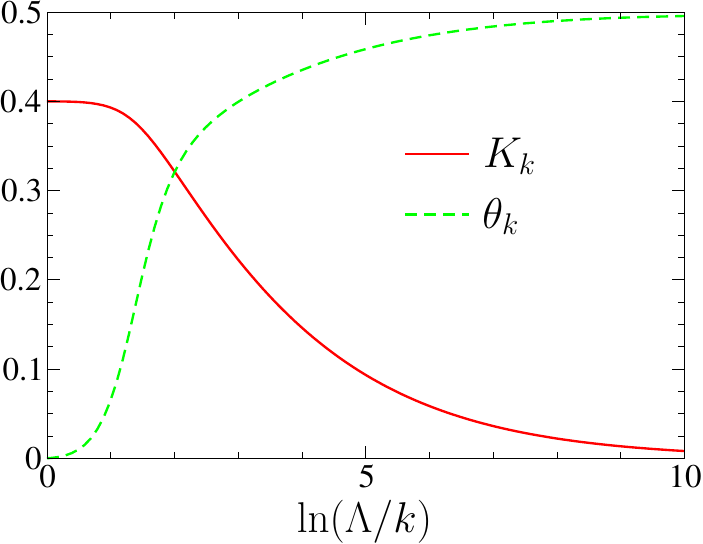}}
\caption{$K_k$ and $\theta_k$ vs $k$ in the Bose-glass phase ($K=0.4$). The value $\theta=\lim_{k\to 0}\theta_k=1/2$ is obtained from a fine tuning of the coefficient $\alpha$ in the definition~(\ref{Rk}) of the cutoff function $R_k$ (see text).} 
\label{fig_Ktheta}
\end{figure}

It is difficult to predict precisely the values of $z$ and $\theta$, which turn out to be sensitive to the RG procedure (see Sec.~\ref{subsubsec_conductivity}). Figure~\ref{fig_Ktheta} is obtained for $\theta=1/2$, i.e $z=3/2$ (this value is obtained by a fine-tuning of the coefficient $\alpha$ in the definition~(\ref{Rk}) of the cutoff function $R_k$). 

\subsection{Disorder correlator: Cusp and QBL} 
\label{subsec_cuspQBL}

\subsubsection{Numerical determination of the QBL} 

The fixed-point potential $\delta^*(u)$ is defined by $\dt\delta^*(u)=0$, i.e.  
\beq
-3 \delta^*(u) + \pi \bar l_2 [ \delta^*{}''(u)(\delta^*(u)-\delta^*(0)) + \delta^*{}'(u)^2]  
= 0 
\label{deltafp}
\eeq
since $K^*=0$. Equation~(\ref{deltafp}) admits a nontrivial $\pi$-periodic solution,
\beq
\delta^*(u) = \frac{1}{2\pi\bar l_2} \left[ \left( u - \frac{\pi}{2} \right)^2 - \frac{\pi^2}{12} \right] , \quad u \in [0,\pi] ,
\label{deltafpsol}
\eeq 
which exhibits cusps at $u=n\pi$ ($n$ integer). 

To study the stability of the fixed-point solution~(\ref{deltafpsol}) we write 
\beq 
\delta_k(u) = \delta^*(u) + g(u) e^{\lamb t} 
\eeq 
with $g(u)$ a $k$-independent $\pi$-periodic function and linearize the flow equation $\dt\delta_k(u)$ (with $K_k=0$) about $\delta^*(u)$, 
\beq
\lamb g(u) = - g(0) -2 g(u) + \half  u(u-\pi) g''(u) + (2u-\pi) g'(u) . 
\label{g1} 
\eeq
We then expand $g(u)=\sum_{p=1}^{p_{\rm max}} g_p\cos(2pu)$ in circular harmonics up to a given order $p_{\rm max}$ so that Eq.~(\ref{g1}) becomes a matrix equation $\lamb g_p = \sum_{p'} M_{pp'}g_{p'}$. It is easily seen that the zeroth-order component $g_0$ is not generated by the flow equation. Diagonalizing numerically the matrix $M$ we find that all eigenvalues $\lamb$ are positive: All perturbations are therefore irrelevant since an infinitesimal $K$ also flows to zero when $k\to 0$ (recall that the RG ``time'' $t$ is negative and the infrared fixed point corresponds to $t=\ln(k/\Lamb)\to-\infty$). The smallest eigenvalue seems to converge to 3 for $p_{\rm max}\to\infty$. 

\begin{figure}
	\centerline{\includegraphics[width=3.7cm,angle=0]{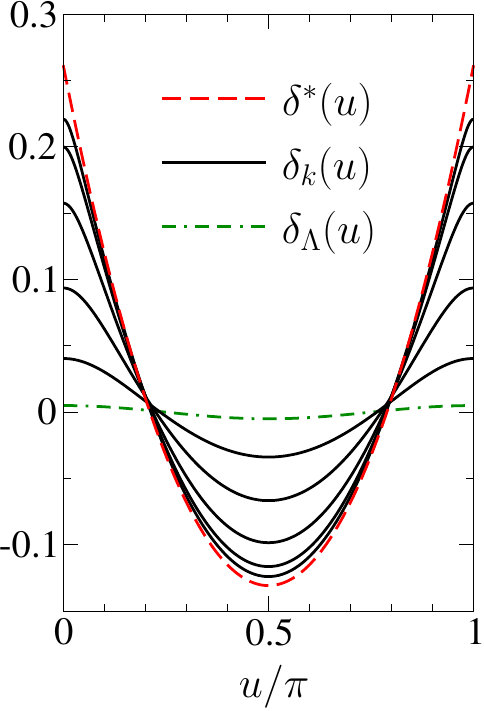}
		\hspace{0.25cm}
		\includegraphics[width=3.85cm,angle=0]{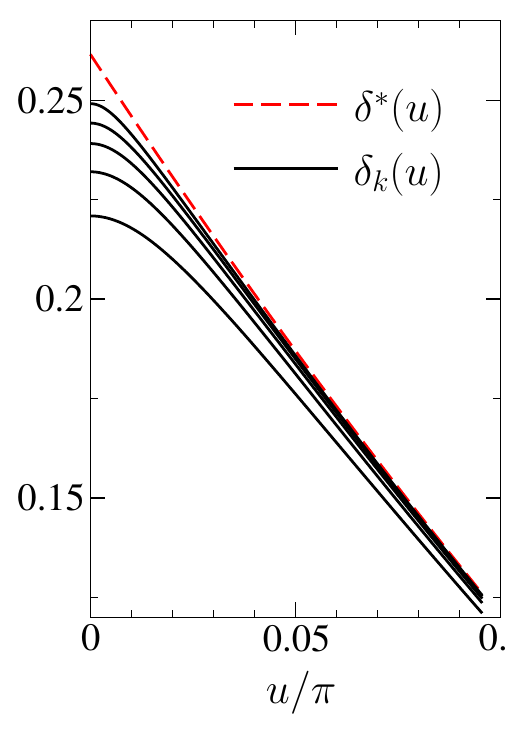}}
	\caption{(Left) Potential $\delta_k(u)=-(K^2/v^2k^3)V_k''(u)$ for various values of $k$ ($K=0.4$ and $\delta_{1,\Lamb}=0.005$). The green curve shows the initial condition $\delta_\Lamb(u)=\delta_{1,\Lamb} \cos(2u)$ and the red one the fixed-point solution~(\ref{deltafpsol}).
		(Right) $\delta_k(u)$ for $u$ near 0 showing the formation of the QBL ($k/\Lamb\simeq 0.050/0.030/0.018/0.011/0.007/0$ from bottom to top).} 
	\label{fig_pot} 
\end{figure}

\begin{figure}
	\centerline{\includegraphics[width=7cm]{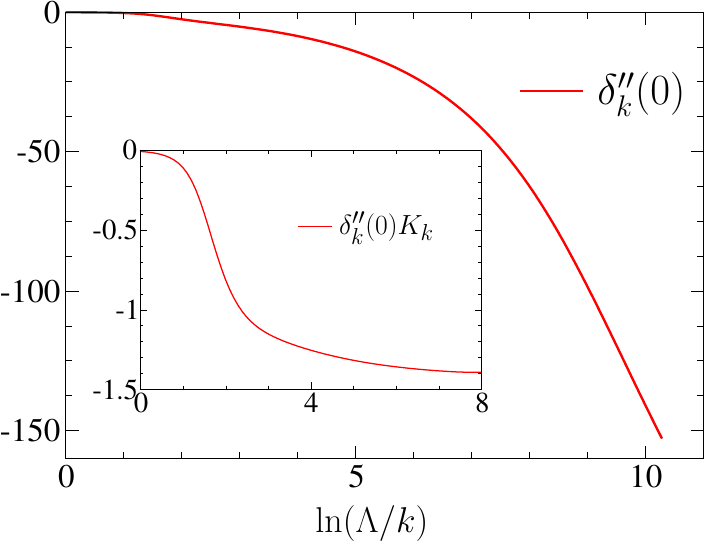}}
	\caption{$\delta_k''(0)$ vs $k$ for $K=0.4$. The inset shows that $\delta_k''(0)\simeq -C/K_k$(with $C>0$) when $k\to 0$.}
	\label{fig_QBL} 
\end{figure}

For any nonzero momentum scale $k$, the cusp singularity is rounded into a QBL as shown in Fig.~\ref{fig_pot}: for $u$ near 0, $\delta_k(u)-\delta_k(0)\propto -|u|$ except in a boundary layer of size $|u|\sim K_k$; as a result the curvature $\delta''_k(0)\simeq -C/K_k \sim 1/K_k\sim k^{-\theta}$ diverges when $k\to 0$ (Fig.~\ref{fig_QBL}).

\subsubsection{Analytic expression of the QBL}

An analytic expression of the QBL can be obtained if we compute the threshold function $l_1$ with the propagator $\tilde P_k$ [Eq.~(\ref{Ptilde})] where the self-energy $\tDelta_k(i\tw)$ is approximated by its low-energy limit $\tw^2$. This derivative-expansion (DE) approximation is justified when $\tDelta_k(i\tw)\simeq \tw^2$ in the frequency range (typically $[0,4]$) selected by the cutoff function $R_k(q,i\w)$. In the present problem, $\tDelta_k(i\tw)\simeq \tw^2$ holds only for $|\tw|\lesssim 1$ in the small-$k$ limit (see Fig.~\ref{fig_tDelta} below), but the DE is nevertheless a good approximation to compute $\delta_k(u)$, $K_k$ and $\theta_k$. To obtain accurately the frequency-dependent self-energies $\tDelta_k(i\tw)$ and $\Delta_k(i\w)$, it is necessary to use the full self-energy $\tDelta(i\tw)$ in the propagator $\tilde P_k$.

The DE approximation makes $l_1$ independent of $k$ ($\bar l_2$ being always $k$ independent), provided that we also approximate $z_k$ by its fixed-point value $z=\lim_{k\to 0}z_k$, which considerably simplifies the flow equation of $\delta_k(u)$. In the small-$k$ limit where $K_k\to 0$, we look for a solution in the form 
\beq
\delta_k(u) = \delta_k(0) + K_k f\left( \frac{u}{K_k} \right) 
\label{deltaQBL}
\eeq 
near $u=0$ but with an arbitrary value of the ratio $u/K_k$. The $k$-independent even function $f(x)$ satisfies $f(0)=f'(0)=0$ and $f''(0)<0$. From~(\ref{rgeq3a}) we obtain 
\beq
\dt \delta_k(0) \simeq -3 \delta_k(0) - l_1 f'' + \pi \bar l_2 (f''f+f'{}^2) 
\eeq 
using $K_k\to 0$. The rhs must be independent of $x=u/K_k$ and equal to $-3 \delta_k(0)-l_1f''(0)$ since $f(0)=f'(0)=0$, i.e., 
\beq 
- l_1 f'' + \pi \bar l_2 (f''f+f'{}^2) = -l_1f''(0) .
\eeq 
This yields 
\beq 
f(x) = \frac{l_1}{\pi\bar l_2} \left[ 1 - \left( 1- \frac{\pi\bar l_2}{l_1} f''(0)x^2\right)^{1/2} \right] . 
\label{fdef}
\eeq 
For $|u|\ll K_k$, $\delta_k(u)-\delta_k(0)=\calO(u^2)$ while 
\beq
\delta_k(u) = \delta_k(0) - \left( \frac{l_1}{\pi\bar l_2}|f''(0)| \right)^{1/2} |u| \quad \mbox{for} \quad  |u|\gg K_k
\eeq 
in agreement with the cusp formation obtained from the numerical solution of the flow equation (Fig.~\ref{fig_pot}). An expression of the (thermal) boundary layer similar to~(\ref{deltaQBL},\ref{fdef}) has been obtained in the random-field Ising model.\cite{Tissier08}

\subsubsection{Classical limit and Larkin length}

In the classical limit, defined by $K\to 0$ with $\calD K^2$ fixed, $\delta_k''(u)$ satisfies the equation 
\begin{align}
\dt \delta_k''(u) ={}& - 3\delta_k''(u) 
+ \pi \bar l_2 \bigl\{ \delta^{(4)}_k(u) [\delta_k(u)-\delta_k(0)] \nonumber \\ 
& + 4 \delta^{(3)}_k(u) \delta'_k(u) + 3 \delta''_k(u)^2 \bigr\} . 
\end{align} 
As long as $\delta_k(u)$ is a regular (even) function of $u$, $\delta'_k(0)=\delta^{(3)}_k(0)=0$ and 
\beq
\dt \delta''_k(0) = - 3  \delta''_k(0) + 3 \pi \bar l_2 \delta''_k(0)^2 , 
\label{dtdeltapp}
\eeq 
with the initial condition $\delta''_{\Lamb}(0)=-32\calD K^2/v^2 \Lamb^3$. Equation~(\ref{dtdeltapp}) admits the solution 
\beq
\delta''_k(0) = \frac{\delta''_{\Lamb}(0)}{\bar k^3-(\bar k^3-1) \pi \bar l_2 \delta''_{\Lamb}(0)} ,
\eeq
where $\bar k=k/\Lamb$. Similarly to what has been observed in zero-temperature classical disordered systems,\cite{Fisher85,Ledoussal04,Tissier08} the curvature at $u=0$ diverges at a finite momentum scale $k_c=1/L_c$ corresponding to the so-called Larkin length\cite{Larkin70,Larkin79}  
\beq 
L_c = \left( \frac{v^2}{32 \pi \bar l_2 \calD K^2} \right)^{1/3}  
\label{Larkinlength}
\eeq 
(we have assumed $\calD K^2/v^2\Lamb^3\ll 1$).\cite{not160,not161} Perturbation theory is valid only at length scales $L\ll L_c$. Beyond the Larkin length, the physics is dominated by the existence of metastable states.\cite{Ledoussal04} When $K$ is nonzero, the cusp singularity appears only for $k=0$ but the metastable states show up in the QBL that forms at finite $k$ as we discuss in the following sections.

\subsubsection{Physical meaning of the cusp and the QBL} 
\label{subsubsec_cuspQBLmeaning}

The physics of the cusp and the QBL can be in part understood from the analogy, pointed out in Sec.~\ref{subsec_model}, between a disordered one-dimensional Bose fluid and a random elastic manifold (classical) system with correlated disorder in one direction, the temperature $T$ in the latter corresponding to the Luttinger parameter $K$ in the former. 

In many classical disordered systems, e.g. charge-density waves, elastic manifolds or the random-field Ising model,\cite{Narayan92,Narayan92a,Balents93,Ledoussal04,Tissier08} disorder leads to a zero-temperature fixed point with a cuspy two-replica potential $\delta^*(u)$. The cusp is known to be related to the existence of many metastable states leading to ``shock'' singularities (or static avalanches):\cite{Balents96,Tissier08,Ledoussal09} When the system is subjected to an external force, the ground state varies discontinuously whenever it becomes degenerate with a metastable state (which then becomes the new ground state). At finite temperatures, the system has a non-negligible probability to be in two distinct (nearly degenerate) configurations and the discontinuity is smeared on a scale given by $T$. This explains why the cusp in the disorder correlator $\delta_k(u)$ is rounded into a thermal boundary layer at nonzero scales $k>0$ where the (renormalized) temperature $T_k$ is nonzero.

A similar interpretation holds in the Bose-glass phase. To understand this point, let us first consider the system in the absence of external source ($J=0$). Since the disorder strength grows under renormalization whereas $K$ decreases, a semiclassical approach is justified when looking at low-energy properties. For a given configuration of disorder, there are infinitely many classical ground states defined by $\varphi(x,\tau)=\varphi_0(x)+p\pi$ ($p$ integer). The quantum tunneling between two of these states, e.g. $\varphi(x,\tau)=\varphi_0(x)$ and $\varphi(x,\tau)=\varphi_0(x)+\pi$, can be taken into account by instantons involving the creation of a soliton-antisoliton (i.e., a kink-antikink) pair. Typical instantons play an important role in the nonlinear dc transport~\cite{Nattermann03,Malinin04} but yield an exponentially small contribution to the low-frequency conductivity $\sigma(\w)$.\cite{Rosenow06} However, in sufficiently large systems, it is possible to find a soliton and an antisoliton at a distance $L$ apart with exceptionally low excitation energy.\cite{Rosenow06,Nattermann07} This implies the existence of low-lying metastable states obtained from a particular classical ground state by shifting $\varphi$ by $\pm\pi$ in the region between the soliton and the antisoliton. Spontaneous formation of instantons, describing quantum tunneling between the ground states and these rare metastable states, is important and must be taken into account.\cite{Rosenow06} In Refs.~\onlinecite{Rosenow06,Fogler02}, it was shown that these rare instantons, which describe the hopping of a soliton over the distance $L$ (i.e., in the particle language, the hopping of a boson), are responsible for the Mott-Halperin conductivity $\Re[\sigma(\w)]\sim \w^2\ln^2(\w)$ of free fermions or hard-core bosons (corresponding to $K=1$),\cite{Mott68,*not220,Giamarchi_book} a result that will be reproduced in Sec.~\ref{subsubsec_conductivity}. In the presence of an arbitrary external source $J$, the invariance of the action under $\varphi(x,\tau)\to\varphi(x,\tau)+p\pi$ is lost and the classical ground state is in general unique. When considering the evolution of the system under a change of the applied source, one expects abrupt switches at a set of discrete, sample-dependent, values of the source, where the classical ground state becomes degenerate with a metastable state (which then becomes the new ground state). A nonzero value of $K$ leads to quantum tunneling between the ground state and the low-lying metastable states ({\it via} the spontaneous formation of instantons discussed above), which smears the discontinuous change of ground state on a scale given by $K$. Thus quantum fluctuations are responsible for the rounding of the cusp in $\delta_{k}(u)$ into a QBL at nonzero scales $k>0$ where the renormalized Luttinger parameter $K_k>0$.

It is interesting to rephrase this discussion in a fully quantum picture. Quantum fluctuations between the classical ground state and the low-lying metastable states give rise to a (renormalized) ground state and low-energy quantum states that can be viewed as consisting of quantum soliton and antisoliton at a distance $L$ apart. The energy of the various quantum states, and in particular the ground state, depends on the external source $J$. If the source Hamiltonian commutes with the system Hamiltonian, there will be discontinuous changes in the ground state, i.e. first-order quantum phase transitions due to level crossings, when the source is varied. It is however clear that in a disordered macroscopic system and for a generic external perturbation the source and system Hamiltonians do not commute. In that case, as the source is varied there will  be avoided level crossings suppressing the cuspy behavior of the ground state energy. 

Let us finally mention that, given the quantum tunneling between the metastable classical states, one expects the existence of ``superfluid'' domains with significant density fluctuations and therefore reduced fluctuations (i.e. a finite rigidity) of the phase $\hat\theta$ of the boson operator $\hat\psi=e^{i\hat\theta}\sqrt{\hat\rho}$. These superfluid regions, which exist at all length scales, provide us with a natural explanation of the fact that the superfluid stiffness $\rho_{s,k}$ and the Drude weight $D_k$ vanish for $k\to 0$, as expected for a localized phase, but are nonzero for any finite scale $k$ [See Eqs.~(\ref{rhosk},\ref{Dk})].

\subsection{Dynamics}
\label{subsec_dynamics} 

The disordered Bose fluid differs from its classical counterparts with regard to the dynamics. In classical systems the study of the dynamics requires to consider an equation of motion, e.g. a Langevin equation, in addition to the Hamiltonian. In quantum systems the equilibrium dynamics can be retrieved from the Euclidean action after a Wick rotation from imaginary to real time. The presence of $\delta_k''(0)$, which diverges for $k\to 0$, in the flow equation~(\ref{rgeq3b}) of the self-energy suggests that the QBL plays an essential role in the dynamics.

\subsubsection{Self-energy}
\label{subsubsec_self_energy}

\begin{figure}
	\centerline{\includegraphics[width=7cm]{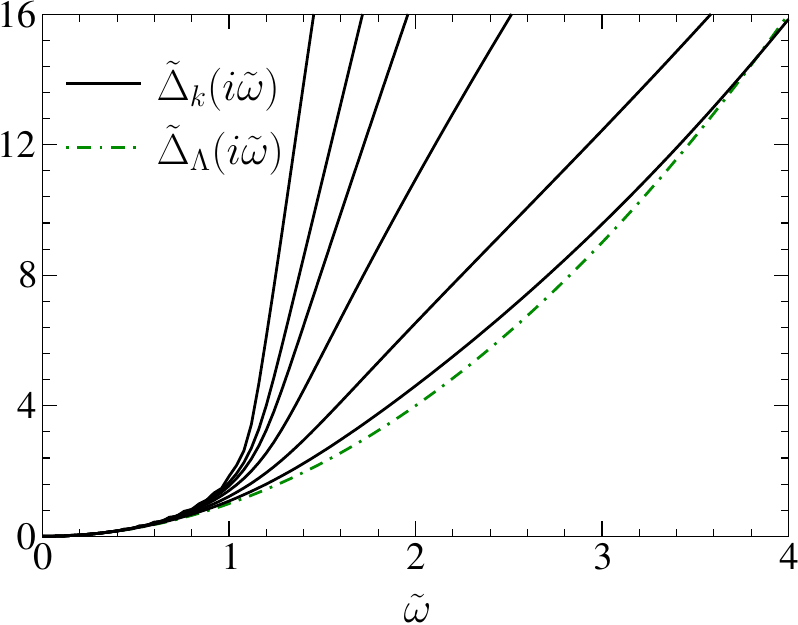}}
	\caption{Dimensionless self-energy $\tDelta_k(i\tw)$ for $k/\Lamb\simeq 1/0.223/0.135/0.082/0.050/0.030/0.011$ (from bottom to top) in the case where $z=1+\theta=3/2$ ($K=0.4$ and $\delta_{1,\Lamb}=0.005$). The green dashed-dotted line shows the initial condition $\tDelta_\Lamb(i\tw)=\tw^2$.}
	\label{fig_tDelta}
\end{figure}

\begin{figure}
	\centerline{\includegraphics[width=7cm]{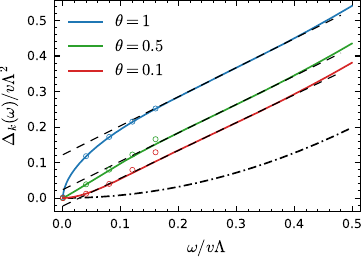}} 
	\caption{Low-frequency behavior of the self-energy $\Delta_k(i\w)$ for $k/\Lamb\simeq e^{-10}$ and $\theta=1/0.5/0.1$. The dash-dotted line shows the initial condition $\Delta_\Lamb(i\w)=\w^2/\pi v K$, the circles correspond to $B'|\w|^{(2-\theta)/z}$ and the dashed lines to $\Delta_k(i\w)=A+B|\w|$.} 
	\label{fig_Delta} 
	\centerline{\includegraphics[width=7cm]{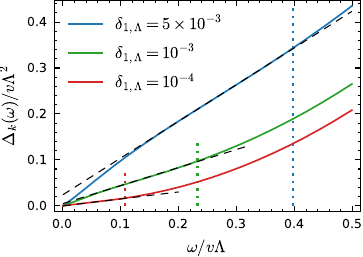}} 
\caption{Same as Fig.~\ref{fig_Delta} but for $\theta=0.5$ and three different values of the disorder. The vertical dotted lines show the pinning frequency $\w_p=v/L_c$ determined by the Larking length~(\ref{Larkinlength}).}
\label{fig_Delta1} 
\end{figure}

Figure~\ref{fig_tDelta} shows the dimensionless self-energy $\tDelta_k(i\tw)$ for various values of $k$ obtained from the numerical solution of~(\ref{rgeq3b}). In the small-frequency limit $|\tw|\ll 1$, $\tDelta_k(i\tw)=\tw^2+\calO(\tw^4)$. This is a consequence of the regulator term~(\ref{DeltaSk}), which ensures that all vertices are regular functions of $\tilde q^2$ and $\tw^2$ in the low-energy limit, and the definition of the dynamical critical exponent $z_k$ which implies $\partial_{\tw^2}\tDelta_k(i\tw)|_{\tw=0}=1$ (Appendix~\ref{app_rgeq:subsec_self}). In standard applications of the FRG approach, the dimensionless two-point vertex significantly differs from $\tilde q^2$ or $\tw^2$ only when $\tilde q^2\gg 1$ or $\tw^2\gg 1$, i.e. for a momentum or frequency range where the threshold functions are exponentially suppressed by the cutoff function~(\ref{Rk}). This is not the case in the Bose-glass phase: For sufficiently small $k$ the self-energy $\tDelta_k(i\tw)$ strongly deviates from $\tw^2$ even for $|\tw|=\calO(1)$. This implies that the approximation $\tDelta_k(i\tw)=\tw^2$ in the calculation of the threshold functions is not fully justified (contrary to the usual case) although its gives correct results for $\delta_k(u)$, $K_k$ and $\theta_k$. For the frequency dependence of the self-energy $\Delta_k(i\w)$ (Fig.~\ref{fig_Delta}), and in particular to determine the localization (or pinning) length, it is much more accurate to compute the threshold functions with the full frequency-dependent $\tDelta_k(i\tw)$. 

Let us consider the self-energy $\Delta_k(i\w)$ for a fixed, small, value of $\w$. As long as $|\w|\lesssim v_kk$, i.e. $k\gtrsim k_c(\w)$, $\Delta_k(i\w)$ is well approximated by $Z_x\w^2/v_k^2$; this follows from $\tDelta_k(i\tw)\simeq \tw^2$ for $|\tw|\lesssim 1$. The crossover scale $k_c(\w)\sim|\w/a|^{1/z}$ is obtained by approximating $v_k\simeq ak^{z-1}$, with $z=\lim_{k\to 0}z_k=1+\theta$, which is justified if $|\w|$ is sufficiently small. Thus $\Delta_{k_c(\w)}(i\w)\simeq Z_x(\w/a)^{2/z}$. When $k\lesssim k_c(\w)$, i.e. $|\tw|\gtrsim 1$, the threshold function $\bar l_1(i\tw)$ can be neglected in~(\ref{rgeq3b}) and we obtain 
\begin{align}
\dt \Delta_k(i\w) 
&\simeq  Z_x k^2 \pi \bar l_1(0) \delta_k''(0) \nonumber \\ 
&\simeq - Z_x \pi \bar l_1(0) C \frac{v k^{2-\theta}}{Ka} ,
\label{dDelta1} 
\end{align}
where we have approximated $\delta_k''(0)$ by $-C/K_k$ (see Sec.~\ref{subsec_cuspQBL}) and used $K_k=(K/v)v_k=(K/v)ak^\theta$. Note that $\bar l_1(0)$ is independent of $k$. Integration of~(\ref{dDelta1}) between $k_c(\w)$ and $k$ yields 
\begin{align}
\Delta_k(i\w) ={}& Z_x|\w/a|^{2/z} \nonumber \\ & + \frac{Z_x \pi \bar l_1(0) C v}{(2-\theta)Ka}  \left[ |\w/a|^{(2-\theta)/z} - k^{2-\theta} \right] .
\label{Delta3} 
\end{align}
For $k\to 0$ the self-energy is of the form $A'|\w|^{2/z}+B'\w^{(2-\theta)/z}$. The term $|\w|^{2/z}$ obtained by naive scaling is subleading wrt $|\w|^{(2-\theta)/z}$ when $\theta>0$. This is due to that the fact that the threshold function $\bar l_1(0)$ is independent of $\tw$ and nonzero so that the flow of the self-energy is not exponentially suppressed when $|\tw|\gg 1$.

The exponent $\theta$ can be expressed entirely in terms of the cutoff function $R_k$. Since the threshold function $\bar m_\tau\equiv\bar m_\tau(\theta_k)$ is a linear function of $\theta_k$ (Appendix~\ref{app_threshold}), Eq.~(\ref{thetak}) gives 
\begin{equation} 
\theta_k = \frac{\pi}{2} \frac{\delta_k''(0) \bar m_\tau(0)}{1-\frac{\pi}{2} \delta_k''(0) [\bar m_\tau(1) - \bar m_\tau(0)]} 
\label{theta1}
\end{equation}
and 
\begin{equation}
\theta = \lim_{k\to 0} \theta_k = \frac{\bar m_\tau(0)}{\bar m_\tau(0) - \bar m_\tau(1)}
\label{theta}
\end{equation}
using $\lim_{k\to 0}\delta_k''(0)=-\infty$. With the cutoff function~(\ref{Rk}) and $r(x)=\alpha/(1-e^x)$ one finds that $\theta$ decreases from 0.76 to 0.26 when $\alpha$ increases from 2 to 3; there is no principle of minimum sensitivity which would allows one to determine the optimal value of $\alpha$. 

This strong dependence on the cutoff function is an unavoidable consequence of~(\ref{theta}) and is in sharp contrast with usual second-order phase transitions where the critical exponents depend on both the threshold functions and the values of the coupling constants at the fixed point. In the latter case one observes that the dependence of the coupling constants on $R_k$ largely compensates that of the threshold functions to make the critical exponents eventually weakly dependent of the cutoff function. It would be interesting to consider a more involved truncation of the effective action, e.g. including the third cumulant $\Gamma_{3,k}$, and see whether this would lead to a more precise estimate of the dynamical critical exponent in the Bose-glass phase.

Equation~(\ref{Delta3}) is confirmed by the numerical solutions of the flow equations. 
We observe that at higher frequencies, up to the pinning frequency $\w_p=v/L_c$ determined by the Larkin length, $\Delta_{k=0}(i\w)\simeq A+B|\w|$ with a coefficient $B$ which is independent of $\theta$ (see Figs.~\ref{fig_Delta} and \ref{fig_Delta1}). We shall see in Sec.~\ref{subsubsec_conductivity} that the Mott-Halperin law for the conductivity when $K=1$ requires the form $\Delta_k(i\w)\simeq A+B|\w|$ to extend down to zero frequency. We therefore expect that the self-energy converges nonuniformly toward a singular solution: 
\begin{equation}
\lim_{k\to 0} \Delta_k(i\w) = \llbrace 
\begin{array}{lll} 
0 & \mbox{if} & \w=0 , 
\\ A+B|\w| & \mbox{if} & \w\neq 0 ,
\end{array}
\right. 
\label{Delta}
\end{equation} 
even though the truncation~(\ref{ansatz}) does not allow to confirm this behavior at very low frequencies. The singular expression~(\ref{Delta}) appears to be necessary to obtain both a finite compressibility and a conductivity $\sig(\w)$ vanishing as $\w^2$ (Secs.~\ref{subsubsec_chirhorho} and \ref{subsubsec_conductivity}).

The self-energy~(\ref{Delta}) has also been obtained from the Gaussian variational method (GVM).\cite{Giamarchi96} In this approach, the Bose-glass phase is described by a replica-symmetry-broken (RSB) solution confined to the $\w=0$ mode. It is known that a cuspy disorder correlator in the FRG approach goes hand in hand with a RSB solution in the GVM.\cite{Ledoussal08,Mouhanna10,Mouhanna16}  
Note however that the GVM takes into account only small fluctuations of the phase field $\varphi$ about its equilibrium position and ignores solitonlike excitations. By contrast both types of excitations are included in the FRG approach as shown in a recent study of the sine-Gordon model.\cite{Daviet19}

With the self-energy~(\ref{Delta}) the connected propagator $G_c\equiv G_{c,k=0}$ [Eq.~(\ref{Gc1})] takes the form 
\begin{equation}
G_c(q,i\w) = \llbrace \begin{array}{lcc} 
\dfrac{1}{Z_x q^2} & \mbox{if} & \w=0 , \\
\dfrac{1}{B(\mathfrak{D} q^2 + 1/\tau_c + |\w|)} & \mbox{if} & \w\neq 0 ,
\end{array} 
\right. 
\label{Gc2} 
\end{equation}   
at low energies, where 
\beq
\xi = \sqrt{\mathfrak{D}\tau_c}, \quad  \tau_c = \frac{B}{A}, \quad \mathfrak{D}=\frac{Z_x}{B}  
\eeq
defines the localization (or pinning) length $\xi$; $\tau_c$ is the associated timescale and $\mathfrak{D}$ a diffusion coefficient. Equation~(\ref{Gc2}) implies 
a $x\sim\tau^{1/2}$ scaling for timescales smaller than $\tau_c$ (see Eq.~(\ref{chirhorho}) below). This defines a ``dynamical critical exponent'' $z=2$, which should however not be confused with the exponent $z=1+\theta$ associated with the $\w=0$ sector and controlling the RG flow toward the scale invariant disorder correlator $\delta^*(u)$.

\subsubsection{Density-density response function} 
\label{subsubsec_chirhorho} 

The density-density correlation function is given by $\chi_{\rho\rho}(q,i\w)=(q/\pi)^2 G_c(q,i\w)$ in the small $q$ limit. For $i\w=0$ we obtain the compressibility 
\beq
\kappa = \lim_{q\to 0} \chi_{\rho\rho}(q,i\w=0) = \frac{1}{\pi^2 Z_x} = \frac{K}{\pi v} 
\eeq 
since $\Delta_{k=0}(i\w=0)=0$. On the other hand the retarded density-density correlation function is obtained from the analytic continuation $i\w\to\w+i0^+$ (with a positive Matsubara frequency),\cite{Negele_book} i.e. 
\begin{equation} 
\chi^R_{\rho\rho}(q,\w)= \frac{q^2}{\pi^2 B(\mathfrak{D}q^2 + 1/\tau_c -i \w)} 
\end{equation}
in the small $(q,\w)$ limit, where $\w$ is a now a real frequency. This gives 
\begin{align}
\chi^R_{\rho\rho}(x,t) &= \intinf \frac{dq}{2\pi} \intinf \frac{d\w}{2\pi}  e^{i(qx-\w t)} \chi^R_{\rho\rho}(q,\w) \nonumber \\ 
&= \Theta(t) \frac{2\mathfrak{D}t-x^2}{8B(\pi \mathfrak{D}t)^{5/2}} e^{-\frac{x^2}{4\mathfrak{D}t}-\frac{t}{\tau_c}} ,
\label{chirhorho}
\end{align}  
where $\Theta(t)$ is the step function. The dynamics is diffusive for timescales smaller than $\tau_c$.

\subsubsection{Conductivity} 
\label{subsubsec_conductivity} 

Let us first consider the $k$-dependent conductivity $\sig_k(i\w)$. Particle number conservation implies that $\sig_k(i\w)$ is related to the density-density correlation function,\cite{Shankar90}
\beq
\sig_k(i\w) = \lim_{q\to 0} \frac{\w}{q^2} \chi_{\rho\rho,k}(q,i\w) = \frac{\w}{\pi^2 \Delta_k(i\w)} .
\label{sigvschi}
\eeq
In the small-frequency limit, we can use $\Delta_k(i\w)\simeq Z_x\w^2/v_k^2$, so that 
\beq
\sig_k(i\w) = \frac{K_kv_k}{\pi\w} \quad \mbox{for} \quad  |\w|\ll v_kk .
\eeq 
After analytic continuation $i\w\to\w+i0^+$ this gives 
\beq
\sig_k(\w) = \frac{i}{\pi} \frac{D_k}{\w+i0^+} = D_k \left[ \delta(\w) + \frac{i}{\pi} \calP \frac{1}{\w} \right] , 
\eeq
where $D_k=v_kK_k=(v/K)K_k^2$ is a scale-dependent Drude weight (or charge stiffness) and $\calP$ denotes the principal part. Thus $D_k\sim k^{2\theta}$ vanishes in the limit $k\to 0$ but is nonzero at any finite scale $k$. 

Consider now the limit $k\to 0$ with $|\w|\gg v_kk$. From~(\ref{sigvschi}) and (\ref{Delta}), one deduces   
\beq
\sig(\w) \equiv \sig_{k=0}(\w) = \xi^2 \kappa (-i\w + \w^2 \tau_c) + \calO(\w^3) . 
\label{cond} 
\eeq 
Equation~(\ref{cond}) agrees, up to logarithmic corrections, with the Mott-Halperin law $\Re[\sig(\w)]\sim \w^2\ln^2|\w|$ when $K=1$ (corresponding to hard-core bosons or free fermions). Note that this is a consequence of the linear behavior~(\ref{Delta}) of the self-energy at low energies.

It is also possible to determine analytically the large-frequency limit of the conductivity. In Appendix~\ref{app_largew}, we show that $\Re[\sig(\w)]\sim 1/|\w|^{4-2K}$ in agreement with Ref.~\onlinecite{Giamarchi_book}. 

In principle one can obtain the full frequency dependence of the conductivity by performing the analytic continuation from the numerically known $\Delta_{k=0}(i\w)$ by means of the resonances-via-Pad\'e method.\cite{Schlessinger68,Vidberg77,Tripolt19} This procedure has been successfully used in many works using the FRG formalism,\cite{Dupuis09b,Sinner10,Schmidt11,Rose15,Rose16a,Rose17a,Rose18,Tripolt17} but turns out to be rather unstable in the present case.

\subsection{Droplet phenomenology}
\label{subsec_droplet} 

The knowledge of the propagator allows us to compute the correlation function 
\begin{align}
C(x,\tau) &= \overline{ \mean{[\varphi(x,\tau)-\varphi(0,0)]^2} } \nonumber \\ 
&= \frac{2}{\beta L} \sum_{q,\w} \calG^{(2)}_{aa,k=0}(q,i\w) [ 1- \cos(qx-\w\tau)] ,
\end{align}
where $\mean{\cdots}$ stands for an average with the action $S[\varphi;\xi]$. $\calG^{(2)}_{aa}=G_c+G_d$ is defined in Appendix~\ref{app_GF}. To obtain $G_{d,k=0}(Q)$ we use Eq.~(\ref{Gd1}) with $V_k''(0)=-(v^2k^3/K^2)\delta^*(0)$ and identify $k$ with $|q|$. The leading contribution to $C(x,0)$ in the large distance limit comes from $G_d$,
\beq
C(x,0) \simeq 2\pi \delta^*(0) \ln |x| . 
\label{Cx}
\eeq 
On the other hand it is easily seen that $C(0,\tau)$ remains finite when $\tau\to\infty$. In agreement with the vanishing of $K_k$ in the limit $k\to 0$, quantum fluctuations remain small and one can view the field $\varphi(x,\tau)$ as a (quasi)classical field with weak temporal fluctuations. It is therefore not surprising that Eq.~(\ref{Cx}) agrees with the result obtained for random periodic elastic manifold (classical) systems (such as charge-density waves or vortex lattices) where the roughness exponent $\zeta$, which defines the scaling dimension of the field, vanishes. (For nonzero $\zeta$, one has $C(x)\sim |x|^{2\zeta}$.)

In classical disordered systems in which the long-distance physics is controlled by a zero-temperature fixed point, the low-energy properties are usually explained in the framework of the droplet picture.\cite{Fisher88b} The latter supposes the existence, at each length scale $L$, of a small number of excitations $\delta\varphi\sim L^\zeta$ above the ground state, drawn from an energy distribution $P_L(E)$ of width $\Delta E\sim L^\theta$ with a constant weight $\sim L^{-\theta}$ near $E=0$. The number of thermally active excitations is therefore $\sim T/L^\theta$, i.e., the system has a probability $\sim T/L^\theta$ to be in two nearly degenerate configurations. Thermal fluctuations are dominated by these rare droplet excitations and one has 
\beq
\overline{ [\mean{\varphi^2}-\mean{\varphi}^2]^p} \sim \frac{T}{L^\theta} L^{2p\zeta} 
\label{chin} 
\eeq 
at length scale $L$. From a theoretical point of view, the core of the connection between the FRG formalism and droplet phenomenology is the existence of a thermal boundary layer in the disorder correlator.\cite{Balents04,Balents05} 

Given the analogy, pointed out in Sec.~\ref{subsec_model} (see also Sec.~\ref{subsubsec_cuspQBLmeaning}), between a disordered one-dimensional Bose fluid and a random elastic manifold (classical) system, we expect the droplet phenomenology to apply also to the Bose-glass phase of a one-dimensional Bose gas with the Luttinger parameter playing the role of the temperature. The droplets are naturally identified with the metastable states, discussed in Sec.~\ref{subsubsec_cuspQBLmeaning} obtained by creating a soliton-antisoliton pair. They are two-dimensional since the action of the field $\varphi(x,\tau)$ is defined in a (1+1)-dimensional spacetime. For a droplet of spatial size $L$, the extension $L_\tau\sim L^z$ in the imaginary-time direction can be interpreted as a quantum coherence time.

To support the droplet picture, let us consider
\begin{align}
\chi_1 &= \overline{ \mean{\varphi(x,\tau)^2}-\mean{\varphi(x,\tau)}^2 } \nonumber \\ 
&= \frac{1}{\beta L} \sum_Q G_{c,k}(Q)
= \int_Q G_{c,k}(Q) , 
\label{chi1a}
\end{align} 
where $\int_Q=\int\frac{dq}{2\pi}\int\frac{d\w}{2\pi}$. We have assumed $L,\beta\to\infty$ and used the fact that in that limit the mode $\w=0$ does not contribute to the Matsubara frequency sum in~(\ref{chi1a}). To probe the behavior of the system at length scale $L$ we simply set $k\sim 1/L$. In the large-$L$ limit, the $L$-dependent part of $\chi_1$ comes from the low-energy limit of $G_{c,k}(Q)$ where the self-energy $\Delta_k(i\w)$ can be approximated by $Z_x(\w/v_k)^2$ and $R_k(Q)$ by $Z_xk^2$. This gives 
\begin{align} 
\chi_1 &\simeq \int_Q \frac{1}{Z_x(q^2+\w^2/v_k^2+k^2)} \nonumber \\ 
&\simeq - \frac{v_k}{2\pi Z_x} \ln k  \nonumber \\ 
&\sim \frac{K}{L^\theta} \ln L , 
\label{chi1} 
\end{align}
in agreement with~(\ref{chin}) when $\zeta=0$. In Appendix~\ref{app_chi2}, we show that 
\begin{equation}
\chi_2 = \overline{ [\mean{\varphi(x,\tau)^2}-\mean{\varphi(x,\tau)}^2]^2 } -\chi_1^2 
\sim \frac{K}{L^\theta} , 
\label{chi2} 
\end{equation} 
again in agreement with~(\ref{chin}) (we expect a more refined analysis to yield a logarithmic factor $\ln L$ as in~(\ref{chi1})). Similarly we can probe the behavior of the system at timescale $L_\tau$ by setting $k\sim 1/L_\tau^{1/z}$ when computing $\chi_1$ and $\chi_2$. One then obtains~(\ref{chi1}) and (\ref{chi2}) with $L$ replaced by $L_\tau^{1/z}$.

The droplet picture can also explain the low-frequency conductivity. When the system is subjected to an electric field $E(\w)$ at frequency $\w$, the power absorbed per unit length is $\sig(\w)|E(\w)|^2$. The contribution of low-energy droplets with size $L$ can be expressed as $\sim \w P_L(\w)/L$, where $\w$ is the energy of the absorbed photon, $P_L(\w)\sim 1/L^\theta$ the probability for the droplets to have an excitation energy $\w$ and the factor $1/L$ comes from the density\cite{Fisher86a} of droplets. Since an external field at frequency $\w$ probes the system at length scale $L\sim \w^{-1/z}$, we finally obtain $\sig(\w)\sim \w^2$. Interestingly this result is independent of $\theta$ in agreement with the numerical solution of the flow equations discussed in Sec.~\ref{subsec_dynamics}.\cite{not230}

Finally we stress that although our FRG approach is justified only in the limit of weak disorder (where bosonization is a valid starting point),\cite{Giamarchi87,Giamarchi88} the low-energy physics of the Bose-glass phase is expected to be independent of the disorder strength\cite{Pollet14} so that the droplet scenario should hold in the entire localized phase.\cite{not130}

\subsection{The need of a nonperturbative RG approach}
\label{subsec_whyNP}

The free replica sum expansion allows us to invert the matrix $\Gamma^{(2)}_k+R_k$ in replica space and obtain the propagator $G_k$ (Appendix~\ref{app_inversion}) but makes the resulting RG equation for $\delta_k$ (and $V_k$) apparently perturbative to the extent that $\dt\delta_k$ is of second order in $\delta_k$ and its derivatives. The nonperturbative aspect is then entirely due to the exponent $\theta_k$ (or, equivalently, the dynamical critical exponent $z_k=1+\theta_k$) which is determined by an equation of the type $\theta_k=A_k+B_k\theta_k$ with $A_k,B_k\propto \delta''_k(0)$  [Eqs.~(\ref{thetak},\ref{theta1})] and is therefore of infinite order in $\delta_k''(0)$. This is a crucial difference with the perturbative FRG (PFRG) approach where the equation for $\delta_k$ is the same as in the nonperturbative FRG (NPFRG) approach but the equation for $\theta_k$ is simply $\theta_k=A_k\propto \delta''_k(0)$ (the PFRG equations are obtained from the nonperturbative ones by replacing $\bar m_\tau(\eta_k)$ by $\bar m_\tau(0)$ in the rhs of~(\ref{thetak}));\cite{not200} the cusp then forms at a nonzero scale $k>0$.\cite{not210} The divergence of $\delta''_k(0)$ associated with the cusp formation, with a concomitant divergence of the dynamical critical exponent $z_k$, clearly calls into question the validity of the perturbative approach.  

The fact that $\theta_k$ (and more generally any anomalous dimension) is determined nonperturbatively is an essential feature of the NPFRG approach. It can be traced back to the term $\dt R_k$ in the rhs of Wetterich's equation~(\ref{eqwet}). To understand this point in more detail, let us consider the simpler case of the $\varphi^4$ theory with action $S=\int_x \{ \half (\nablabf\varphi)^2 + r_0 \varphi^2 + u_0\varphi^4\}$ and infrared regulator term $\Delta S_k=\half \int_{x,x'} \varphi(x) R_k(x-x')\varphi(x')$. The cutoff function must be of the form $R_k(q)=Z_kq^2 r(q^2/k^2)$ with a prefactor $Z_k$ given by the field renormalization factor $Z_k$; the latter enters the propagator as $G(q,\phi)=[Z_k q^2 + U_k''(\phi) + R_k(q)]^{-1}$ where $U_k(\phi)$ is the effective potential and we neglect a possible field dependence of $Z_k$. This form is necessary to allow for a scaling solution of the flow equation and therefore the existence of a fixed-point solution when the system is critical.\cite{Delamotte16a} The equation determining the anomalous dimension $\eta_k=-\dt\ln Z_k$, of the form $\eta_k=A_k+B_k\eta_k$, then follows from $\dt R_k=-Z_kq^2(\eta_k r+2(q/k)^2r')$. Omitting the prefactor $Z_k$ in the cutoff function would violate scale invariance and introduce an additional, unphysical, scale in the flow equations. From a more technical point of view, a proper definition of $R_k$ is necessary to eliminate any reference to microscopic scales and write the RG equations solely in terms of dimensionless variables (i.e. variables expressed in units of $k$). It has been shown that the prefactor $Z_k$ in $R_k$ is necessary to recover two-loop (and higher-order) contributions to $\eta$ from the exact flow equation~(\ref{eqwet}).\cite{Papenbrock95,[{For a concrete example, see, e.g., }]Reuter94}

In the Bose-glass case, there are two anomalous dimensions: $\eta_x=-\dt\ln Z_x$ which vanishes due to the statistical tilt symmetry and $\eta_{\tau,k}=-\dt \ln(Z_x/v_k^2)=2\theta_k$. In the NPFRG approach, the nonperturbative equation $\theta_k=A_k+B_k\theta_k$ allows for a scale-invariant, cuspy, solution $\delta^*(u)$ that is reached only for $k=0$ with a finite dynamical critical exponent $z_k$.  

There are other examples where the nonperturbative aspect comes from a nontrivial equation for the anomalous dimension while the beta functions $\dt g_{i,k}=\beta_i(\{g_{j,k}\},\eta_k)$ are perturbative in the $g_{j,k}$'s. For instance, in the flat phase of polymerized phantom membranes, the NPFRG yields perturbative (second-order) beta functions for the two coupling constants $u_k$ and $v_k$ but a nonpolynomial running anomalous dimension $\eta_k\equiv\eta_k(u_k,v_k)$.\cite{Kownacki09} The value of the anomalous dimension $\eta$ is in very good agreement with numerical results.\cite{Essafi14}

\section{Conclusion}

We have reported the first application of the nonperturbative FRG approach to a quantum disordered system. We find strong similarities between the Bose-glass phase of a one-dimensional Bose fluid and 
classical disordered systems in which the long-distance physics is controlled by a zero-temperature fixed point. This can be partially understood from the analogy between the imaginary-time action of the disordered boson system and that of elastic manifolds in a disordered medium (with a random potential that is perfectly correlated in one direction). 

The main result of the manuscript is that the Bose-glass phase is described by a strong-disorder fixed point with a vanishing Luttinger parameter. A key feature is the cuspy functional form of the disorder correlator which reveals the existence of metastable states and the ensuing glassy properties (pinning and shocks). The QBL rounding the cusp at nonzero momentum scale $k$, and encoding the quantum tunneling between the ground state and the low-lying metastable states, is associated with the existence of (rare) superfluid regions and is responsible for the $\w^2$ behavior of the conductivity at low frequencies. These results can be understood within the droplet picture of glassy systems.\cite{Fisher88b} 

We believe that the success of the nonperturbative FRG approach in describing the Bose glass phase comes from its ability to take into account the metastable states consisting of soliton-antisoliton pairs with exceptionally low excitation energy, a feature which is reminiscent of its ability to describe excited states (solitons and antisolitons as well as their bound states, i.e. breathers) in the sine-Gordon model.\cite{Daviet19}

Our FRG approach opens up the possibility to study other one-dimensional disordered systems. For instance one could address the effects of long-range confining interactions\cite{Chou18,Dupuis20a} or a periodic lattice potential\cite{Orignac99,Giamarchi01} on the Bose-glass phase. The study of disordered bosons in higher dimensions, where bosonization is not possible, might also be a promising avenue of research.

\begin{acknowledgments}
ND is indebted to P. Azaria, A. Fedorenko, N. Laflorencie, P. Le Doussal, G. Lemari\'e, D. Mouhanna, E. Orignac, A. Ran\c{c}on, B. Svistunov, G. Tarjus, M. Tissier, C. Wetterich and K. Wiese for discussions and/or correspondence.   
\end{acknowledgments} 

\appendix

\section{Statistical tilt symmetry}
\label{app_sts}

Let us consider the replicated action~(\ref{action}).
With the new field $\varphi'_a(x,\tau)=\varphi_a(x,\tau)+w(x)$, one has
\beq
S[\varphiset] = S[\varphipset] + \frac{n}{2} \beta Z_x \int_x (\dx w)^2 + Z_x \int_{x,\tau} \sum_a \varphi'_a \dx^2 w 
\eeq 
with $Z_x=v/\pi K$, since the disorder part of the action~(\ref{action}) in invariant when the field is shifted by a time-independent (but otherwise arbitrary) function $w(x)$. This allows us to rewrite the partition function~(\ref{partition}) as 
\beq
\calZ[\Jset] = \calZ[\Jpset] e^{ \frac{n}{2} \beta Z_x \int_x (\dx w)^2 - \int_{x,\tau} \sum_a J'_a w }
\eeq 
where 
\beq
J'_a(x,\tau) = J_a(x,\tau) - Z_x \dx^2 w .
\eeq 
This leads to the effective action 
\begin{align}
\Gamma[\phiset] ={}& - \ln \calZ[\Jset] +  \int_{x,\tau} \sum_a J_a \phi_a \nonumber \\ 
={}& - \ln \calZ[\Jpset] + \int_{x,\tau} \sum_a J'_a \phi'_a - \frac{n}{2} \beta Z_x \int_x (\dx w)^2 
\nonumber \\ & + Z_x \int_{x,\tau} \sum_a \phi_a \dx^2 w ,
\label{app6} 
\end{align}
where 
\beq
\begin{split} 
	\phi_a(x,\tau) &= \frac{\delta\ln \calZ[\Jfset]}{\delta J_a(x,\tau)} = \phi'_a(x,\tau) - w(x) , \\  
	\phi'_a(x,\tau) &= \frac{\delta\ln \calZ[\Jfpset]}{\delta J'_a(x,\tau)} .  
\end{split}
\eeq 
From~(\ref{app6}) we then deduce 
\begin{align}
\Gamma[\phiset] ={}& \Gamma[\{\phi_a+w\}] - \frac{n}{2} \beta Z_x \int_x (\dx w)^2 \nonumber \\ & 
- Z_x \int_{x,\tau} \sum_a (\dx w) (\dx \phi_a) . 
\label{app7} 
\end{align} 
Equation~(\ref{app7}) has important consequences. First it shows that $\Gamma[\phiset]$ is invariant under a constant (i.e. time-independent and uniform) shift of the field, $\phi_a(x,\tau)\to \phi_a(x,\tau)+w$. This implies that the self-energy $\Delta_k(i\w)$ in~(\ref{ansatz}) must vanish for $\w=0$. Second $\Gamma_1[\phi_a]$ must include the term $\half Z_x\int_{x,\tau} (\dx\phi_a)^2$ and no higher-order space derivatives are allowed. Third $\Gamma_i[\phi_{a_1},\cdots,\phi_{a_i}]$ is invariant in the shift $\phi_a(x,\tau)\to \phi_a(x,\tau)+w(x)$ when $i\geq 2$. This implies in particular that the second-order disorder cumulant, 
\beq
\Gamma_{2,k}[\phi_a,\phi_b] = \int_{x,\tau,x',\tau'} V_k(\phi_a(x,\tau),\phi_b(x',\tau')) , 
\eeq 
is necessarily of the form given in~(\ref{ansatz}).

\subsubsection*{STS invariant regulator} 

The preceding conclusions hold for the scale-dependent effective action $\Gamma_k[\phiset]$ only if the regulator term $\Delta S_k[\varphiset]$ is invariant in the transformation $\varphi'_a(x,\tau)=\varphi_a(x,\tau)+w(x)$. The most general form, quadratic in the field and consistent with the invariance of the action under permutation of the replica indices, reads 
\beq
\Delta S_k[\varphiset] = \half \sum_{q,\w\atop a,b} \varphi_a(-q,-i\w) R_{k,ab}(q,i\w) \varphi_b(q,i\w) ,
\eeq 
where 
\beq
R_{k,ab}=\delta_{ab}\hat R_k + \tilde R_k . 
\eeq 
In the shift $\varphi_a(q,i\w)\to\varphi_a(q,i\w)+\sqrt{\beta} \delta_{\w,0} w(q)$, the regulator term varies by the amount 
\begin{align}
& \frac{\beta}{2} \sum_{a,q} \Bigl[ w(-q)w(q) + \frac{2}{\sqrt{\beta}} \varphi_a(-q,0)w(q) \Bigr]
\nonumber \\ & \times  
\bigl[ \hat R_k(q,0) + n \tilde R(q,0) \bigr] ,
\end{align}
where $n$ is the number of replicas. 

An STS invariant regulator must therefore satisfy 
\beq
\hat R_k(q,0) + n \tilde R_k(q,0) = 0 , 
\eeq 
which shows that $\tilde R_k$ is necessary nonzero (given that any decent cutoff function $\hat R_k(q,0)$ depends on $q$). The simplest choice is to take a time-independent cutoff function $\tilde R_k$ defined by
\beq
\tilde R_k(q,i\w) = - \frac{1}{n} \delta_{\w,0} \hat R_k(q,0) .
\eeq 
This cutoff function does not contribute in the zero-temperature limit since a prefactor $\beta$ is missing in front of the Kronecker delta $\delta_{\w,0}$. This can be explicitly seen by noting that the sole effect of $\tilde R_k$ is to replace $\Gamma_k^{(11)}[\phi_a,\phi_b]$ by $\Gamma_k^{(11)}[\phi_a,\phi_b]-\tilde R_k$ [see Eq.~(\ref{app8})]. With the ansatz~(\ref{ansatz}), this means that $\tilde R_k(q,i\w)$ is always involved in the combination 
\beq 
\beta \delta_{\w,0} \Bigl[  V^{(11)}(\phi_a,\phi_b) - \frac{1}{\beta} \tilde R_k(q,0) \Bigr]
\eeq 
and therefore does not play any role in the limit $\beta\to\infty$ where $\beta \delta_{\w,0}\to 2\pi \delta(\w)$.

\section{Matrix inversion in the replica formalism}
\label{app_inversion} 

In this section we briefly review how to invert a matrix using the free replica sum expansion and consider the particular case of $\Gamma^{(2)}_{k,ab}[\phifset]+R_{k,ab}$.\cite{Tarjus08} 

\subsection{Matrix inversion}

Any matrix $A_{ab}[\phifset]$ can be written in the general form 
\beq
A_{ab}[\phifset] = \delta_{ab} \hat A_a[\phifset] + \tilde A_{ab}[\phifset] , 
\eeq
with 
\beq
\begin{split}
	\hat A_a[\phifset] &= \hat A^{[0]}[\phi_a] + \sum_c \hat A^{[1]}[\phi_a|\phi_c] + \cdots \\ 
	\tilde A_{ab}[\phifset] &= \tilde A^{[0]}[\phi_a,\phi_b] + \sum_c \tilde A^{[1]}[\phi_a,\phi_b|\phi_c] + \cdots  
\end{split}
\eeq 
where the superscripts in square brackets denote the order in the free replica sum expansion. A similar expansion holds for the matrix $B=A^{-1}$. The term-by-term identification of the condition $AB=\mathbb{1}$ leads to 
\beq 
\begin{split}
	\hat B^{[0]}[\phi_a] &=  \hat A^{[0]}[\phi_a]^{-1} , \\ 
	\tilde B^{[0]}[\phi_a,\phi_b] &=  - \hat B^{[0]}[\phi_a] \tilde A^{[0]}[\phi_a,\phi_b] \hat B^{[0]}[\phi_b]
\end{split}
\eeq
and  
\beq 
\begin{split}
	\hat B^{[1]}[\phi_a|\phi_c] ={}& - \hat B^{[0]}[\phi_a] \hat A^{[1]}[\phi_a|\phi_c] \hat B^{[0]}[\phi_a] , \\ 
	\tilde B^{[1]}[\phi_a,\phi_b|\phi_c] ={}& - \hat B^{[0]}[\phi_a] \Bigl\{ 
	\tilde A^{[1]}[\phi_a,\phi_b|\phi_c] \\ &\hspace{-1.5cm} 
	- \hat A^{[1]}[\phi_a|\phi_c] \hat B^{[0]}[\phi_a] \tilde A^{[0]}[\phi_a,\phi_b] \\ & \hspace{-1.5cm} 
	- \tilde A^{[0]}[\phi_a,\phi_b] \hat B^{[0]}[\phi_b] \hat A^{[1]}[\phi_b|\phi_c] \\ & \hspace{-1.5cm} 
	- \tilde A^{[0]}[\phi_a,\phi_c] \hat B^{[0]}[\phi_c] \tilde A^{[0]}[\phi_c,\phi_b] 
	\Bigr\} \hat B^{[0]}[\phi_b] .
\end{split}
\eeq

\subsection{Propagator and flow equations}

For the matrix $A_{ab}[\phifset]=\Gamma^{(2)}_{k,ab}[\phifset]+R_{k,ab}$, considering a cutoff function $R_{k,ab}=\delta_{ab}\hat R_k + \tilde R_k$, one has 
\beq
\begin{split}
	\hat A^{[0]}[\phi_a] &= \Gamma_{1,k}^{(2)}[\phi_a] + \hat R_k , \\
	\hat A^{[1]}[\phi_a|\phi_c] &= - \Gamma_{2,k}^{(20)}[\phi_a,\phi_c] , \\ 
	\tilde A^{[0]}[\phi_a,\phi_b] &= - \Gamma_{2,k}^{(11)}[\phi_a,\phi_b] + \tilde R_k , \\  
	\tilde A^{[1]}[\phi_a,\phi_b|\phi_c] &= 0 ,
	\label{app8}
\end{split}
\eeq
ignoring cumulants $\Gamma_{i,k}$ with $i\geq 3$ and considering only the terms that are needed for the derivation of the flow equations of $\Gamma_{1,k}$ and $\Gamma_{2,k}$. The propagator $G_k[\phiset]=(\Gamma^{(2)}_k[\phiset]+R_k)^{-1}$ is thus defined by 
\beq 
\begin{split}
	\hat G_k^{[0]}[\phi_a] &= P_k[\phi_a] \\ 
	\hat G_k^{[1]}[\phi_a|\phi_c] &= P_k[\phi_a] \Gamma_{2,k}^{(20)}[\phi_a,\phi_c] P_k[\phi_a] , \\ 
\end{split} 
\label{app2}
\eeq 
and  
\beq 
\begin{split}
	\tilde G_k^{[0]}[\phi_a,\phi_b] &= P_k[\phi_a] \Gamma_{2,k}^{(11)}[\phi_a,\phi_b] P_k[\phi_b] \\ 
	\tilde G_k^{[1]}[\phi_a,\phi_b|\phi_c] &= P_k[\phi_a] \Bigl\{
	\Gamma_{2,k}^{(20)}[\phi_a,\phi_c]  P_k[\phi_a] \Gamma_{2,k}^{(11)}[\phi_a,\phi_b] \\ &
	+ \Gamma_{2,k}^{(11)}[\phi_a,\phi_b] P_k[\phi_b] \Gamma_{2,k}^{(20)}[\phi_b,\phi_c] \\ &
	+ \Gamma_{2,k}^{(11)}[\phi_a,\phi_c] P_k[\phi_c] \Gamma_{2,k}^{(11)}[\phi_c,\phi_b]
	\Bigr\} P_k[\phi_b]  , 
\end{split} 
\label{app3}
\eeq 
where $P_k[\phi_a] = \bigl( \Gamma_{1,k}^{(2)}[\phi_a] + \hat R_k \bigr)^{-1}$ is the propagator obtained from $\Gamma_{1,k}[\phi_a]$.

The exact flow equation~(\ref{eqwet}) can now be written as 
\begin{multline}
\dt \Gamma_k[\phifset] = \half \tr\Bigl\{ 
\sum_a \Bigl( \dt(\hat R_k+\tilde R_k) \hat G_{k,a}[\phifset] \\
+ \dt \hat R_k \tilde G_{k,aa}[\phifset] \Bigr) 
+ \sum_{a,b} \dt \tilde R_k \tilde G_{k,ba}[\phifset] \Bigr\} , 
\label{app1}
\end{multline}
where the trace is now over spacetime variables $(x,\tau)$ or $(q,i\w)$ only. From~(\ref{app1}) one deduces 
\beq
\dt\Gamma_{1,k}[\phi_a] = \half \tr \Bigl\{  \dt(\hat R_k+\tilde R_k) \hat G^{[0]}_{k}[\phi_a] 
+ \dt \hat R_k  \tilde G^{[0]}_{k}[\phi_a,\phi_a] \Bigr\}
\label{app4}
\eeq 
and 
\begin{align}
\dt\Gamma_{2,k}[\phi_a,\phi_b] ={}& - \half \tr \Bigl\{ 
\dt(\hat R_k+\tilde R_k) \hat G^{[1]}_{k}[\phi_a|\phi_b] \nonumber \\ &
+ \dt \hat R_k \tilde G^{[1]}_{k}[\phi_a,\phi_a|\phi_b] \nonumber \\ & 
+ \dt \tilde R_k \tilde G^{[0]}_{k}[\phi_a,\phi_b] \Bigr\} + \mbox{perm}(a,b) ,
\label{app5}
\end{align}
${\rm perm}(a,b)$ denotes all terms obtained by exchanging the replica indices $a$ and $b$. Equations~(\ref{rgeq1}) and (\ref{rgeq2}) follow from~(\ref{app2}-\ref{app3}) and (\ref{app4}-\ref{app5}) when the cutoff function $R_{k,ab}=\delta_{ab}R_k$ is diagonal in the replica indices (i.e. $\tilde R_k=0$).

\section{Green functions}
\label{app_GF} 

In Sec.~\ref{subsec_scaledepgamma} the propagators $G_c$ and $G_d$ were defined directly from the cumulants $W_1[J_a]$ and $W_2[J_a,J_b]$ of the random functional $W[J;\xi]$. Alternatively one can define Green functions from the functional $Z[\Jset]$ [Eq.~(\ref{partition})],
\beq 
\calG^{(p)}_{a_1\cdots a_p}(X_1\cdots X_p) = \frac{1}{Z[\Jset]} \frac{\delta^p Z[\Jset]}{\delta J_{a_1}(X_1)\cdots \delta J_{a_p}(X_p)} \biggl|_{J_{a_i}=0} ,
\eeq
where $X=(x,\tau)$. In the formal limit where the number of replicas $n\to 0$, one finds 
\beq
\begin{split}
\calG^{(1)}_a(X) &= \overline{\mean{\varphi(X)}} , \\ 
\calG^{(2)}_{ab}(X,X') &= \llbrace 
\begin{array}{ccc}
\overline{ \mean{\varphi(X) \varphi(X')} } & \mbox{if} & a=b , \\ 
\overline{ \mean{\varphi(X) } } \overline{ \mean{\varphi(X')} } & \mbox{if} & a\neq b ,
\end{array} 
\right. 
\end{split}
\eeq 
etc., where $\mean{\cdots}$ stands for an average with the action $S[\varphi;\xi]$. It is also useful to introduce the connected Green functions 
\beq 
W^{(p)}_{a_1\cdots a_p}(X_1\cdots X_p) = \frac{\delta^p W[\Jset]}{\delta J_{a_1}(X_1)\cdots \delta J_{a_p}(X_p)} \biggl|_{J_{a_i}=0} ,
\eeq
where $W[\Jset]=\ln Z[\Jset]$, i.e., 
\beq
\begin{split}
W^{(1)}_a(X) &= \calG^{(1)}_a(X) , \\ 
W^{(2)}_{ab}(X,X') &= \calG^{(2)}_{ab}(X,X') - \calG^{(1)}_a(X) \calG^{(1)}_b(X') ,
\end{split}
\eeq
etc. The $W^{(p)}$'s can be related to the vertices\cite{Zinn_book}  
\beq
\Gamma^{(p)}_{a_1\cdots a_p}(X_1\cdots X_p) =  \frac{\delta^p \Gamma[\phiset]}{\delta \phi_{a_1}(X_1) \cdots \delta \phi_{a_p}(X_p)} \biggl|_{\phi_{a}(X)=\phi} 
\eeq 
calculated in a constant (arbitrary) field configuration. For instance, one has $W^{(2)}=\Gamma^{(2)}{}^{-1}$ (where the inverse must be understood in a matrix sense). As for $W^{(4)}$, one finds
\begin{multline} 
W^{(4)}_{abcd}(Q_1,Q_2,Q_3,Q_4) = - \sum_{a',b',c',d'} W^{(2)}_{aa'}(Q_1) W^{(2)}_{bb'}(Q_2)
\\ 
\times  W^{(2)}_{cc'}(Q_3) W^{(2)}_{dd'}(Q_4) \Gamma^{(4)}_{a'b'c'd'}(Q_1,Q_2,Q_3,Q_4) ,
\label{W4}
\end{multline} 
since $\Gamma^{(3)}$ vanishes when $J_a=0$.

\section{Flow equations} 
\label{app_rgeq} 

\subsection{Two-replica potential}

For uniform fields, $\phi_a(x,\tau)=\phi_a$, $\Gamma_{2,k}[\phi_a,\phi_b]=\beta^2 L  V_k(\phi_a-\phi_b)$ where $L$ is the length of the system. By using the expression of the vertices given in Appendix~\ref{app_vertices}, one finds
\begin{align} 
\dt V_{k,ab} =&{} \frac{1}{2\beta} \tdt \int_Q \Bigl\{ 
P_k(Q) \beta \bigl[ V^{(20)}_{k,ab} + V^{(20)}_{k,ba} \bigr] \nonumber \\ &
+ \beta^2 \delta_{\w,0} P_k(Q)^2 \bigl[ V^{(20)}_{k,ab} V^{(11)}_{k,aa} + V^{(20)}_{k,ba} V^{(11)}_{bb} \nonumber \\ &
+ V^{(11)}_{k,ab} V^{(11)}_{k,ba}\bigr] 
\Bigr\} ,
\end{align}
where we use the notations $V_{k,ab}=V_k(\phi_a-\phi_b)$ and $V_{k,ab}^{(nm)}=\partial^n_{\phi_a}\partial^m_{\phi_b}V_{k,ab}$. 
Introducing the dimensionless potential 
\beq 
\tilde V_{k,ab}= \frac{K^2}{v^2k^3} V_{k,ab} , 
\eeq 
we obtain 
\begin{align}
\dt \tilde V_{k,ab} ={}& -3 \tilde V_{k,ab} 
- \frac{K_k}{2} l_1 \bigl[ \tilde V^{(20)}_{k,ab} + \tilde V^{(20)}_{k,ba} \bigr] \nonumber \\ &
- \frac{\pi}{2} \bar l_2 \bigl[ \tilde V^{(20)}_{k,ab} \tilde V^{(11)}_{k,aa}  + \tilde V^{(20)}_{k,ba} \tilde V^{(11)}_{bb}  \nonumber \\ &
+ \tilde V^{(11)}_{k,ab} \tilde V^{(11)}_{k,ba}\bigr] ,
\end{align}
where the threshold functions $l_1$ and $\bar l_2$ are defined in Appendix~\ref{app_threshold}. This equation can be rewritten as a flow equation for $\delta_k(u)$ [Eq.~(\ref{dimvara})] using 
\beq
\begin{split}
\delta_k(\phi_a-\phi_b) &= \tilde V^{(11)}_{k,ab} , \\ 
\delta'_k(\phi_a-\phi_b) &= \tilde V^{(21)}_{k,ab} = - \tilde V^{(12)}_{k,ab} , \\  
\delta''_k(\phi_a-\phi_b) &= - \tilde V^{(22)}_{k,ab} = \tilde V^{(31)}_{k,ab} = \tilde V^{(13)}_{k,ab} ,
\end{split}
\eeq
which leads to Eq.~(\ref{rgeq3a}).

\subsection{Self-energy} 
\label{app_rgeq:subsec_self}

\begin{figure}
	\centerline{\includegraphics[width=7cm]{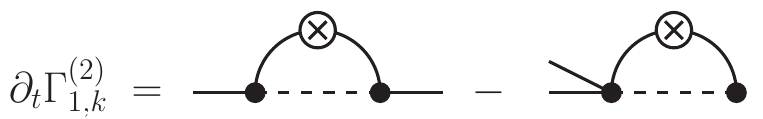}} 
	\caption{Diagrammatic representation of the flow equation~\ref{app9} satisfied by $\Gamma_{1,k}^{(2)}$. (See the caption of Fig.~\ref{fig_rgeq} for the meaning of the various lines.)}
	\label{fig_rgeq2} 
\end{figure}

From~(\ref{rgeq1}) and the fact that $P_k$ is field independent with the ansatz~(\ref{ansatz}) we deduce the flow equation 
\begin{multline}
\dt \Gamma_{1,k}^{(2)}[X,X';\phi_a] = - \half \tdt \tr\Bigl\{ 
P_k \bigl( \Gamma_{2,k}^{(31)}[X,X',\phi_a;\phi_a] \\
+ \Gamma_{2,k}^{(22)}[X,\phi_a;X',\phi_a]
+ \Gamma_{2,k}^{(22)}[X',\phi_a;X,\phi_a] \\ 
+ \Gamma_{2,k}^{(13)}[\phi_a;X,X',\phi_a] \bigr) \Bigr\} . 
\end{multline} 
In constant fields, using the results of Appendix~\ref{app_vertices}, we obtain 
\beq
\dt \Gamma_{1,k}^{(2)}(P) = - V_{k,aa}^{(22)} \tdt \int_q [ P_k(q,i\w) - P_k(q,0) ] ,
\label{app9}
\eeq
where $P=(p,i\w)$. $\dt \Gamma_{1,k}^{(2)}(P)$ is independent of $p$, in agreement with the STS. Equation~(\ref{app9}) is shown diagrammatically in Fig.~\ref{fig_rgeq2}. The flow equation for the self-energy is simply $\dt\Delta_k(i\w)=\dt\Gamma_{1,k}^{(2)}(0,i\w)$. In dimensionless variables we thus obtain~(\ref{rgeq3b}). The dynamical critical exponent $z_k$ is defined by $\dt v_k=(z_k-1)v_k$ and can be determined from the condition $\tilde\Delta_k(i\tw)=\tw^2+\calO(\tw^4)$, i.e.
\beq
\dt \frac{\partial^2 \tilde\Delta_k(i\tw)}{\partial\tw^2} \biggl|_{\tw=0} = 0  . 
\eeq
Expanding Eq.~(\ref{rgeq3b}) about $\tw=0$ then gives~(\ref{thetak}) where the threshold function 
$\bar m_\tau$ is given in Appendix~\ref{app_threshold}. 

\subsection{Vertices in a uniform field}
\label{app_vertices}

With the ansatz~(\ref{ansatz}) the propagator 
\begin{align}
P_k(Q) \equiv P_k[Q,\phi_a] 
&= \frac{1}{\Gamma_{1,k}^{(2)}(Q,-Q)+R_k(Q)} \nonumber \\ 
&= \frac{1}{Z_x q^2 + \Delta_k(i\w)+R_k(Q) } 
\end{align} 
is field independent for any configuration of the field $\phi_a(x,\tau)$. The other vertices, which are necessary for the flow equations of $\Gamma_{1,k}$ and $\Gamma_{2,k}$, are given by
\beq
\begin{split} 
	\Gamma_{2,k}^{(11)}(Q,\phi_a;-Q,\phi_b) &= \beta \delta_{\w,0} V^{(11)}_{k,ab} , \\ 
	\Gamma_{2,k}^{(20)}(Q,-Q,\phi_a;\phi_b) &= \beta V^{(20)}_{ab} , \\ 
	\Gamma_{2,k}^{(21)}(Q_1,Q_2,\phi_a;Q_3,\phi_b) &= \frac{\delta_{\sum_i Q_i,0}}{\sqrt{\beta L}}  \beta \delta_{\w_3,0} V^{(21)}_{k,ab} , \\ 
	\Gamma_{2,k}^{(31)}(Q_1,Q_2,Q_3,\phi_a;Q_4,\phi_b) &= \frac{\delta_{\sum_i Q_i,0}}{\beta L}  \beta \delta_{\w_4,0} V^{(31)}_{k,ab} , \\ 
	\Gamma_{2,k}^{(22)}(Q_1,Q_2,\phi_a;Q_3,Q_4,\phi_b) &= \frac{\delta_{\sum_i Q_i,0}}{\beta L}  \beta \delta_{\w_1+\w_2,0} V^{(22)}_{k,ab} 
\end{split}
\eeq
for constant (i.e. uniform and time-independent) fields $\phi_a(x,\tau)=\phi_a$. We use the notation 
\beq
\begin{split}
\Gamma_{2,k}^{(11)}(Q,\phi_a;-Q,\phi_b) &= \frac{\delta^2\Gamma_{2,k}[\phi_a,\phi_b]}{\delta\phi_a(-Q)\delta \phi_b(Q)}\biggl|_{\phifset =\rm const} , \\  
\Gamma_{2,k}^{(20)}(Q,-Q,\phi_a;\phi_b) &= \frac{\delta^2\Gamma_{2,k}[\phi_a,\phi_b]}{\delta\phi_a(-Q)\delta \phi_a(Q)}\biggl|_{\phifset =\rm const} ,
\end{split}
\eeq
etc. 

\subsection{Threshold functions} 
\label{app_threshold} 

The threshold functions are defined by 
\begin{equation}
\begin{gathered}
l_1 = \int_0^\infty d\tilde q \int_{-\infty}^\infty \frac{d\tw}{2\pi} \frac{\dt R_k(\tilde q,i\tw)}{Z_xk^2} 
\tilde P_k(\tilde q,i\tw)^{2} , \\ 
\bar l_1(i\tw) = \int_0^\infty d\tilde q \frac{\dt R_k(\tilde q,i\tw)}{Z_xk^2} 
\tilde P_k(\tilde q,i\tw)^{2} , \\  
\bar l_2 = 2 \int_0^\infty d\tilde q \frac{\dt R_k(\tilde q,0)}{Z_xk^2} 
\tilde P_k(\tilde q,0)^{3} , \\  
\bar m_\tau = \partial_{\tw^2} \bar l_1(i\tw) \bigl|_{\tw=0} ,
\end{gathered}
\end{equation}
where 
\beq
\tilde P_k(\tilde q,i\tw) = Z_xk^2 P_k(q,i\w) = \frac{1}{(\tilde q^2+\tilde\Delta_k)(1+r)} , 
\label{Ptilde} 
\eeq
is the dimensionless propagator and 
\begin{multline}  
\frac{\dt R_k(\tilde q,i\tw)}{Z_xk^2}  = 2 \tilde\Delta_k r- 2\tilde q^2  (\tilde q^2+\tilde\Delta_k)r' \\
+ (\dt\tilde\Delta_k|_\tw - z_k \tw \partial_\tw \tilde\Delta_k)[r + (\tilde q^2+\tilde\Delta_k)r'] ,
\end{multline}
with $r\equiv r(\tilde q^2+\tilde\Delta_k)$ and $\tilde\Delta_k\equiv\tilde\Delta_k(i\tw)$. For $\theta_k=0$ and $\tilde\Delta_k(i\tw)=\tw^2$, the threshold function $l_1=1/2$ is universal, i.e. independent of the function $r(x)$ provided that the latter satisfies $r(0)=\infty$ and $r(\infty)=0$. 

\section{Large-frequency behavior}
\label{app_largew}

The large-frequency behavior of the self-energy comes from large values of the running momentum scale $k$ such that $|\w|\ll vk$. Values of $k$ smaller than $|\w|/v$ correspond to a large $\tw$, and therefore a negligible threshold function $\bar l_1(i\tw)$, and contribute only a subleading frequency-independent term. For $k\gg |\w|/v$ (and $|\w|$ large) the flow is still in the perturbative regime and it is possible to use the approximation defined by Eqs.~(\ref{flow2}), i.e. 
\beq
\begin{split}
\dt \tilde\calD_k &= (-3+2K_k) \tilde\calD_k , \\ 
\dt K_k &= -24 \pi \tilde\calD_k \bar m_\tau , 
\end{split}
\label{rgeq4}
\eeq 
where $\tilde\calD_k=(K^2/v^2k^3)\calD_k=\delta_{1,k}/8$ and $\bar m_\tau\equiv\bar m_\tau|_{\theta_k=0}<0$. In the limit of weak disorder, the renormalization of $K_k$ is very small in the initial stage of the flow, so that the first equation gives 
\beq
\tilde\calD_k \simeq \tilde\calD_\Lamb \left(\frac{k}{\Lamb} \right)^{-3+2K} ,
\eeq 
The second one can be rewritten as $\dt K_k = a k^{-3+2K}$, i.e. 
\begin{align}
K_k &= K - b (k^{-3+2K} - \Lamb^{-3+2K}) \nonumber \\
&\equiv K_\Lamb^R - b k^{-3+2K} , 
\end{align}
with $a$ and $b=-a/(2K-3)$ positive constants (in the Bose-glass phase, i.e., for $K<3/2$) and $K^R_\Lamb=K+b\Lamb^{-3+2K}$. This yields the self-energy 
\beq
\Delta_k(i\w) = \frac{Z_x\w^2}{v_k^2} 
\simeq \w^2 \frac{K}{\pi v (K^R_\Lamb)^2} \left( 1 + \frac{2b}{K^R_\Lamb} k^{2K-3} \right) 
\eeq 
to quadratic order in $\w$ and for $b\propto\calD_\Lamb\to 0$. Since this expression is valid for $k\gg |\w|/v$ while $\dt\Delta_k(i\w)\simeq 0$ for $k\ll |\w|/v$, we can obtain the large-frequency behavior by identifying $k$ with $|\w|/v$, which gives 
\beq 
\Delta_{k=0}(i\w) \simeq \w^2 \frac{K}{\pi v (K^R_\Lamb)^2} \left[ 1 + \frac{2b}{K^R_\Lamb} \left( \frac{|\w|}{v} \right)^{2K-3} \right]  
\eeq 
and in turn 
\beq 
\sig_{k=0}(i\w) \propto \frac{1}{\w} \left[ 1 - \frac{2b}{K^R_\Lamb} \left( \frac{|\w|}{v} \right)^{2K-3} \right] .
\label{app10} 
\eeq 
After analytic continuation $i\w\to\w+i0^+$ the first term in~(\ref{app10}) produces a Dirac peak $\sim \delta(\w)$ (which is outside the domain of validity of the large-frequency perturbative expansion) in the real part of the conductivity while the second one gives a high-frequency tail $\propto 1/|\w|^{4-2K}$ in agreement with results from perturbative RG.\cite{Giamarchi_book}

\section{Computation of $\chi_2$}
\label{app_chi2} 

From the definition~(\ref{chi2}) of $\chi_2$ and the Green functions defined in Appendix~\ref{app_GF}, one finds  
\begin{align}
\chi_2 ={}&  \calG^{(4)}_{abcd,k} - 2 \calG^{(4)}_{aabc,k} + \calG^{(4)}_{aabb,k} - \chi_1^2 \nonumber \\ 
={}& W^{(4)}_{abcd,k} - 2 W^{(4)}_{aabc,k} + W^{(4)}_{aabb,k} ,
\label{chi2bis}
\end{align}
where $\calG^{(4)}$ and $W^{(4)}$ have all their arguments equal to $X=(x,\tau)$. Distinct replica indices mean distinct replicas (no summation over repeated indices is implied). $W^{(4)}$ can be obtained from~(\ref{W4}) and the expression
\begin{multline}
\Gamma^{(4)}_{abcd,k}(Q_1,Q_2,Q_3,Q_4) = \frac{\delta_{\sum_i Q_i,0}}{L} \delta_k''(0) \frac{v^2k^3}{K^2} \\ 
\times [ \delta_{ab} \delta_{cd} \delta_{\w_1+\w_2,0} + \delta_{ac} \delta_{bd} \delta_{\w_1+\w_3,0} + \delta_{ad} \delta_{bc} \delta_{\w_1+\w_4,0} \\ 
- \delta_{abc} \delta_{\w_4,0} - \delta_{abd} \delta_{\w_3,0} - \delta_{acd} \delta_{\w_2,0} - \delta_{bcd} \delta_{\w_1,0} ]
\end{multline} 
obtained from the effective action~(\ref{ansatz}). Here $\delta_{abc}=\delta_{ab}\delta_{bc}$. Using $W^{(2)}_{ab}(Q)=\delta_{ab}G_c(Q)+G_d(Q)$, one finds that the disconnected propagator $G_d$ does not contribute to $W^{(4)}$,
\begin{multline}
W^{(4)}_{abcd,k}(Q_1,Q_2,Q_3,Q_4) = - G_{c,k}(Q_1) G_{c,k}(Q_2) \\
\times  G_{c,k}(Q_3) G_{c,k}(Q_4) \Gamma^{(4)}_{abcd,k}(Q_1,Q_2,Q_3,Q_4) .
\end{multline}
From~(\ref{chi2bis}), one then obtains
\begin{align} 
\chi_2 ={}& -  \delta_k''(0) \frac{v^2k^3}{K^2} \int_{q_1,q_2,q_3} \int_{\w_1,\w_3} 
G_{c,k}(q_1,i\w_1) \nonumber \\ 
&\times G_{c,k}(q_2,-i\w_1) G_{c,k}(q_3,i\w_3)G_{c,k}(q_4,-i\w_3)
\label{chi2ter}
\end{align}
where $q_4=-q_1-q_2-q_3$ and we use the notation $\int_q=\int\frac{dq}{2\pi}$ and $\int_\w=\int\frac{d\w}{2\pi}$. As in Sec.~\ref{subsec_droplet} we identify $k$ with $1/L$ and approximate the self-energy $\Delta_k(i\w)$ by $Z_x(\w/v_k)^2$. The integral in~(\ref{chi2ter}) gives $\sim v_k^2/Z_x^4k^3$. Since $v_k=vK_k/K$ and $\delta_k''(0)\sim -1/K_k$ with $K_k\sim Kk^\theta$, one finally obtains 
\beq
\chi_2 \sim K k^\theta ,
\eeq 
which leads to~(\ref{chi2}).



%

\end{document}

%% file: definition.tex

\def\rhoeq{\hat\rho_{\rm eq}}

\newcommand{\marge}[1]{\marginpar{\scriptsize #1}}
\newcommand{\remarque}[1]{\marginpar{\scriptsize Remarque}{\it [#1]}}
\newcommand{\new}[1]{{\bf #1}}
\newlength{\textlarg}
\newcommand{\redbar}[1]{\textcolor{red}{\st{#1}}} 
\newcommand{\bluebar}[1]{\textcolor{blue}{\st{#1}}} 

\newcommand{\beq}{\begin{equation}}
\newcommand{\eeq}{\end{equation}}
\newcommand{\bfig}{\begin{figure}}
\newcommand{\efig}{\end{figure}}
\newcommand{\bline}{\begin{multline}}
\newcommand{\eline}{\end{multline}}
\newcommand{\bremark}{\begin{quotation} \noindent \small }
\newcommand{\eremark}{\end{quotation}}
\newcommand{\llbrace}{\left\lbrace}  
\newcommand{\rrbrace}{\right\rbrace}
\newcommand{\lbraket}{\left[}
\newcommand{\rbraket}{\right]}
\newcommand{\llangle}{\left\langle}
\newcommand{\rrangle}{\right\rangle} 

\newcommand{\Tr}{{\rm Tr}} 
\newcommand{\tr}{{\rm tr}} 
\newcommand{\sgn}{\,{\rm sgn}} 
\newcommand{\mean}[1]{\langle #1 \rangle}
\newcommand{\commu}[2]{[#1,#2]} 
\newcommand{\bra}[1]{\langle#1|}
\newcommand{\ket}[1]{|#1\rangle}
\newcommand{\braket}[2]{\langle #1|#2\rangle}
\newcommand{\ketbra}[2]{|#1\rangle\langle#2|}
\newcommand{\dbraket}[3]{\langle #1|#2|#3\rangle}
\newcommand{\tens}[1]{\overleftrightarrow{#1}}  
\newcommand{\vac}{|{\rm vac}\rangle} 
\newcommand{\bravac}{\langle{\rm vac}|}
\newcommand{\const}{{\rm const}} 
\newcommand{\atanh}{\,{\rm atanh}}
\newcommand{\cotanh}{\,{\rm cotanh}}

\newcommand{\ie}{i.e.\xspace}
\newcommand{\iet}{i.e.}
\newcommand{\eg}{e.g.\xspace}
\newcommand{\cc}{{\rm c.c.}} 
\newcommand{\hc}{{\rm h.c.}} 
\newcommand{\etal}{{\it et al. }}
\newcommand\eme{$^{\mbox{\footnotesize ème}}$\xspace}

\newcommand{\jhatbf}{\hat {\textbf \jold}} 
\newcommand{\Jhatbf}{\hat {\textbf \J}} 
\newcommand{\jhat}{\hat {\jmath}} 
\newcommand{\Jhat}{\hat {J}} 
\newcommand{\jbf}{\textbf j}
\newcommand{\Jbf}{\textbf J}

\def\chibf{\boldsymbol{\chi}}
\def\down{\downarrow}
\def\eps{\epsilon}
\def\gam{\gamma} 
\def\alphabf{\boldsymbol{\alpha}}
\def\phibf{\boldsymbol{\phi}}
\def\varphibf{\boldsymbol{\varphi}}
\def\varphibfs{\boldsymbol{\varphi}_<}
\def\varphibfl{\boldsymbol{\varphi}_>}
\def\varphis{\varphi_{<}}
\def\varphil{\varphi_{>}}
\def\psibf{\boldsymbol{\psi}}
\def\thetabf{\boldsymbol{\theta}}
\def\Ome{\Omega}
\def\omeD{{\omega_D}} 
\def\bfOme{\boldsymbol{\Omega}} 
\def\Omebf{\boldsymbol{\Omega}} 
\def\lamb{\lambda}
\def\Lamb{\Lambda}
\def\sig{\sigma}
\def\Sig{\Sigma}
\def\sigp{{\sigma'}} 
\def\bfsig{\boldsymbol{\sigma}} 
\def\sigbf{\boldsymbol{\sigma}} 
\def\bfSig{\boldsymbol{\Sigma}} 
\def\The{\Theta} 
\def\up{\uparrow}

\def\epsk{\epsilon_{\bf k}} 
\def\xik{\xi_{\bf k}} 
\def\txik{\tilde\xi_{\bf k}} 
\def\xip{\xi_{\bf p}} 
\def\xiq{\xi_{\bf q}} 
\def\xikq{\xi_{{\bf k}+{\bf q}}} 
\def\Ek{E_{\bf k}} 
\def\Ep{E_{\bf p}}
\def\Eq{E_{\bf q}}
\def\Heff{\hat H_{\rm eff}}
\def\Hem{\hat H_{\rm em}}
\def\Hint{\hat H_{\rm int}}
\def\Hloc{\hat H_{\rm loc}}
\def\HMF{\hat H_{\rm MF}}
\def\Sem{S_{\rm em}}
\def\SMF{S_{\rm MF}} 
\def\SHF{S_{\rm HF}} 
\def\SRPA{S_{\rm RPA}} 
\def\Sint{S_{\rm int}} 
\def\Sloc{S_{\rm loc}}
\def\TN{T_{\rm N}} 
\def\TNHF{T^{\rm HF}_{\rm N}} 
\def\Zloc{Z_{\rm loc}} 
\def\ZMF{Z_{\rm MF}} 
\def\ZHF{Z_{\rm HF}} 
\def\ZRPA{Z_{\rm RPA}} 
\def\RPA{{\rm RPA}}
\def\loc{{\rm loc}} 
\def\pp{{\rm pp}}
\def\ph{{\rm ph}} 
\def\ch{{\rm ch}}
\def\sp{{\rm sp}} 
\def\qtf{q_{\rm TF}}
\def\epstf{\eps^{}_{\rm TF}} 
\def\epsrpa{\eps^{}_{\rm RPA}} 
\def\chinnzpp{\chi_{nn}^{0}{}\!\!\!''}

\def\half{\frac{1}{2}}
\def\dhalf{\dfrac{1}{2}}
\def\third{\frac{1}{3}} 
\def\quarter{\frac{1}{4}}

\def\qr{{\bf q}\cdot{\bf r}}
\def\wt{\omega t} 

\def\a{{\bf a}}
\def\b{{\bf b}}
\newcommand{\cv}{{\bf c}} 
\def\e{{\bf e}}
\def\f{{\bf f}}
\def\g{{\bf g}}
\def\h{{\bf h}}
\def\jold{\char"11}
\def\j{{\bf j}}
\def\k{{\bf k}}
\def\l{{\bf l}}
\def\m{{\bf m}}
\def\n{{\bf n}} 
\def\p{{\bf p}} 
\def\q{{\bf q}}
\def\r{{\bf r}}
\def\t{{\bf t}}
\def\u{{\bf u}}
\newcommand{\vv}{{\bf v}}
\def\x{{\bf x}}
\def\y{{\bf y}} 
\def\z{{\bf z}} 
\def\A{{\bf A}}
\def\B{{\bf B}}
\def\D{{\bf D}} 
\def\E{{\bf E}} 
\def\F{{\bf F}} 
\def\H{{\bf H}}  
\def\J{{\bf J}}
\def\K{{\bf K}} 

\def\G{{\bf G}}
\def\L{{\bf L}}
\def\M{{\bf M}}  
\def\O{{\bf O}} 
\def\P{{\bf P}} 
\def\Q{{\bf Q}} 
\def\R{{\bf R}}
\def\S{{\bf S}}
\def\U{{\bf U}} 
\def\V{{\bf V}} 
\def\X{{\bf X}} 
\def\Y{{\bf Y}} 
\def\epsbf{\boldsymbol{\epsilon}}
\def\betabf{\boldsymbol{\beta}}
\def\deltabf{\boldsymbol{\delta}}
\def\mubf{\boldsymbol{\mu}}
\def\nablabf{\boldsymbol{\nabla}}
\def\rhobf{\boldsymbol{\rho}}
\def\sigmabf{\boldsymbol{\sigma}} 
\def\Pibf{\boldsymbol{\Pi}}
\def\pibf{\boldsymbol{\pi}}

\def\para{\parallel}
\def\kpara{{k_\parallel}}
\def\kperp{{k_\perp}} 
\def\kperpp{{k_\perp'}} 
\def\qperp{{q_\perp}} 
\def\tperp{{t_\perp}} 

\def\w{\omega}
\def\wn{\omega_n}
\def\wm{\omega_m}
\def\wnu{\omega_\nu}
\def\wp{\omega_p} 
\def\dmu{{\partial_\mu}}
\def\dnu{{\partial_\nu}}
\def\dl{{\partial_l}}  
\def\dt{\partial_t} 
\def\tdt{\tilde\partial_t}
\def\dk{\partial_k}
\def\tdk{\tilde\partial_k}
\def\dx{\partial_x}
\def\dy{\partial_y} 
\def\dtau{{\partial_\tau}}  
\def\det{{\rm det}} 
\def\Pf{{\rm Pf}}
\def\diag{{\rm diag}}

\def\dsum{\displaystyle \sum}
\def\dint{\displaystyle \int} 
\def\intt{\int_{-\infty}^\infty dt} 
\def\inttp{\int_{-\infty}^\infty dt'} 
\def\intk{\int_{\bf k}} 
\def\intkd{\int \frac{d^dk}{(2\pi)^d}}
\def\intq{\int_{\bf q}} 
\def\intr{\int d^dr}  
\def\dintr{\displaystyle \int d^dr} 
\def\intrp{\int d^dr'}
\def\dinttau{\displaystyle \int_0^\beta d\tau}
\def\dinttaup{\displaystyle \int_0^\beta d\tau'}
\def\inttau{\int_0^\beta d\tau}
\def\inttaup{\int_0^\beta d\tau'}
\def\intx{\int d^{d+1}x} 
\def\inttaur{\int_0^\beta d\tau \int d^dr}
\def\intinf{\int_{-\infty}^\infty}
\def\dinttaur{\displaystyle \int_0^\beta d\tau \int d^dr}
\def\dintinf{\displaystyle \int_{-\infty}^\infty}
\def\intw{\int_{-\infty}^\infty \frac{d\w}{2\pi}}
\def\sumr{\sum_{\bf r}} 

\def\calA{{\cal A}}
\def\calAbf{\bm{{\cal A}}}
\def\calB{{\cal B}} 
\def\calC{{\cal C}} 
\def\dt{\partial_t}
\def\calD{{\cal D}}
\def\calE{{\cal E}}
\def\calF{{\cal F}} 
\def\calFbf{\bm{{\cal F}}}
\def\calG{{\cal G}}
\def\calH{{\cal H}}
\def\calI{{\cal I}}
\def\calJ{{\cal J}}
\def\calK{{\cal K}}
\def\calL{{\cal L}} 
\def\calM{{\cal M}} 
\def\calN{{\cal N}}
\def\calO{{\cal O}}
\def\calP{{\cal P}}  
\def\calR{{\cal R}} 
\def\calS{{\cal S}}
\def\calT{{\cal T}}
\def\calU{{\cal U}}
\def\calV{{\cal V}}
\def\calX{{\cal X}} 
\def\calY{{\cal Y}} 
\def\calZ{{\cal Z}} 

\def\calbfB{{\bf \cal B}}
\def\calbfF{{\bf \cal F}}

\def\tT{{\tilde T}}
\def\talpha{{\tilde\alpha}}
\def\tbeta{{\tilde\beta}}
\def\tchi{{\tilde\chi}}
\def\tdelta{{\tilde\delta}}
\def\tDelta{{\tilde\Delta}}
\def\teta{{\tilde\eta}} 
\def\tlamb{{\tilde\lambda}}
\def\tmu{{\tilde\mu}}
\def\tphibf{{\tilde\phibf}}
\def\trho{{\tilde\rho}}
\def\tvarphibf{{\tilde\varphibf}} 
\def\tw{{\tilde\omega}}
\def\twn{{\tilde\omega_n}}
\def\twnu{{\tilde\omega_\nu}}

\def\asinh{{\rm asinh}} 

%% file: bg19b.bbl
\begin{thebibliography}{112}%
	\makeatletter
	\providecommand \@ifxundefined [1]{%
		\@ifx{#1\undefined}
	}%
	\providecommand \@ifnum [1]{%
		\ifnum #1\expandafter \@firstoftwo
		\else \expandafter \@secondoftwo
		\fi
	}%
	\providecommand \@ifx [1]{%
		\ifx #1\expandafter \@firstoftwo
		\else \expandafter \@secondoftwo
		\fi
	}%
	\providecommand \natexlab [1]{#1}%
	\providecommand \enquote  [1]{``#1''}%
	\providecommand \bibnamefont  [1]{#1}%
	\providecommand \bibfnamefont [1]{#1}%
	\providecommand \citenamefont [1]{#1}%
	\providecommand \href@noop [0]{\@secondoftwo}%
	\providecommand \href [0]{\begingroup \@sanitize@url \@href}%
	\providecommand \@href[1]{\@@startlink{#1}\@@href}%
	\providecommand \@@href[1]{\endgroup#1\@@endlink}%
	\providecommand \@sanitize@url [0]{\catcode `\\12\catcode `\$12\catcode
		`\&12\catcode `\#12\catcode `\^12\catcode `\_12\catcode `\%12\relax}%
	\providecommand \@@startlink[1]{}%
	\providecommand \@@endlink[0]{}%
	\providecommand \url  [0]{\begingroup\@sanitize@url \@url }%
	\providecommand \@url [1]{\endgroup\@href {#1}{\urlprefix }}%
	\providecommand \urlprefix  [0]{URL }%
	\providecommand \Eprint [0]{\href }%
	\providecommand \doibase [0]{http://dx.doi.org/}%
	\providecommand \selectlanguage [0]{\@gobble}%
	\providecommand \bibinfo  [0]{\@secondoftwo}%
	\providecommand \bibfield  [0]{\@secondoftwo}%
	\providecommand \translation [1]{[#1]}%
	\providecommand \BibitemOpen [0]{}%
	\providecommand \bibitemStop [0]{}%
	\providecommand \bibitemNoStop [0]{.\EOS\space}%
	\providecommand \EOS [0]{\spacefactor3000\relax}%
	\providecommand \BibitemShut  [1]{\csname bibitem#1\endcsname}%
	\let\auto@bib@innerbib\@empty
	\bibitem [{\citenamefont {Anderson}(1958)}]{Anderson58a}%
	\BibitemOpen
	\bibfield  {author} {\bibinfo {author} {\bibfnamefont {P.~W.}\ \bibnamefont
			{Anderson}},\ }\bibfield  {title} {\enquote {\bibinfo {title} {{Absence of
					Diffusion in Certain Random Lattices}},}\ }\href {\doibase
		10.1103/PhysRev.109.1492} {\bibfield  {journal} {\bibinfo  {journal} {Phys.
				Rev.}\ }\textbf {\bibinfo {volume} {109}},\ \bibinfo {pages} {1492} (\bibinfo
		{year} {1958})}\BibitemShut {NoStop}%
	\bibitem [{\citenamefont {Abrahams}(2010)}]{Anderson_50years}%
	\BibitemOpen
	\bibinfo {editor} {\bibfnamefont {E.}~\bibnamefont {Abrahams}},\ ed.,\ \href
	{\doibase 10.1142/7663} {\emph {\bibinfo {title} {50 Years of Anderson
				Localization}}}\ (\bibinfo  {publisher} {World Scientific},\ \bibinfo {year}
	{2010})\BibitemShut {NoStop}%
	\bibitem [{\citenamefont {Aspect}\ and\ \citenamefont
		{Inguscio}(2009)}]{Aspect09}%
	\BibitemOpen
	\bibfield  {author} {\bibinfo {author} {\bibfnamefont {A.}~\bibnamefont
			{Aspect}}\ and\ \bibinfo {author} {\bibfnamefont {M.}~\bibnamefont
			{Inguscio}},\ }\bibfield  {title} {\enquote {\bibinfo {title} {Anderson
				localization of ultracold atoms},}\ }\href {\doibase 10.1063/1.3206092}
	{\bibfield  {journal} {\bibinfo  {journal} {Physics Today}\ ,\ \bibinfo
			{pages} {30}} (\bibinfo {year} {2009})}\BibitemShut {NoStop}%
	\bibitem [{\citenamefont {Giamarchi}\ and\ \citenamefont
		{Schulz}(1987)}]{Giamarchi87}%
	\BibitemOpen
	\bibfield  {author} {\bibinfo {author} {\bibfnamefont {T.}~\bibnamefont
			{Giamarchi}}\ and\ \bibinfo {author} {\bibfnamefont {H.~J.}\ \bibnamefont
			{Schulz}},\ }\bibfield  {title} {\enquote {\bibinfo {title} {{Localization
					and Interaction in One-Dimensional Quantum Fluids}},}\ }\href {\doibase
		http://dx.doi.org/10.1209/0295-5075/3/12/007} {\bibfield  {journal} {\bibinfo
			{journal} {Europhys. Lett.}\ }\textbf {\bibinfo {volume} {3}},\ \bibinfo
		{pages} {1287} (\bibinfo {year} {1987})}\BibitemShut {NoStop}%
	\bibitem [{\citenamefont {Giamarchi}\ and\ \citenamefont
		{Schulz}(1988)}]{Giamarchi88}%
	\BibitemOpen
	\bibfield  {author} {\bibinfo {author} {\bibfnamefont {T.}~\bibnamefont
			{Giamarchi}}\ and\ \bibinfo {author} {\bibfnamefont {H.~J.}\ \bibnamefont
			{Schulz}},\ }\bibfield  {title} {\enquote {\bibinfo {title} {{Anderson
					localization and interactions in one-dimensional metals}},}\ }\href {\doibase
		10.1103/PhysRevB.37.325} {\bibfield  {journal} {\bibinfo  {journal} {Phys.
				Rev. B}\ }\textbf {\bibinfo {volume} {37}},\ \bibinfo {pages} {325} (\bibinfo
		{year} {1988})}\BibitemShut {NoStop}%
	\bibitem [{\citenamefont {Fisher}\ \emph {et~al.}(1989)\citenamefont {Fisher},
		\citenamefont {Weichman}, \citenamefont {Grinstein},\ and\ \citenamefont
		{Fisher}}]{Fisher89}%
	\BibitemOpen
	\bibfield  {author} {\bibinfo {author} {\bibfnamefont {M.~P.~A.}\
			\bibnamefont {Fisher}}, \bibinfo {author} {\bibfnamefont {P.~B.}\
			\bibnamefont {Weichman}}, \bibinfo {author} {\bibfnamefont {G.}~\bibnamefont
			{Grinstein}}, \ and\ \bibinfo {author} {\bibfnamefont {D.~S.}\ \bibnamefont
			{Fisher}},\ }\bibfield  {title} {\enquote {\bibinfo {title} {{Boson
					localization and the superfluid-insulator transition}},}\ }\href {\doibase
		10.1103/PhysRevB.40.546} {\bibfield  {journal} {\bibinfo  {journal} {Phys.
				Rev. B}\ }\textbf {\bibinfo {volume} {40}},\ \bibinfo {pages} {546} (\bibinfo
		{year} {1989})}\BibitemShut {NoStop}%
	\bibitem [{\citenamefont {Billy}\ \emph {et~al.}(2008)\citenamefont {Billy},
		\citenamefont {Josse}, \citenamefont {Zuo}, \citenamefont {Bernard},
		\citenamefont {Hambrecht}, \citenamefont {Lugan}, \citenamefont
		{Cl{\'e}ment}, \citenamefont {Sanchez-Palencia}, \citenamefont {Bouyer},\
		and\ \citenamefont {Aspect}}]{Billy08}%
	\BibitemOpen
	\bibfield  {author} {\bibinfo {author} {\bibfnamefont {J.}~\bibnamefont
			{Billy}}, \bibinfo {author} {\bibfnamefont {V.}~\bibnamefont {Josse}},
		\bibinfo {author} {\bibfnamefont {Z.}~\bibnamefont {Zuo}}, \bibinfo {author}
		{\bibfnamefont {A.}~\bibnamefont {Bernard}}, \bibinfo {author} {\bibfnamefont
			{B.}~\bibnamefont {Hambrecht}}, \bibinfo {author} {\bibfnamefont
			{P.}~\bibnamefont {Lugan}}, \bibinfo {author} {\bibfnamefont
			{D.}~\bibnamefont {Cl{\'e}ment}}, \bibinfo {author} {\bibfnamefont
			{L.}~\bibnamefont {Sanchez-Palencia}}, \bibinfo {author} {\bibfnamefont
			{P.}~\bibnamefont {Bouyer}}, \ and\ \bibinfo {author} {\bibfnamefont
			{A.}~\bibnamefont {Aspect}},\ }\bibfield  {title} {\enquote {\bibinfo {title}
			{{Direct observation of Anderson localization of matter waves in a controlled
					disorder}},}\ }\href {\doibase 10.1038/nature07000} {\bibfield  {journal}
		{\bibinfo  {journal} {Nature}\ }\textbf {\bibinfo {volume} {453}},\ \bibinfo
		{pages} {891} (\bibinfo {year} {2008})}\BibitemShut {NoStop}%
	\bibitem [{\citenamefont {Roati}\ \emph {et~al.}(2008)\citenamefont {Roati},
		\citenamefont {D'Errico}, \citenamefont {Fallani}, \citenamefont {Fattori},
		\citenamefont {Fort}, \citenamefont {Zaccanti}, \citenamefont {Modugno},
		\citenamefont {Modugno},\ and\ \citenamefont {Inguscio}}]{Roati08}%
	\BibitemOpen
	\bibfield  {author} {\bibinfo {author} {\bibfnamefont {G.}~\bibnamefont
			{Roati}}, \bibinfo {author} {\bibfnamefont {C.}~\bibnamefont {D'Errico}},
		\bibinfo {author} {\bibfnamefont {L.}~\bibnamefont {Fallani}}, \bibinfo
		{author} {\bibfnamefont {M.}~\bibnamefont {Fattori}}, \bibinfo {author}
		{\bibfnamefont {C.}~\bibnamefont {Fort}}, \bibinfo {author} {\bibfnamefont
			{M.}~\bibnamefont {Zaccanti}}, \bibinfo {author} {\bibfnamefont
			{G.}~\bibnamefont {Modugno}}, \bibinfo {author} {\bibfnamefont
			{M.}~\bibnamefont {Modugno}}, \ and\ \bibinfo {author} {\bibfnamefont
			{M.}~\bibnamefont {Inguscio}},\ }\bibfield  {title} {\enquote {\bibinfo
			{title} {{Anderson localization of a non-interacting Bose--Einstein
					condensate}},}\ }\href {\doibase 10.1038/nature07071} {\bibfield  {journal}
		{\bibinfo  {journal} {Nature}\ }\textbf {\bibinfo {volume} {453}},\ \bibinfo
		{pages} {895} (\bibinfo {year} {2008})}\BibitemShut {NoStop}%
	\bibitem [{\citenamefont {Pasienski}\ \emph {et~al.}(2010)\citenamefont
		{Pasienski}, \citenamefont {McKay}, \citenamefont {White},\ and\
		\citenamefont {DeMarco}}]{Pasienski10}%
	\BibitemOpen
	\bibfield  {author} {\bibinfo {author} {\bibfnamefont {M.}~\bibnamefont
			{Pasienski}}, \bibinfo {author} {\bibfnamefont {D.}~\bibnamefont {McKay}},
		\bibinfo {author} {\bibfnamefont {M.}~\bibnamefont {White}}, \ and\ \bibinfo
		{author} {\bibfnamefont {B.}~\bibnamefont {DeMarco}},\ }\bibfield  {title}
	{\enquote {\bibinfo {title} {A disordered insulator in an optical lattice},}\
	}\href {\doibase 10.1038/nphys1726} {\bibfield  {journal} {\bibinfo
			{journal} {Nat. Phys.}\ }\textbf {\bibinfo {volume} {6}},\ \bibinfo {pages}
		{677} (\bibinfo {year} {2010})}\BibitemShut {NoStop}%
	\bibitem [{\citenamefont {Hong}\ \emph {et~al.}(2010)\citenamefont {Hong},
		\citenamefont {Zheludev}, \citenamefont {Manaka},\ and\ \citenamefont
		{Regnault}}]{Hong10}%
	\BibitemOpen
	\bibfield  {author} {\bibinfo {author} {\bibfnamefont {Tao}\ \bibnamefont
			{Hong}}, \bibinfo {author} {\bibfnamefont {A.}~\bibnamefont {Zheludev}},
		\bibinfo {author} {\bibfnamefont {H.}~\bibnamefont {Manaka}}, \ and\ \bibinfo
		{author} {\bibfnamefont {L.-P.}\ \bibnamefont {Regnault}},\ }\bibfield
	{title} {\enquote {\bibinfo {title} {{Evidence of a magnetic Bose glass in
					${({\text{CH}}_{3})}_{2}{\text{CHNH}}_{3}\text{Cu}{({\text{Cl}}_{{0.95}}{\text{Br}}_{{0.05}})}_{3}$
					from neutron diffraction}},}\ }\href {\doibase 10.1103/PhysRevB.81.060410}
	{\bibfield  {journal} {\bibinfo  {journal} {Phys. Rev. B}\ }\textbf {\bibinfo
			{volume} {81}},\ \bibinfo {pages} {060410(R)} (\bibinfo {year}
		{2010})}\BibitemShut {NoStop}%
	\bibitem [{\citenamefont {Yamada}\ \emph {et~al.}(2011)\citenamefont {Yamada},
		\citenamefont {Tanaka}, \citenamefont {Ono},\ and\ \citenamefont
		{Nojiri}}]{Yamada11}%
	\BibitemOpen
	\bibfield  {author} {\bibinfo {author} {\bibfnamefont {F.}~\bibnamefont
			{Yamada}}, \bibinfo {author} {\bibfnamefont {H.}~\bibnamefont {Tanaka}},
		\bibinfo {author} {\bibfnamefont {T.}~\bibnamefont {Ono}}, \ and\ \bibinfo
		{author} {\bibfnamefont {H.}~\bibnamefont {Nojiri}},\ }\bibfield  {title}
	{\enquote {\bibinfo {title} {{Transition from Bose glass to a condensate of
					triplons in Tl${}_{1\ensuremath{-}x}$K${}_{x}$CuCl${}_{3}$}},}\ }\href
	{\doibase 10.1103/PhysRevB.83.020409} {\bibfield  {journal} {\bibinfo
			{journal} {Phys. Rev. B}\ }\textbf {\bibinfo {volume} {83}},\ \bibinfo
		{pages} {020409(R)} (\bibinfo {year} {2011})}\BibitemShut {NoStop}%
	\bibitem [{\citenamefont {Zheludev}\ and\ \citenamefont
		{Roscilde}(2013)}]{Zheludev13}%
	\BibitemOpen
	\bibfield  {author} {\bibinfo {author} {\bibfnamefont {A.}~\bibnamefont
			{Zheludev}}\ and\ \bibinfo {author} {\bibfnamefont {T.}~\bibnamefont
			{Roscilde}},\ }\bibfield  {title} {\enquote {\bibinfo {title} {Dirty-boson
				physics with magnetic insulators},}\ }\href {\doibase
		https://doi.org/10.1016/j.crhy.2013.10.001} {\bibfield  {journal} {\bibinfo
			{journal} {C. R. Phys.}\ }\textbf {\bibinfo {volume} {14}},\ \bibinfo {pages}
		{740} (\bibinfo {year} {2013})}\BibitemShut {NoStop}%
	\bibitem [{\citenamefont {Giamarchi}(2004)}]{Giamarchi_book}%
	\BibitemOpen
	\bibfield  {author} {\bibinfo {author} {\bibfnamefont {T.}~\bibnamefont
			{Giamarchi}},\ }\href@noop {} {\emph {\bibinfo {title} {{Quantum physics in
					one dimension}}}}\ (\bibinfo  {publisher} {Oxford University Press},\
	\bibinfo {address} {Oxford},\ \bibinfo {year} {2004})\BibitemShut {NoStop}%
	\bibitem [{\citenamefont {Fukuyama}\ and\ \citenamefont
		{Lee}(1978)}]{Fukuyama78}%
	\BibitemOpen
	\bibfield  {author} {\bibinfo {author} {\bibfnamefont {H.}~\bibnamefont
			{Fukuyama}}\ and\ \bibinfo {author} {\bibfnamefont {P.~A.}\ \bibnamefont
			{Lee}},\ }\bibfield  {title} {\enquote {\bibinfo {title} {{Dynamics of the
					charge-density wave. I. Impurity pinning in a single chain}},}\ }\href
	{\doibase 10.1103/PhysRevB.17.535} {\bibfield  {journal} {\bibinfo  {journal}
			{Phys. Rev. B}\ }\textbf {\bibinfo {volume} {17}},\ \bibinfo {pages} {535}
		(\bibinfo {year} {1978})}\BibitemShut {NoStop}%
	\bibitem [{\citenamefont {Houzet}\ and\ \citenamefont
		{Glazman}(2019)}]{Houzet19}%
	\BibitemOpen
	\bibfield  {author} {\bibinfo {author} {\bibfnamefont {Manuel}\ \bibnamefont
			{Houzet}}\ and\ \bibinfo {author} {\bibfnamefont {Leonid~I.}\ \bibnamefont
			{Glazman}},\ }\bibfield  {title} {\enquote {\bibinfo {title} {Microwave
				spectroscopy of a weakly pinned charge density wave in a superinductor},}\
	}\href {\doibase 10.1103/PhysRevLett.122.237701} {\bibfield  {journal}
		{\bibinfo  {journal} {Phys. Rev. Lett.}\ }\textbf {\bibinfo {volume} {122}},\
		\bibinfo {pages} {237701} (\bibinfo {year} {2019})}\BibitemShut {NoStop}%
	\bibitem [{\citenamefont {Giamarchi}\ and\ \citenamefont
		{Le~Doussal}(1996)}]{Giamarchi96}%
	\BibitemOpen
	\bibfield  {author} {\bibinfo {author} {\bibfnamefont {T.}~\bibnamefont
			{Giamarchi}}\ and\ \bibinfo {author} {\bibfnamefont {P.}~\bibnamefont
			{Le~Doussal}},\ }\bibfield  {title} {\enquote {\bibinfo {title} {{Variational
					theory of elastic manifolds with correlated disorder and localization of
					interacting quantum particles}},}\ }\href {\doibase
		10.1103/PhysRevB.53.15206} {\bibfield  {journal} {\bibinfo  {journal} {Phys.
				Rev. B}\ }\textbf {\bibinfo {volume} {53}},\ \bibinfo {pages} {15206}
		(\bibinfo {year} {1996})}\BibitemShut {NoStop}%
	\bibitem [{\citenamefont {Ristivojevic}\ \emph {et~al.}(2012)\citenamefont
		{Ristivojevic}, \citenamefont {Petkovi\'c}, \citenamefont {Le~Doussal},\ and\
		\citenamefont {Giamarchi}}]{Ristivojevic12}%
	\BibitemOpen
	\bibfield  {author} {\bibinfo {author} {\bibfnamefont {Z.}~\bibnamefont
			{Ristivojevic}}, \bibinfo {author} {\bibfnamefont {A.}~\bibnamefont
			{Petkovi\'c}}, \bibinfo {author} {\bibfnamefont {P.}~\bibnamefont
			{Le~Doussal}}, \ and\ \bibinfo {author} {\bibfnamefont {T.}~\bibnamefont
			{Giamarchi}},\ }\bibfield  {title} {\enquote {\bibinfo {title} {{Phase
					Transition of Interacting Disordered Bosons in One Dimension}},}\ }\href
	{\doibase 10.1103/PhysRevLett.109.026402} {\bibfield  {journal} {\bibinfo
			{journal} {Phys. Rev. Lett.}\ }\textbf {\bibinfo {volume} {109}},\ \bibinfo
		{pages} {026402} (\bibinfo {year} {2012})}\BibitemShut {NoStop}%
	\bibitem [{\citenamefont {Ristivojevic}\ \emph {et~al.}(2014)\citenamefont
		{Ristivojevic}, \citenamefont {Petkovi\'c}, \citenamefont {Le~Doussal},\ and\
		\citenamefont {Giamarchi}}]{Ristivojevic14}%
	\BibitemOpen
	\bibfield  {author} {\bibinfo {author} {\bibfnamefont {Z.}~\bibnamefont
			{Ristivojevic}}, \bibinfo {author} {\bibfnamefont {A.}~\bibnamefont
			{Petkovi\'c}}, \bibinfo {author} {\bibfnamefont {P.}~\bibnamefont
			{Le~Doussal}}, \ and\ \bibinfo {author} {\bibfnamefont {T.}~\bibnamefont
			{Giamarchi}},\ }\bibfield  {title} {\enquote {\bibinfo {title}
			{{Superfluid/Bose-glass transition in one dimension}},}\ }\href {\doibase
		10.1103/PhysRevB.90.125144} {\bibfield  {journal} {\bibinfo  {journal} {Phys.
				Rev. B}\ }\textbf {\bibinfo {volume} {90}},\ \bibinfo {pages} {125144}
		(\bibinfo {year} {2014})}\BibitemShut {NoStop}%
	\bibitem [{\citenamefont {Altman}\ \emph {et~al.}(2004)\citenamefont {Altman},
		\citenamefont {Kafri}, \citenamefont {Polkovnikov},\ and\ \citenamefont
		{Refael}}]{Altman04}%
	\BibitemOpen
	\bibfield  {author} {\bibinfo {author} {\bibfnamefont {Ehud}\ \bibnamefont
			{Altman}}, \bibinfo {author} {\bibfnamefont {Yariv}\ \bibnamefont {Kafri}},
		\bibinfo {author} {\bibfnamefont {Anatoli}\ \bibnamefont {Polkovnikov}}, \
		and\ \bibinfo {author} {\bibfnamefont {Gil}\ \bibnamefont {Refael}},\
	}\bibfield  {title} {\enquote {\bibinfo {title} {{Phase Transition in a
					System of One-Dimensional Bosons with Strong Disorder}},}\ }\href {\doibase
		10.1103/PhysRevLett.93.150402} {\bibfield  {journal} {\bibinfo  {journal}
			{Phys. Rev. Lett.}\ }\textbf {\bibinfo {volume} {93}},\ \bibinfo {pages}
		{150402} (\bibinfo {year} {2004})}\BibitemShut {NoStop}%
	\bibitem [{\citenamefont {Altman}\ \emph {et~al.}(2008)\citenamefont {Altman},
		\citenamefont {Kafri}, \citenamefont {Polkovnikov},\ and\ \citenamefont
		{Refael}}]{Altman08}%
	\BibitemOpen
	\bibfield  {author} {\bibinfo {author} {\bibfnamefont {Ehud}\ \bibnamefont
			{Altman}}, \bibinfo {author} {\bibfnamefont {Yariv}\ \bibnamefont {Kafri}},
		\bibinfo {author} {\bibfnamefont {Anatoli}\ \bibnamefont {Polkovnikov}}, \
		and\ \bibinfo {author} {\bibfnamefont {Gil}\ \bibnamefont {Refael}},\
	}\bibfield  {title} {\enquote {\bibinfo {title} {Insulating phases and
				superfluid-insulator transition of disordered boson chains},}\ }\href
	{\doibase 10.1103/PhysRevLett.100.170402} {\bibfield  {journal} {\bibinfo
			{journal} {Phys. Rev. Lett.}\ }\textbf {\bibinfo {volume} {100}},\ \bibinfo
		{pages} {170402} (\bibinfo {year} {2008})}\BibitemShut {NoStop}%
	\bibitem [{\citenamefont {Altman}\ \emph {et~al.}(2010)\citenamefont {Altman},
		\citenamefont {Kafri}, \citenamefont {Polkovnikov},\ and\ \citenamefont
		{Refael}}]{Altman10}%
	\BibitemOpen
	\bibfield  {author} {\bibinfo {author} {\bibfnamefont {Ehud}\ \bibnamefont
			{Altman}}, \bibinfo {author} {\bibfnamefont {Yariv}\ \bibnamefont {Kafri}},
		\bibinfo {author} {\bibfnamefont {Anatoli}\ \bibnamefont {Polkovnikov}}, \
		and\ \bibinfo {author} {\bibfnamefont {Gil}\ \bibnamefont {Refael}},\
	}\bibfield  {title} {\enquote {\bibinfo {title} {Superfluid-insulator
				transition of disordered bosons in one dimension},}\ }\href {\doibase
		10.1103/PhysRevB.81.174528} {\bibfield  {journal} {\bibinfo  {journal} {Phys.
				Rev. B}\ }\textbf {\bibinfo {volume} {81}},\ \bibinfo {pages} {174528}
		(\bibinfo {year} {2010})}\BibitemShut {NoStop}%
	\bibitem [{\citenamefont {Pielawa}\ and\ \citenamefont
		{Altman}(2013)}]{Pielawa13}%
	\BibitemOpen
	\bibfield  {author} {\bibinfo {author} {\bibfnamefont {Susanne}\ \bibnamefont
			{Pielawa}}\ and\ \bibinfo {author} {\bibfnamefont {Ehud}\ \bibnamefont
			{Altman}},\ }\bibfield  {title} {\enquote {\bibinfo {title} {Numerical
				evidence for strong randomness scaling at a superfluid-insulator transition
				of one-dimensional bosons},}\ }\href {\doibase 10.1103/PhysRevB.88.224201}
	{\bibfield  {journal} {\bibinfo  {journal} {Phys. Rev. B}\ }\textbf {\bibinfo
			{volume} {88}},\ \bibinfo {pages} {224201} (\bibinfo {year}
		{2013})}\BibitemShut {NoStop}%
	\bibitem [{\citenamefont {Pollet}\ \emph {et~al.}(2013)\citenamefont {Pollet},
		\citenamefont {Prokof'ev},\ and\ \citenamefont {Svistunov}}]{Pollet13}%
	\BibitemOpen
	\bibfield  {author} {\bibinfo {author} {\bibfnamefont {Lode}\ \bibnamefont
			{Pollet}}, \bibinfo {author} {\bibfnamefont {Nikolay~V.}\ \bibnamefont
			{Prokof'ev}}, \ and\ \bibinfo {author} {\bibfnamefont {Boris~V.}\
			\bibnamefont {Svistunov}},\ }\bibfield  {title} {\enquote {\bibinfo {title}
			{{Classical-field renormalization flow of one-dimensional disordered
					bosons}},}\ }\href {\doibase 10.1103/PhysRevB.87.144203} {\bibfield
		{journal} {\bibinfo  {journal} {Phys. Rev. B}\ }\textbf {\bibinfo {volume}
			{87}},\ \bibinfo {pages} {144203} (\bibinfo {year} {2013})}\BibitemShut
	{NoStop}%
	\bibitem [{\citenamefont {Pollet}\ \emph {et~al.}(2014)\citenamefont {Pollet},
		\citenamefont {Prokof'ev},\ and\ \citenamefont {Svistunov}}]{Pollet14}%
	\BibitemOpen
	\bibfield  {author} {\bibinfo {author} {\bibfnamefont {Lode}\ \bibnamefont
			{Pollet}}, \bibinfo {author} {\bibfnamefont {Nikolay~V.}\ \bibnamefont
			{Prokof'ev}}, \ and\ \bibinfo {author} {\bibfnamefont {Boris~V.}\
			\bibnamefont {Svistunov}},\ }\bibfield  {title} {\enquote {\bibinfo {title}
			{Asymptotically exact scenario of strong-disorder criticality in
				one-dimensional superfluids},}\ }\href {\doibase 10.1103/PhysRevB.89.054204}
	{\bibfield  {journal} {\bibinfo  {journal} {Phys. Rev. B}\ }\textbf {\bibinfo
			{volume} {89}},\ \bibinfo {pages} {054204} (\bibinfo {year}
		{2014})}\BibitemShut {NoStop}%
	\bibitem [{\citenamefont {Yao}\ \emph {et~al.}(2016)\citenamefont {Yao},
		\citenamefont {Pollet}, \citenamefont {Prokof'ev},\ and\ \citenamefont
		{Svistunov}}]{Yao16}%
	\BibitemOpen
	\bibfield  {author} {\bibinfo {author} {\bibfnamefont {Zhiyuan}\ \bibnamefont
			{Yao}}, \bibinfo {author} {\bibfnamefont {Lode}\ \bibnamefont {Pollet}},
		\bibinfo {author} {\bibfnamefont {N.}~\bibnamefont {Prokof'ev}}, \ and\
		\bibinfo {author} {\bibfnamefont {B.}~\bibnamefont {Svistunov}},\ }\bibfield
	{title} {\enquote {\bibinfo {title} {Superfluid-insulator transition in
				strongly disordered one-dimensional systems},}\ }\href {\doibase
		10.1088/1367-2630/18/4/045018} {\bibfield  {journal} {\bibinfo  {journal}
			{New J. Phys.}\ }\textbf {\bibinfo {volume} {18}},\ \bibinfo {pages} {045018}
		(\bibinfo {year} {2016})}\BibitemShut {NoStop}%
	\bibitem [{\citenamefont {Doggen}\ \emph {et~al.}(2017)\citenamefont {Doggen},
		\citenamefont {Lemari\'e}, \citenamefont {Capponi},\ and\ \citenamefont
		{Laflorencie}}]{Doggen17}%
	\BibitemOpen
	\bibfield  {author} {\bibinfo {author} {\bibfnamefont {E.~V.~H.}\
			\bibnamefont {Doggen}}, \bibinfo {author} {\bibfnamefont {G.}~\bibnamefont
			{Lemari\'e}}, \bibinfo {author} {\bibfnamefont {S.}~\bibnamefont {Capponi}},
		\ and\ \bibinfo {author} {\bibfnamefont {N.}~\bibnamefont {Laflorencie}},\
	}\bibfield  {title} {\enquote {\bibinfo {title} {{Weak- versus
					strong-disorder superfluid--Bose glass transition in one dimension}},}\
	}\href {\doibase 10.1103/PhysRevB.96.180202} {\bibfield  {journal} {\bibinfo
			{journal} {Phys. Rev. B}\ }\textbf {\bibinfo {volume} {96}},\ \bibinfo
		{pages} {180202(R)} (\bibinfo {year} {2017})}\BibitemShut {NoStop}%
	\bibitem [{\citenamefont {Fisher}(1985)}]{Fisher85}%
	\BibitemOpen
	\bibfield  {author} {\bibinfo {author} {\bibfnamefont {D.~S.}\ \bibnamefont
			{Fisher}},\ }\bibfield  {title} {\enquote {\bibinfo {title} {{Random fields,
					random anisotropies, nonlinear \ensuremath{\sigma} models, and dimensional
					reduction}},}\ }\href {\doibase 10.1103/PhysRevB.31.7233} {\bibfield
		{journal} {\bibinfo  {journal} {Phys. Rev. B}\ }\textbf {\bibinfo {volume}
			{31}},\ \bibinfo {pages} {7233} (\bibinfo {year} {1985})}\BibitemShut
	{NoStop}%
	\bibitem [{\citenamefont {Narayan}\ and\ \citenamefont
		{Fisher}(1992{\natexlab{a}})}]{Narayan92}%
	\BibitemOpen
	\bibfield  {author} {\bibinfo {author} {\bibfnamefont {O.}~\bibnamefont
			{Narayan}}\ and\ \bibinfo {author} {\bibfnamefont {D.~S.}\ \bibnamefont
			{Fisher}},\ }\bibfield  {title} {\enquote {\bibinfo {title} {{Dynamics of
					sliding charge-density waves in 4-\ensuremath{\epsilon} dimensions}},}\
	}\href {\doibase 10.1103/PhysRevLett.68.3615} {\bibfield  {journal} {\bibinfo
			{journal} {Phys. Rev. Lett.}\ }\textbf {\bibinfo {volume} {68}},\ \bibinfo
		{pages} {3615} (\bibinfo {year} {1992}{\natexlab{a}})}\BibitemShut {NoStop}%
	\bibitem [{\citenamefont {{T. Nattermann}}\ \emph {et~al.}(1992)\citenamefont
		{{T. Nattermann}}, \citenamefont {{S. Stepanow}}, \citenamefont {{L.-H.
				Tang}},\ and\ \citenamefont {{H. Leschhorn}}}]{Nattermann92}%
	\BibitemOpen
	\bibfield  {author} {\bibinfo {author} {\bibnamefont {{T. Nattermann}}},
		\bibinfo {author} {\bibnamefont {{S. Stepanow}}}, \bibinfo {author}
		{\bibnamefont {{L.-H. Tang}}}, \ and\ \bibinfo {author} {\bibnamefont {{H.
					Leschhorn}}},\ }\bibfield  {title} {\enquote {\bibinfo {title} {{Dynamics of
					interface depinning in a disordered medium}},}\ }\href {\doibase
		10.1051/jp2:1992214} {\bibfield  {journal} {\bibinfo  {journal} {J. Phys. II
				France}\ }\textbf {\bibinfo {volume} {2}},\ \bibinfo {pages} {1483} (\bibinfo
		{year} {1992})}\BibitemShut {NoStop}%
	\bibitem [{\citenamefont {Balents}(1993)}]{Balents93}%
	\BibitemOpen
	\bibfield  {author} {\bibinfo {author} {\bibfnamefont {L}~\bibnamefont
			{Balents}},\ }\bibfield  {title} {\enquote {\bibinfo {title} {{Localization
					of Elastic Layers by Correlated Disorder}},}\ }\href {\doibase
		10.1209/0295-5075/24/6/011} {\bibfield  {journal} {\bibinfo  {journal}
			{Europhys. Lett.}\ }\textbf {\bibinfo {volume} {24}},\ \bibinfo {pages} {489}
		(\bibinfo {year} {1993})}\BibitemShut {NoStop}%
	\bibitem [{\citenamefont {Chauve}\ \emph {et~al.}(2001)\citenamefont {Chauve},
		\citenamefont {Doussal},\ and\ \citenamefont {Wiese}}]{Chauve01}%
	\BibitemOpen
	\bibfield  {author} {\bibinfo {author} {\bibfnamefont {P.}~\bibnamefont
			{Chauve}}, \bibinfo {author} {\bibfnamefont {P.~Le}\ \bibnamefont {Doussal}},
		\ and\ \bibinfo {author} {\bibfnamefont {K.~J.}\ \bibnamefont {Wiese}},\
	}\bibfield  {title} {\enquote {\bibinfo {title} {{Renormalization of Pinned
					Elastic Systems: How Does It Work Beyond One Loop?}}}\ }\href {\doibase
		10.1103/physrevlett.86.1785} {\bibfield  {journal} {\bibinfo  {journal}
			{Phys. Rev. Lett.}\ }\textbf {\bibinfo {volume} {86}},\ \bibinfo {pages}
		{1785} (\bibinfo {year} {2001})}\BibitemShut {NoStop}%
	\bibitem [{\citenamefont {Le~Doussal}\ \emph {et~al.}(2004)\citenamefont
		{Le~Doussal}, \citenamefont {Wiese},\ and\ \citenamefont
		{Chauve}}]{Ledoussal04}%
	\BibitemOpen
	\bibfield  {author} {\bibinfo {author} {\bibfnamefont {P.}~\bibnamefont
			{Le~Doussal}}, \bibinfo {author} {\bibfnamefont {K.~J.}\ \bibnamefont
			{Wiese}}, \ and\ \bibinfo {author} {\bibfnamefont {P.}~\bibnamefont
			{Chauve}},\ }\bibfield  {title} {\enquote {\bibinfo {title} {{Functional
					renormalization group and the field theory of disordered elastic systems}},}\
	}\href {\doibase 10.1103/PhysRevE.69.026112} {\bibfield  {journal} {\bibinfo
			{journal} {Phys. Rev. E}\ }\textbf {\bibinfo {volume} {69}},\ \bibinfo
		{pages} {026112} (\bibinfo {year} {2004})}\BibitemShut {NoStop}%
	\bibitem [{\citenamefont {Tarjus}\ and\ \citenamefont
		{Tissier}(2004)}]{Tarjus04}%
	\BibitemOpen
	\bibfield  {author} {\bibinfo {author} {\bibfnamefont {G.}~\bibnamefont
			{Tarjus}}\ and\ \bibinfo {author} {\bibfnamefont {M.}~\bibnamefont
			{Tissier}},\ }\bibfield  {title} {\enquote {\bibinfo {title}
			{{Nonperturbative Functional Renormalization Group for Random-Field Models:
					The Way Out of Dimensional Reduction}},}\ }\href {\doibase
		10.1103/PhysRevLett.93.267008} {\bibfield  {journal} {\bibinfo  {journal}
			{Phys. Rev. Lett.}\ }\textbf {\bibinfo {volume} {93}},\ \bibinfo {pages}
		{267008} (\bibinfo {year} {2004})}\BibitemShut {NoStop}%
	\bibitem [{\citenamefont {Le~Doussal}(2010)}]{Ledoussal10}%
	\BibitemOpen
	\bibfield  {author} {\bibinfo {author} {\bibfnamefont {P.}~\bibnamefont
			{Le~Doussal}},\ }\bibfield  {title} {\enquote {\bibinfo {title} {{Exact
					results and open questions in first principle functional {RG}}},}\ }\href
	{\doibase 10.1016/j.aop.2009.10.010} {\bibfield  {journal} {\bibinfo
			{journal} {Ann. Phys.}\ }\textbf {\bibinfo {volume} {325}},\ \bibinfo {pages}
		{49} (\bibinfo {year} {2010})}\BibitemShut {NoStop}%
	\bibitem [{\citenamefont {Berges}\ \emph {et~al.}(2002)\citenamefont {Berges},
		\citenamefont {Tetradis},\ and\ \citenamefont {Wetterich}}]{Berges02}%
	\BibitemOpen
	\bibfield  {author} {\bibinfo {author} {\bibfnamefont {J.}~\bibnamefont
			{Berges}}, \bibinfo {author} {\bibfnamefont {N.}~\bibnamefont {Tetradis}}, \
		and\ \bibinfo {author} {\bibfnamefont {C.}~\bibnamefont {Wetterich}},\
	}\bibfield  {title} {\enquote {\bibinfo {title} {{Non-perturbative
					renormalization flow in quantum field theory and statistical physics}},}\
	}\href {\doibase doi:10.1016/S0370-1573(01)00098-9} {\bibfield  {journal}
		{\bibinfo  {journal} {Phys. Rep.}\ }\textbf {\bibinfo {volume} {363}},\
		\bibinfo {pages} {223} (\bibinfo {year} {2002})}\BibitemShut {NoStop}%
	\bibitem [{\citenamefont {Delamotte}(2012)}]{Delamotte12}%
	\BibitemOpen
	\bibfield  {author} {\bibinfo {author} {\bibfnamefont {B.}~\bibnamefont
			{Delamotte}},\ }\bibfield  {title} {\enquote {\bibinfo {title} {{An
					Introduction to the Nonperturbative Renormalization Group}},}\ }in\ \href
	{\doibase 10.1007/978-3-642-27320-9_2} {\emph {\bibinfo {booktitle}
			{{Renormalization Group and Effective Field Theory Approaches to Many-Body
					Systems}}}},\ \bibinfo {series} {Lecture Notes in Physics}, Vol.\ \bibinfo
	{volume} {852},\ \bibinfo {editor} {edited by\ \bibinfo {editor}
		{\bibfnamefont {A.}~\bibnamefont {Schwenk}}\ and\ \bibinfo {editor}
		{\bibfnamefont {J.}~\bibnamefont {Polonyi}}}\ (\bibinfo  {publisher}
	{Springer Berlin Heidelberg},\ \bibinfo {year} {2012})\ pp.\ \bibinfo {pages}
	{49--132}\BibitemShut {NoStop}%
	\bibitem [{\citenamefont {Kopietz}\ \emph {et~al.}(2010)\citenamefont
		{Kopietz}, \citenamefont {Bartosch},\ and\ \citenamefont
		{Sch\"utz}}]{Kopietz_book}%
	\BibitemOpen
	\bibfield  {author} {\bibinfo {author} {\bibfnamefont {P.}~\bibnamefont
			{Kopietz}}, \bibinfo {author} {\bibfnamefont {L.}~\bibnamefont {Bartosch}}, \
		and\ \bibinfo {author} {\bibfnamefont {F.}~\bibnamefont {Sch\"utz}},\ }\href
	{\doibase 10.1007/978-3-642-05094-7} {\emph {\bibinfo {title} {{Introduction
					to the Functional Renormalization Group}}}}\ (\bibinfo  {publisher}
	{Springer},\ \bibinfo {address} {Berlin},\ \bibinfo {year}
	{2010})\BibitemShut {NoStop}%
	\bibitem [{\citenamefont {Tarjus}\ and\ \citenamefont
		{Tissier}(2008)}]{Tarjus08}%
	\BibitemOpen
	\bibfield  {author} {\bibinfo {author} {\bibfnamefont {G.}~\bibnamefont
			{Tarjus}}\ and\ \bibinfo {author} {\bibfnamefont {M.}~\bibnamefont
			{Tissier}},\ }\bibfield  {title} {\enquote {\bibinfo {title}
			{{Nonperturbative functional renormalization group for random field models
					and related disordered systems. I. Effective average action formalism}},}\
	}\href {\doibase 10.1103/PhysRevB.78.024203} {\bibfield  {journal} {\bibinfo
			{journal} {Phys. Rev. B}\ }\textbf {\bibinfo {volume} {78}},\ \bibinfo
		{pages} {024203} (\bibinfo {year} {2008})}\BibitemShut {NoStop}%
	\bibitem [{\citenamefont {Tissier}\ and\ \citenamefont
		{Tarjus}(2008)}]{Tissier08}%
	\BibitemOpen
	\bibfield  {author} {\bibinfo {author} {\bibfnamefont {M.}~\bibnamefont
			{Tissier}}\ and\ \bibinfo {author} {\bibfnamefont {G.}~\bibnamefont
			{Tarjus}},\ }\bibfield  {title} {\enquote {\bibinfo {title} {{Nonperturbative
					functional renormalization group for random field models and related
					disordered systems. II. Results for the random field $O(N)$ model}},}\ }\href
	{\doibase 10.1103/PhysRevB.78.024204} {\bibfield  {journal} {\bibinfo
			{journal} {Phys. Rev. B}\ }\textbf {\bibinfo {volume} {78}},\ \bibinfo
		{pages} {024204} (\bibinfo {year} {2008})}\BibitemShut {NoStop}%
	\bibitem [{\citenamefont {Tissier}\ and\ \citenamefont
		{Tarjus}(2012{\natexlab{a}})}]{Tissier12}%
	\BibitemOpen
	\bibfield  {author} {\bibinfo {author} {\bibfnamefont {M.}~\bibnamefont
			{Tissier}}\ and\ \bibinfo {author} {\bibfnamefont {G.}~\bibnamefont
			{Tarjus}},\ }\bibfield  {title} {\enquote {\bibinfo {title} {{Nonperturbative
					functional renormalization group for random field models and related
					disordered systems. III. Superfield formalism and ground-state dominance}},}\
	}\href {\doibase 10.1103/PhysRevB.85.104202} {\bibfield  {journal} {\bibinfo
			{journal} {Phys. Rev. B}\ }\textbf {\bibinfo {volume} {85}},\ \bibinfo
		{pages} {104202} (\bibinfo {year} {2012}{\natexlab{a}})}\BibitemShut
	{NoStop}%
	\bibitem [{\citenamefont {Tissier}\ and\ \citenamefont
		{Tarjus}(2012{\natexlab{b}})}]{Tissier12a}%
	\BibitemOpen
	\bibfield  {author} {\bibinfo {author} {\bibfnamefont {M.}~\bibnamefont
			{Tissier}}\ and\ \bibinfo {author} {\bibfnamefont {G.}~\bibnamefont
			{Tarjus}},\ }\bibfield  {title} {\enquote {\bibinfo {title} {{Nonperturbative
					functional renormalization group for random field models and related
					disordered systems. IV. Supersymmetry and its spontaneous breaking}},}\
	}\href {\doibase 10.1103/PhysRevB.85.104203} {\bibfield  {journal} {\bibinfo
			{journal} {Phys. Rev. B}\ }\textbf {\bibinfo {volume} {85}},\ \bibinfo
		{pages} {104203} (\bibinfo {year} {2012}{\natexlab{b}})}\BibitemShut
	{NoStop}%
	\bibitem [{\citenamefont {{Tarjus}}\ and\ \citenamefont
		{{Tissier}}(2019)}]{Tarjus19}%
	\BibitemOpen
	\bibfield  {author} {\bibinfo {author} {\bibfnamefont {Gilles}\ \bibnamefont
			{{Tarjus}}}\ and\ \bibinfo {author} {\bibfnamefont {Matthieu}\ \bibnamefont
			{{Tissier}}},\ }\href@noop {} {\enquote {\bibinfo {title} {{Random-field
					Ising and $O(N)$ models: Theoretical description through the functional
					renormalization group}},}\ } (\bibinfo {year} {2019}),\ \Eprint
	{http://arxiv.org/abs/1910.03530} {arXiv:1910.03530 [cond-mat.dis-nn]}
	\BibitemShut {NoStop}%
	\bibitem [{\citenamefont {Balog}\ \emph {et~al.}(2019)\citenamefont {Balog},
		\citenamefont {Tarjus},\ and\ \citenamefont {Tissier}}]{Balog19a}%
	\BibitemOpen
	\bibfield  {author} {\bibinfo {author} {\bibfnamefont {Ivan}\ \bibnamefont
			{Balog}}, \bibinfo {author} {\bibfnamefont {Gilles}\ \bibnamefont {Tarjus}},
		\ and\ \bibinfo {author} {\bibfnamefont {Matthieu}\ \bibnamefont {Tissier}},\
	}\bibfield  {title} {\enquote {\bibinfo {title} {Benchmarking the
				nonperturbative functional renormalization group approach on the random
				elastic manifold model in and out of equilibrium},}\ }\href {\doibase
		10.1088/1742-5468/ab3da5} {\bibfield  {journal} {\bibinfo  {journal} {J.
				Stat. Mech: Theory Exp.}\ }\textbf {\bibinfo {volume} {2019}},\ \bibinfo
		{pages} {103301} (\bibinfo {year} {2019})}\BibitemShut {NoStop}%
	\bibitem [{\citenamefont {Dupuis}(2019)}]{Dupuis19}%
	\BibitemOpen
	\bibfield  {author} {\bibinfo {author} {\bibfnamefont {Nicolas}\ \bibnamefont
			{Dupuis}},\ }\bibfield  {title} {\enquote {\bibinfo {title} {Glassy
				properties of the Bose-glass phase of a one-dimensional disordered Bose
				fluid},}\ }\href {\doibase 10.1103/PhysRevE.100.030102} {\bibfield  {journal}
		{\bibinfo  {journal} {Phys. Rev. E}\ }\textbf {\bibinfo {volume} {100}},\
		\bibinfo {pages} {030102(R)} (\bibinfo {year} {2019})}\BibitemShut {NoStop}%
	\bibitem [{\citenamefont {Balents}\ \emph {et~al.}(1996)\citenamefont
		{Balents}, \citenamefont {Bouchaud},\ and\ \citenamefont
		{M{\'{e}}zard}}]{Balents96}%
	\BibitemOpen
	\bibfield  {author} {\bibinfo {author} {\bibfnamefont {L.}~\bibnamefont
			{Balents}}, \bibinfo {author} {\bibfnamefont {J.-P.}\ \bibnamefont
			{Bouchaud}}, \ and\ \bibinfo {author} {\bibfnamefont {M.}~\bibnamefont
			{M{\'{e}}zard}},\ }\bibfield  {title} {\enquote {\bibinfo {title} {{The Large
					Scale Energy Landscape of Randomly Pinned Objects}},}\ }\href {\doibase
		10.1051/jp1:1996112} {\bibfield  {journal} {\bibinfo  {journal} {J. Phys. I}\
		}\textbf {\bibinfo {volume} {6}},\ \bibinfo {pages} {1007} (\bibinfo {year}
		{1996})}\BibitemShut {NoStop}%
	\bibitem [{\citenamefont {Fisher}\ and\ \citenamefont
		{Huse}(1988)}]{Fisher88b}%
	\BibitemOpen
	\bibfield  {author} {\bibinfo {author} {\bibfnamefont {D.~S.}\ \bibnamefont
			{Fisher}}\ and\ \bibinfo {author} {\bibfnamefont {D.~A.}\ \bibnamefont
			{Huse}},\ }\bibfield  {title} {\enquote {\bibinfo {title} {{Equilibrium
					behavior of the spin-glass ordered phase}},}\ }\href {\doibase
		10.1103/PhysRevB.38.386} {\bibfield  {journal} {\bibinfo  {journal} {Phys.
				Rev. B}\ }\textbf {\bibinfo {volume} {38}},\ \bibinfo {pages} {386} (\bibinfo
		{year} {1988})}\BibitemShut {NoStop}%
	\bibitem [{\citenamefont {Haldane}(1981)}]{Haldane81}%
	\BibitemOpen
	\bibfield  {author} {\bibinfo {author} {\bibfnamefont {F.~D.~M.}\
			\bibnamefont {Haldane}},\ }\bibfield  {title} {\enquote {\bibinfo {title}
			{{Effective Harmonic-Fluid Approach to Low-Energy Properties of
					One-Dimensional Quantum Fluids}},}\ }\href {\doibase
		10.1103/PhysRevLett.47.1840} {\bibfield  {journal} {\bibinfo  {journal}
			{Phys. Rev. Lett.}\ }\textbf {\bibinfo {volume} {47}},\ \bibinfo {pages}
		{1840} (\bibinfo {year} {1981})}\BibitemShut {NoStop}%
	\bibitem [{\citenamefont {Cazalilla}\ \emph {et~al.}(2011)\citenamefont
		{Cazalilla}, \citenamefont {Citro}, \citenamefont {Giamarchi}, \citenamefont
		{Orignac},\ and\ \citenamefont {Rigol}}]{Cazalilla11}%
	\BibitemOpen
	\bibfield  {author} {\bibinfo {author} {\bibfnamefont {M.~A.}\ \bibnamefont
			{Cazalilla}}, \bibinfo {author} {\bibfnamefont {R.}~\bibnamefont {Citro}},
		\bibinfo {author} {\bibfnamefont {T.}~\bibnamefont {Giamarchi}}, \bibinfo
		{author} {\bibfnamefont {E.}~\bibnamefont {Orignac}}, \ and\ \bibinfo
		{author} {\bibfnamefont {M.}~\bibnamefont {Rigol}},\ }\bibfield  {title}
	{\enquote {\bibinfo {title} {{One dimensional bosons: From condensed matter
					systems to ultracold gases}},}\ }\href {\doibase 10.1103/RevModPhys.83.1405}
	{\bibfield  {journal} {\bibinfo  {journal} {Rev. Mod. Phys.}\ }\textbf
		{\bibinfo {volume} {83}},\ \bibinfo {pages} {1405} (\bibinfo {year}
		{2011})}\BibitemShut {NoStop}%
	\bibitem [{\citenamefont {M\'ezard}\ \emph {et~al.}(1987)\citenamefont
		{M\'ezard}, \citenamefont {Parisi},\ and\ \citenamefont
		{Virasoo}}]{Mezard87}%
	\BibitemOpen
	\bibfield  {author} {\bibinfo {author} {\bibfnamefont {M.}~\bibnamefont
			{M\'ezard}}, \bibinfo {author} {\bibfnamefont {G.}~\bibnamefont {Parisi}}, \
		and\ \bibinfo {author} {\bibfnamefont {M.~A.}\ \bibnamefont {Virasoo}},\
	}\href@noop {} {\emph {\bibinfo {title} {Spin glass theory and beyond}}},\
	World Scientific lecture notes in physics 9\ (\bibinfo  {publisher} {World
		Scientific},\ \bibinfo {year} {1987})\BibitemShut {NoStop}%
	\bibitem [{\citenamefont {Fisher}(1986)}]{Fisher86}%
	\BibitemOpen
	\bibfield  {author} {\bibinfo {author} {\bibfnamefont {D.~S.}\ \bibnamefont
			{Fisher}},\ }\bibfield  {title} {\enquote {\bibinfo {title} {{Interface
					Fluctuations in Disordered Systems: $5-\epsilon$ Expansion and Failure of
					Dimensional Reduction}},}\ }\href {\doibase 10.1103/physrevlett.56.1964}
	{\bibfield  {journal} {\bibinfo  {journal} {Phys. Rev. Lett.}\ }\textbf
		{\bibinfo {volume} {56}},\ \bibinfo {pages} {1964} (\bibinfo {year}
		{1986})}\BibitemShut {NoStop}%
	\bibitem [{\citenamefont {Chauve}\ \emph {et~al.}(2000)\citenamefont {Chauve},
		\citenamefont {Giamarchi},\ and\ \citenamefont {Le~Doussal}}]{Chauve00}%
	\BibitemOpen
	\bibfield  {author} {\bibinfo {author} {\bibfnamefont {P.}~\bibnamefont
			{Chauve}}, \bibinfo {author} {\bibfnamefont {T.}~\bibnamefont {Giamarchi}}, \
		and\ \bibinfo {author} {\bibfnamefont {P.}~\bibnamefont {Le~Doussal}},\
	}\bibfield  {title} {\enquote {\bibinfo {title} {{Creep and depinning in
					disordered media}},}\ }\href {\doibase 10.1103/PhysRevB.62.6241} {\bibfield
		{journal} {\bibinfo  {journal} {Phys. Rev. B}\ }\textbf {\bibinfo {volume}
			{62}},\ \bibinfo {pages} {6241} (\bibinfo {year} {2000})}\BibitemShut
	{NoStop}%
	\bibitem [{\citenamefont {Le~Doussal}\ \emph {et~al.}(2002)\citenamefont
		{Le~Doussal}, \citenamefont {Wiese},\ and\ \citenamefont
		{Chauve}}]{Ledoussal02a}%
	\BibitemOpen
	\bibfield  {author} {\bibinfo {author} {\bibfnamefont {P.}~\bibnamefont
			{Le~Doussal}}, \bibinfo {author} {\bibfnamefont {K.~J.}\ \bibnamefont
			{Wiese}}, \ and\ \bibinfo {author} {\bibfnamefont {P.}~\bibnamefont
			{Chauve}},\ }\bibfield  {title} {\enquote {\bibinfo {title} {{Two-loop
					functional renormalization group theory of the depinning transition}},}\
	}\href {\doibase 10.1103/PhysRevB.66.174201} {\bibfield  {journal} {\bibinfo
			{journal} {Phys. Rev. B}\ }\textbf {\bibinfo {volume} {66}},\ \bibinfo
		{pages} {174201} (\bibinfo {year} {2002})}\BibitemShut {NoStop}%
	\bibitem [{\citenamefont {Le~Doussal}(2006)}]{Ledoussal06a}%
	\BibitemOpen
	\bibfield  {author} {\bibinfo {author} {\bibfnamefont {P.}~\bibnamefont
			{Le~Doussal}},\ }\bibfield  {title} {\enquote {\bibinfo {title}
			{{Finite-temperature functional RG, droplets and decaying Burgers
					turbulence}},}\ }\href {\doibase 10.1209/epl/i2006-10295-1} {\bibfield
		{journal} {\bibinfo  {journal} {Europhys. Lett.}\ }\textbf {\bibinfo {volume}
			{76}},\ \bibinfo {pages} {457} (\bibinfo {year} {2006})}\BibitemShut
	{NoStop}%
	\bibitem [{\citenamefont {Balents}\ and\ \citenamefont
		{Le~Doussal}(2004)}]{Balents04}%
	\BibitemOpen
	\bibfield  {author} {\bibinfo {author} {\bibfnamefont {L.}~\bibnamefont
			{Balents}}\ and\ \bibinfo {author} {\bibfnamefont {P.}~\bibnamefont
			{Le~Doussal}},\ }\bibfield  {title} {\enquote {\bibinfo {title} {{Field
					theory of statics and dynamics of glasses: Rare events and barrier
					distributions}},}\ }\href {\doibase
		https://doi.org/10.1209/epl/i2003-10170-7} {\bibfield  {journal} {\bibinfo
			{journal} {Europhys. Lett.}\ }\textbf {\bibinfo {volume} {65}},\ \bibinfo
		{pages} {685} (\bibinfo {year} {2004})}\BibitemShut {NoStop}%
	\bibitem [{\citenamefont {Fedorenko}(2008)}]{Fedorenko08}%
	\BibitemOpen
	\bibfield  {author} {\bibinfo {author} {\bibfnamefont {A.~A.}\ \bibnamefont
			{Fedorenko}},\ }\bibfield  {title} {\enquote {\bibinfo {title} {{Elastic
					systems with correlated disorder: Response to tilt and application to surface
					growth}},}\ }\href {\doibase 10.1103/PhysRevB.77.094203} {\bibfield
		{journal} {\bibinfo  {journal} {Phys. Rev. B}\ }\textbf {\bibinfo {volume}
			{77}},\ \bibinfo {pages} {094203} (\bibinfo {year} {2008})}\BibitemShut
	{NoStop}%
	\bibitem [{\citenamefont {Wetterich}(1993)}]{Wetterich93}%
	\BibitemOpen
	\bibfield  {author} {\bibinfo {author} {\bibfnamefont {C.}~\bibnamefont
			{Wetterich}},\ }\bibfield  {title} {\enquote {\bibinfo {title} {{Exact
					evolution equation for the effective potential}},}\ }\href {\doibase
		doi:10.1016/0370-2693(93)90726-X} {\bibfield  {journal} {\bibinfo  {journal}
			{Phys. Lett. B}\ }\textbf {\bibinfo {volume} {301}},\ \bibinfo {pages} {90}
		(\bibinfo {year} {1993})}\BibitemShut {NoStop}%
	\bibitem [{\citenamefont {Ellwanger}(1994)}]{Ellwanger94}%
	\BibitemOpen
	\bibfield  {author} {\bibinfo {author} {\bibfnamefont {Ulrich}\ \bibnamefont
			{Ellwanger}},\ }\bibfield  {title} {\enquote {\bibinfo {title} {Flow
				equations for $n$ point functions and bound states},}\ }\href {\doibase
		10.1007/BF01555911} {\bibfield  {journal} {\bibinfo  {journal} {Z. Phys. C}\
		}\textbf {\bibinfo {volume} {62}},\ \bibinfo {pages} {503} (\bibinfo {year}
		{1994})}\BibitemShut {NoStop}%
	\bibitem [{\citenamefont {Morris}(1994)}]{Morris94}%
	\BibitemOpen
	\bibfield  {author} {\bibinfo {author} {\bibfnamefont {T.~R.}\ \bibnamefont
			{Morris}},\ }\bibfield  {title} {\enquote {\bibinfo {title} {{The exact
					renormalization group and approximate solutions}},}\ }\href {\doibase
		10.1142/S0217751X94000972} {\bibfield  {journal} {\bibinfo  {journal} {Int.
				J. Mod. Phys. A}\ }\textbf {\bibinfo {volume} {09}},\ \bibinfo {pages} {2411}
		(\bibinfo {year} {1994})}\BibitemShut {NoStop}%
	\bibitem [{\citenamefont {Schulz}\ \emph {et~al.}(1988)\citenamefont {Schulz},
		\citenamefont {Villain}, \citenamefont {Br{\'e}zin},\ and\ \citenamefont
		{Orland}}]{Schulz88}%
	\BibitemOpen
	\bibfield  {author} {\bibinfo {author} {\bibfnamefont {U.}~\bibnamefont
			{Schulz}}, \bibinfo {author} {\bibfnamefont {J.}~\bibnamefont {Villain}},
		\bibinfo {author} {\bibfnamefont {E.}~\bibnamefont {Br{\'e}zin}}, \ and\
		\bibinfo {author} {\bibfnamefont {H.}~\bibnamefont {Orland}},\ }\bibfield
	{title} {\enquote {\bibinfo {title} {{Thermal fluctuations in some random
					field models}},}\ }\href {\doibase 10.1007/BF01015318} {\bibfield  {journal}
		{\bibinfo  {journal} {J. Stat. Phys.}\ }\textbf {\bibinfo {volume} {51}},\
		\bibinfo {pages} {1} (\bibinfo {year} {1988})}\BibitemShut {NoStop}%
	\bibitem [{not({\natexlab{a}})}]{not88}%
	\BibitemOpen
	\href@noop {} {} \bibinfo {note} {This quadratic
		approximation involving time derivatives to inifinite order is known as the
		LPA'', see Refs.~\onlinecite{Hasselmann12,*Rose18}.}\BibitemShut {Stop}%
	\bibitem [{not({\natexlab{b}})}]{not140}%
	\BibitemOpen
	\href@noop {} {}  \bibinfo {note} {This follows from a
		dimensional analysis of the action~(\ref{action}) and the fact that
		$[e^{i2\varphi}]=K$ in the Luttinger liquid.}\BibitemShut {Stop}%
	\bibitem [{\citenamefont {Chaikin}\ and\ \citenamefont
		{Lubensky}(1995)}]{Chaikin_book}%
	\BibitemOpen
	\bibfield  {author} {\bibinfo {author} {\bibfnamefont {P.~M.}\ \bibnamefont
			{Chaikin}}\ and\ \bibinfo {author} {\bibfnamefont {T.~C.}\ \bibnamefont
			{Lubensky}},\ }\href@noop {} {\emph {\bibinfo {title} {{Principles of
					Condensed Matter Physics}}}}\ (\bibinfo  {publisher} {Cambridge University
		Press},\ \bibinfo {year} {1995})\BibitemShut {NoStop}%
	\bibitem [{\citenamefont {Larkin}(1970)}]{Larkin70}%
	\BibitemOpen
	\bibfield  {author} {\bibinfo {author} {\bibfnamefont {A~I.}\ \bibnamefont
			{Larkin}},\ }\href@noop {} {\bibfield  {journal} {\bibinfo  {journal} {Sov.
				Phys. JETP}\ }\textbf {\bibinfo {volume} {31}},\ \bibinfo {pages} {784}
		(\bibinfo {year} {1970})}\BibitemShut {NoStop}%
	\bibitem [{\citenamefont {Larkin}\ and\ \citenamefont
		{Ovchinnikov}(1979)}]{Larkin79}%
	\BibitemOpen
	\bibfield  {author} {\bibinfo {author} {\bibfnamefont {A.~I.}\ \bibnamefont
			{Larkin}}\ and\ \bibinfo {author} {\bibfnamefont {Yu.~N.}\ \bibnamefont
			{Ovchinnikov}},\ }\bibfield  {title} {\enquote {\bibinfo {title} {Pinning in
				type II superconductors},}\ }\href {\doibase 10.1007/BF00117160} {\bibfield
		{journal} {\bibinfo  {journal} {J. Low Temp. Phys.}\ }\textbf {\bibinfo
			{volume} {34}},\ \bibinfo {pages} {409} (\bibinfo {year} {1979})}\BibitemShut
	{NoStop}%
	\bibitem [{not({\natexlab{c}})}]{not160}%
	\BibitemOpen
	\href@noop {} {}  \bibinfo {note} {$L_c$ can be simply
		obtained from the Hamiltonian $\hat H_0+\hat H_{\rm dis}$
		[Eqs.~(\ref{H0},\ref{Hdis})] ignoring quantum fluctuations and comparing the
		typical ``elastic'' energy $v/KL$ with the typical potential energy
		$\sqrt{\calD L}$ associated with fluctuations of the field $\varphi$ at
		length scale $L$.}\BibitemShut {Stop}%
	\bibitem [{not({\natexlab{d}})}]{not161}%
	\BibitemOpen
	\href@noop {} {} \bibinfo {note} {Perturbative RG yields
		the characteristic length~(\ref{Larkinlength}) with the exponent $1/3$
		replaced by $1/(3-2K)$.}\BibitemShut {Stop}%
	\bibitem [{\citenamefont {Narayan}\ and\ \citenamefont
		{Fisher}(1992{\natexlab{b}})}]{Narayan92a}%
	\BibitemOpen
	\bibfield  {author} {\bibinfo {author} {\bibfnamefont {O.}~\bibnamefont
			{Narayan}}\ and\ \bibinfo {author} {\bibfnamefont {D.~S.}\ \bibnamefont
			{Fisher}},\ }\bibfield  {title} {\enquote {\bibinfo {title} {{Critical
					behavior of sliding charge-density waves in $4-\epsilon$ dimensions}},}\
	}\href {\doibase 10.1103/PhysRevB.46.11520} {\bibfield  {journal} {\bibinfo
			{journal} {Phys. Rev. B}\ }\textbf {\bibinfo {volume} {46}},\ \bibinfo
		{pages} {11520} (\bibinfo {year} {1992}{\natexlab{b}})}\BibitemShut {NoStop}%
	\bibitem [{\citenamefont {Le~Doussal}\ and\ \citenamefont
		{Wiese}(2009)}]{Ledoussal09}%
	\BibitemOpen
	\bibfield  {author} {\bibinfo {author} {\bibfnamefont {Pierre}\ \bibnamefont
			{Le~Doussal}}\ and\ \bibinfo {author} {\bibfnamefont {Kay~J\"org}\
			\bibnamefont {Wiese}},\ }\bibfield  {title} {\enquote {\bibinfo {title}
			{{Size distributions of shocks and static avalanches from the functional
					renormalization group}},}\ }\href {\doibase 10.1103/PhysRevE.79.051106}
	{\bibfield  {journal} {\bibinfo  {journal} {Phys. Rev. E}\ }\textbf {\bibinfo
			{volume} {79}},\ \bibinfo {pages} {051106} (\bibinfo {year}
		{2009})}\BibitemShut {NoStop}%
	\bibitem [{\citenamefont {Nattermann}\ \emph {et~al.}(2003)\citenamefont
		{Nattermann}, \citenamefont {Giamarchi},\ and\ \citenamefont
		{Le~Doussal}}]{Nattermann03}%
	\BibitemOpen
	\bibfield  {author} {\bibinfo {author} {\bibfnamefont {T.}~\bibnamefont
			{Nattermann}}, \bibinfo {author} {\bibfnamefont {T.}~\bibnamefont
			{Giamarchi}}, \ and\ \bibinfo {author} {\bibfnamefont {P.}~\bibnamefont
			{Le~Doussal}},\ }\bibfield  {title} {\enquote {\bibinfo {title}
			{{Variable-Range Hopping and Quantum Creep in One Dimension}},}\ }\href
	{\doibase 10.1103/PhysRevLett.91.056603} {\bibfield  {journal} {\bibinfo
			{journal} {Phys. Rev. Lett.}\ }\textbf {\bibinfo {volume} {91}},\ \bibinfo
		{pages} {056603} (\bibinfo {year} {2003})}\BibitemShut {NoStop}%
	\bibitem [{\citenamefont {Malinin}\ \emph {et~al.}(2004)\citenamefont
		{Malinin}, \citenamefont {Nattermann},\ and\ \citenamefont
		{Rosenow}}]{Malinin04}%
	\BibitemOpen
	\bibfield  {author} {\bibinfo {author} {\bibfnamefont {S.~V.}\ \bibnamefont
			{Malinin}}, \bibinfo {author} {\bibfnamefont {T.}~\bibnamefont {Nattermann}},
		\ and\ \bibinfo {author} {\bibfnamefont {B.}~\bibnamefont {Rosenow}},\
	}\bibfield  {title} {\enquote {\bibinfo {title} {{Quantum creep and
					variable-range hopping of one-dimensional interacting electrons}},}\ }\href
	{\doibase 10.1103/PhysRevB.70.235120} {\bibfield  {journal} {\bibinfo
			{journal} {Phys. Rev. B}\ }\textbf {\bibinfo {volume} {70}},\ \bibinfo
		{pages} {235120} (\bibinfo {year} {2004})}\BibitemShut {NoStop}%
	\bibitem [{\citenamefont {Rosenow}\ and\ \citenamefont
		{Nattermann}(2006)}]{Rosenow06}%
	\BibitemOpen
	\bibfield  {author} {\bibinfo {author} {\bibfnamefont {Bernd}\ \bibnamefont
			{Rosenow}}\ and\ \bibinfo {author} {\bibfnamefont {Thomas}\ \bibnamefont
			{Nattermann}},\ }\bibfield  {title} {\enquote {\bibinfo {title} {{Nonlinear
					ac conductivity of interacting one-dimensional electron systems}},}\ }\href
	{\doibase 10.1103/PhysRevB.73.085103} {\bibfield  {journal} {\bibinfo
			{journal} {Phys. Rev. B}\ }\textbf {\bibinfo {volume} {73}},\ \bibinfo
		{pages} {085103} (\bibinfo {year} {2006})}\BibitemShut {NoStop}%
	\bibitem [{\citenamefont {Nattermann}\ \emph {et~al.}(2007)\citenamefont
		{Nattermann}, \citenamefont {Petkovi\ifmmode~\acute{c}\else \'{c}\fi{}},
		\citenamefont {Ristivojevic},\ and\ \citenamefont
		{Sch\"utze}}]{Nattermann07}%
	\BibitemOpen
	\bibfield  {author} {\bibinfo {author} {\bibfnamefont {T.}~\bibnamefont
			{Nattermann}}, \bibinfo {author} {\bibfnamefont {A.}~\bibnamefont
			{Petkovi\ifmmode~\acute{c}\else \'{c}\fi{}}}, \bibinfo {author}
		{\bibfnamefont {Z.}~\bibnamefont {Ristivojevic}}, \ and\ \bibinfo {author}
		{\bibfnamefont {F.}~\bibnamefont {Sch\"utze}},\ }\bibfield  {title} {\enquote
		{\bibinfo {title} {{Absence of the Mott Glass Phase in 1D Disordered
					Fermionic Systems}},}\ }\href {\doibase 10.1103/PhysRevLett.99.186402}
	{\bibfield  {journal} {\bibinfo  {journal} {Phys. Rev. Lett.}\ }\textbf
		{\bibinfo {volume} {99}},\ \bibinfo {pages} {186402} (\bibinfo {year}
		{2007})}\BibitemShut {NoStop}%
	\bibitem [{\citenamefont {Fogler}(2002)}]{Fogler02}%
	\BibitemOpen
	\bibfield  {author} {\bibinfo {author} {\bibfnamefont {Michael~M.}\
			\bibnamefont {Fogler}},\ }\bibfield  {title} {\enquote {\bibinfo {title} {Low
				frequency dynamics of disordered $\mathit{XY}$ spin chains and pinned density
				waves: From localized spin waves to soliton tunneling},}\ }\href {\doibase
		10.1103/PhysRevLett.88.186402} {\bibfield  {journal} {\bibinfo  {journal}
			{Phys. Rev. Lett.}\ }\textbf {\bibinfo {volume} {88}},\ \bibinfo {pages}
		{186402} (\bibinfo {year} {2002})}\BibitemShut {NoStop}%
	\bibitem [{\citenamefont {Mott}(1968)}]{Mott68}%
	\BibitemOpen
	\bibfield  {author} {\bibinfo {author} {\bibfnamefont {N.~F.}\ \bibnamefont
			{Mott}},\ }\bibfield  {title} {\enquote {\bibinfo {title} {Conduction in
				non-crystalline systems i. localized electronic states in disordered
				systems},}\ }\href {\doibase 10.1080/14786436808223200} {\bibfield  {journal}
		{\bibinfo  {journal} {Philos. Mag.}\ }\textbf {\bibinfo {volume} {17}},\
		\bibinfo {pages} {1259} (\bibinfo {year} {1968})}\BibitemShut {NoStop}%
	\bibitem [{not({\natexlab{e}})}]{not220}%
	\BibitemOpen
	\href@noop {} {} \bibinfo {note} {{B.~I. Halperin, as
			cited by Mott.}}\BibitemShut {Stop}%
	\bibitem [{\citenamefont {Le~Doussal}\ \emph {et~al.}(2008)\citenamefont
		{Le~Doussal}, \citenamefont {M\"uller},\ and\ \citenamefont
		{Wiese}}]{Ledoussal08}%
	\BibitemOpen
	\bibfield  {author} {\bibinfo {author} {\bibfnamefont {P.}~\bibnamefont
			{Le~Doussal}}, \bibinfo {author} {\bibfnamefont {M.}~\bibnamefont
			{M\"uller}}, \ and\ \bibinfo {author} {\bibfnamefont {K.~J.}\ \bibnamefont
			{Wiese}},\ }\bibfield  {title} {\enquote {\bibinfo {title} {{Cusps and shocks
					in the renormalized potential of glassy random manifolds: How functional
					renormalization group and replica symmetry breaking fit together}},}\ }\href
	{\doibase 10.1103/PhysRevB.77.064203} {\bibfield  {journal} {\bibinfo
			{journal} {Phys. Rev. B}\ }\textbf {\bibinfo {volume} {77}},\ \bibinfo
		{pages} {064203} (\bibinfo {year} {2008})}\BibitemShut {NoStop}%
	\bibitem [{\citenamefont {Mouhanna}\ and\ \citenamefont
		{Tarjus}(2010)}]{Mouhanna10}%
	\BibitemOpen
	\bibfield  {author} {\bibinfo {author} {\bibfnamefont {D.}~\bibnamefont
			{Mouhanna}}\ and\ \bibinfo {author} {\bibfnamefont {G.}~\bibnamefont
			{Tarjus}},\ }\bibfield  {title} {\enquote {\bibinfo {title} {{Spontaneous
					versus explicit replica symmetry breaking in the theory of disordered
					systems}},}\ }\href {\doibase 10.1103/PhysRevE.81.051101} {\bibfield
		{journal} {\bibinfo  {journal} {Phys. Rev. E}\ }\textbf {\bibinfo {volume}
			{81}},\ \bibinfo {pages} {051101} (\bibinfo {year} {2010})}\BibitemShut
	{NoStop}%
	\bibitem [{\citenamefont {Mouhanna}\ and\ \citenamefont
		{Tarjus}(2016)}]{Mouhanna16}%
	\BibitemOpen
	\bibfield  {author} {\bibinfo {author} {\bibfnamefont {D.}~\bibnamefont
			{Mouhanna}}\ and\ \bibinfo {author} {\bibfnamefont {G.}~\bibnamefont
			{Tarjus}},\ }\bibfield  {title} {\enquote {\bibinfo {title} {{Phase diagram
					and criticality of the random anisotropy model in the large-$N$ limit}},}\
	}\href {\doibase 10.1103/PhysRevB.94.214205} {\bibfield  {journal} {\bibinfo
			{journal} {Phys. Rev. B}\ }\textbf {\bibinfo {volume} {94}},\ \bibinfo
		{pages} {214205} (\bibinfo {year} {2016})}\BibitemShut {NoStop}%
	\bibitem [{\citenamefont {Daviet}\ and\ \citenamefont
		{Dupuis}(2019)}]{Daviet19}%
	\BibitemOpen
	\bibfield  {author} {\bibinfo {author} {\bibfnamefont {R.}~\bibnamefont
			{Daviet}}\ and\ \bibinfo {author} {\bibfnamefont {N.}~\bibnamefont
			{Dupuis}},\ }\bibfield  {title} {\enquote {\bibinfo {title} {{Nonperturbative
					functional renormalization-group approach to the sine-Gordon model and the
					Lukyanov-Zamolodchikov conjecture}},}\ }\href {\doibase
		10.1103/PhysRevLett.122.155301} {\bibfield  {journal} {\bibinfo  {journal}
			{Phys. Rev. Lett.}\ }\textbf {\bibinfo {volume} {122}},\ \bibinfo {pages}
		{155301} (\bibinfo {year} {2019})}\BibitemShut {NoStop}%
	\bibitem [{\citenamefont {Negele}\ and\ \citenamefont
		{Orland}(1998)}]{Negele_book}%
	\BibitemOpen
	\bibfield  {author} {\bibinfo {author} {\bibfnamefont {John~W.}\ \bibnamefont
			{Negele}}\ and\ \bibinfo {author} {\bibfnamefont {Henri}\ \bibnamefont
			{Orland}},\ }\href@noop {} {\emph {\bibinfo {title} {{Quantum Many-particle
					Systems}}}}\ (\bibinfo  {publisher} {Westview Press},\ \bibinfo {year}
	{1998})\BibitemShut {NoStop}%
	\bibitem [{\citenamefont {Shankar}(1990)}]{Shankar90}%
	\BibitemOpen
	\bibfield  {author} {\bibinfo {author} {\bibfnamefont {R.}~\bibnamefont
			{Shankar}},\ }\bibfield  {title} {\enquote {\bibinfo {title} {{Solvable model
					of a metal-insulator transition}},}\ }\href {\doibase
		10.1142/S0217979290001121} {\bibfield  {journal} {\bibinfo  {journal} {Int.
				J. Mod. Phys. B}\ }\textbf {\bibinfo {volume} {04}},\ \bibinfo {pages} {2371}
		(\bibinfo {year} {1990})}\BibitemShut {NoStop}%
	\bibitem [{\citenamefont {Schlessinger}(1968)}]{Schlessinger68}%
	\BibitemOpen
	\bibfield  {author} {\bibinfo {author} {\bibfnamefont {L.}~\bibnamefont
			{Schlessinger}},\ }\bibfield  {title} {\enquote {\bibinfo {title} {Use of
				analyticity in the calculation of nonrelativistic scattering amplitudes},}\
	}\href {\doibase 10.1103/PhysRev.167.1411} {\bibfield  {journal} {\bibinfo
			{journal} {Phys. Rev.}\ }\textbf {\bibinfo {volume} {167}},\ \bibinfo {pages}
		{1411} (\bibinfo {year} {1968})}\BibitemShut {NoStop}%
	\bibitem [{\citenamefont {Vidberg}\ and\ \citenamefont
		{Serene}(1977)}]{Vidberg77}%
	\BibitemOpen
	\bibfield  {author} {\bibinfo {author} {\bibfnamefont {H.~J.}\ \bibnamefont
			{Vidberg}}\ and\ \bibinfo {author} {\bibfnamefont {J.~W.}\ \bibnamefont
			{Serene}},\ }\bibfield  {title} {{\selectlanguage {English}\enquote {\bibinfo
				{title} {{Solving the Eliashberg equations by means of $N$-point Pad\'e
						approximants}},}\ }}\href {\doibase 10.1007/BF00655090} {\bibfield  {journal}
		{\bibinfo  {journal} {J. Low Temp. Phys.}\ }\textbf {\bibinfo {volume}
			{29}},\ \bibinfo {pages} {179} (\bibinfo {year} {1977})}\BibitemShut
	{NoStop}%
	\bibitem [{\citenamefont {Tripolt}\ \emph {et~al.}(2019)\citenamefont
		{Tripolt}, \citenamefont {Gubler}, \citenamefont {Ulybyshev},\ and\
		\citenamefont {von Smekal}}]{Tripolt19}%
	\BibitemOpen
	\bibfield  {author} {\bibinfo {author} {\bibfnamefont {Ralf-Arno}\
			\bibnamefont {Tripolt}}, \bibinfo {author} {\bibfnamefont {Philipp}\
			\bibnamefont {Gubler}}, \bibinfo {author} {\bibfnamefont {Maksim}\
			\bibnamefont {Ulybyshev}}, \ and\ \bibinfo {author} {\bibfnamefont {Lorenz}\
			\bibnamefont {von Smekal}},\ }\bibfield  {title} {\enquote {\bibinfo {title}
			{{Numerical analytic continuation of Euclidean data}},}\ }\href {\doibase
		https://doi.org/10.1016/j.cpc.2018.11.012} {\bibfield  {journal} {\bibinfo
			{journal} {Comput. Phys. Commun.}\ }\textbf {\bibinfo {volume} {237}},\
		\bibinfo {pages} {129} (\bibinfo {year} {2019})}\BibitemShut {NoStop}%
	\bibitem [{\citenamefont {Dupuis}(2009)}]{Dupuis09b}%
	\BibitemOpen
	\bibfield  {author} {\bibinfo {author} {\bibfnamefont {N.}~\bibnamefont
			{Dupuis}},\ }\bibfield  {title} {\enquote {\bibinfo {title} {{Infrared
					behavior and spectral function of a Bose superfluid at zero temperature}},}\
	}\href {\doibase 10.1103/PhysRevA.80.043627} {\bibfield  {journal} {\bibinfo
			{journal} {Phys. Rev. A}\ }\textbf {\bibinfo {volume} {80}},\ \bibinfo
		{pages} {043627} (\bibinfo {year} {2009})}\BibitemShut {NoStop}%
	\bibitem [{\citenamefont {Sinner}\ \emph {et~al.}(2010)\citenamefont {Sinner},
		\citenamefont {Hasselmann},\ and\ \citenamefont {Kopietz}}]{Sinner10}%
	\BibitemOpen
	\bibfield  {author} {\bibinfo {author} {\bibfnamefont {Andreas}\ \bibnamefont
			{Sinner}}, \bibinfo {author} {\bibfnamefont {Nils}\ \bibnamefont
			{Hasselmann}}, \ and\ \bibinfo {author} {\bibfnamefont {Peter}\ \bibnamefont
			{Kopietz}},\ }\bibfield  {title} {\enquote {\bibinfo {title} {{Functional
					renormalization-group approach to interacting bosons at zero temperature}},}\
	}\href {\doibase 10.1103/PhysRevA.82.063632} {\bibfield  {journal} {\bibinfo
			{journal} {Phys. Rev. A}\ }\textbf {\bibinfo {volume} {82}},\ \bibinfo
		{pages} {063632} (\bibinfo {year} {2010})}\BibitemShut {NoStop}%
	\bibitem [{\citenamefont {Schmidt}\ and\ \citenamefont
		{Enss}(2011)}]{Schmidt11}%
	\BibitemOpen
	\bibfield  {author} {\bibinfo {author} {\bibfnamefont {Richard}\ \bibnamefont
			{Schmidt}}\ and\ \bibinfo {author} {\bibfnamefont {Tilman}\ \bibnamefont
			{Enss}},\ }\bibfield  {title} {\enquote {\bibinfo {title} {{Excitation
					spectra and rf response near the polaron-to-molecule transition from the
					functional renormalization group}},}\ }\href {\doibase
		10.1103/PhysRevA.83.063620} {\bibfield  {journal} {\bibinfo  {journal} {Phys.
				Rev. A}\ }\textbf {\bibinfo {volume} {83}},\ \bibinfo {pages} {063620}
		(\bibinfo {year} {2011})}\BibitemShut {NoStop}%
	\bibitem [{\citenamefont {Rose}\ \emph {et~al.}(2015)\citenamefont {Rose},
		\citenamefont {L\'eonard},\ and\ \citenamefont {Dupuis}}]{Rose15}%
	\BibitemOpen
	\bibfield  {author} {\bibinfo {author} {\bibfnamefont {F.}~\bibnamefont
			{Rose}}, \bibinfo {author} {\bibfnamefont {F.}~\bibnamefont {L\'eonard}}, \
		and\ \bibinfo {author} {\bibfnamefont {N.}~\bibnamefont {Dupuis}},\
	}\bibfield  {title} {\enquote {\bibinfo {title} {{Higgs amplitude mode in the
					vicinity of a $(2+1)$-dimensional quantum critical point: A nonperturbative
					renormalization-group approach}},}\ }\href {\doibase
		10.1103/PhysRevB.91.224501} {\bibfield  {journal} {\bibinfo  {journal} {Phys.
				Rev. B}\ }\textbf {\bibinfo {volume} {91}},\ \bibinfo {pages} {224501}
		(\bibinfo {year} {2015})}\BibitemShut {NoStop}%
	\bibitem [{\citenamefont {Rose}\ \emph {et~al.}(2016)\citenamefont {Rose},
		\citenamefont {Benitez}, \citenamefont {L\'eonard},\ and\ \citenamefont
		{Delamotte}}]{Rose16a}%
	\BibitemOpen
	\bibfield  {author} {\bibinfo {author} {\bibfnamefont {F.}~\bibnamefont
			{Rose}}, \bibinfo {author} {\bibfnamefont {F.}~\bibnamefont {Benitez}},
		\bibinfo {author} {\bibfnamefont {F.}~\bibnamefont {L\'eonard}}, \ and\
		\bibinfo {author} {\bibfnamefont {B.}~\bibnamefont {Delamotte}},\ }\bibfield
	{title} {\enquote {\bibinfo {title} {{Bound states of the
					${{\ensuremath{{\phi}}}}^{{4}}$ model via the nonperturbative renormalization
					group}},}\ }\href {\doibase 10.1103/PhysRevD.93.125018} {\bibfield  {journal}
		{\bibinfo  {journal} {Phys. Rev. D}\ }\textbf {\bibinfo {volume} {93}},\
		\bibinfo {pages} {125018} (\bibinfo {year} {2016})}\BibitemShut {NoStop}%
	\bibitem [{\citenamefont {Rose}\ and\ \citenamefont {Dupuis}(2017)}]{Rose17a}%
	\BibitemOpen
	\bibfield  {author} {\bibinfo {author} {\bibfnamefont {F.}~\bibnamefont
			{Rose}}\ and\ \bibinfo {author} {\bibfnamefont {N.}~\bibnamefont {Dupuis}},\
	}\bibfield  {title} {\enquote {\bibinfo {title} {{Superuniversal transport
					near a $(2+1)$-dimensional quantum critical point}},}\ }\href {\doibase
		10.1103/PhysRevB.96.100501} {\bibfield  {journal} {\bibinfo  {journal} {Phys.
				Rev. B}\ }\textbf {\bibinfo {volume} {96}},\ \bibinfo {pages} {100501(R)}
		(\bibinfo {year} {2017})}\BibitemShut {NoStop}%
	\bibitem [{\citenamefont {Rose}\ and\ \citenamefont {Dupuis}(2018)}]{Rose18}%
	\BibitemOpen
	\bibfield  {author} {\bibinfo {author} {\bibfnamefont {F.}~\bibnamefont
			{Rose}}\ and\ \bibinfo {author} {\bibfnamefont {N.}~\bibnamefont {Dupuis}},\
	}\bibfield  {title} {\enquote {\bibinfo {title} {{Nonperturbative
					renormalization-group approach preserving the momentum dependence of
					correlation functions}},}\ }\href {\doibase 10.1103/PhysRevB.97.174514}
	{\bibfield  {journal} {\bibinfo  {journal} {Phys. Rev. B}\ }\textbf {\bibinfo
			{volume} {97}},\ \bibinfo {pages} {174514} (\bibinfo {year}
		{2018})}\BibitemShut {NoStop}%
	\bibitem [{\citenamefont {Tripolt}\ \emph {et~al.}(2017)\citenamefont
		{Tripolt}, \citenamefont {Haritan}, \citenamefont {Wambach},\ and\
		\citenamefont {Moiseyev}}]{Tripolt17}%
	\BibitemOpen
	\bibfield  {author} {\bibinfo {author} {\bibfnamefont {R.-A.}\ \bibnamefont
			{Tripolt}}, \bibinfo {author} {\bibfnamefont {I.}~\bibnamefont {Haritan}},
		\bibinfo {author} {\bibfnamefont {J.}~\bibnamefont {Wambach}}, \ and\
		\bibinfo {author} {\bibfnamefont {N.}~\bibnamefont {Moiseyev}},\ }\bibfield
	{title} {\enquote {\bibinfo {title} {{Threshold energies and poles for hadron
					physical problems by a model-independent universal algorithm}},}\ }\href
	{\doibase https://doi.org/10.1016/j.physletb.2017.10.001} {\bibfield
		{journal} {\bibinfo  {journal} {Phys. Lett. B}\ }\textbf {\bibinfo {volume}
			{774}},\ \bibinfo {pages} {411} (\bibinfo {year} {2017})}\BibitemShut
	{NoStop}%
	\bibitem [{\citenamefont {Balents}\ and\ \citenamefont
		{Doussal}(2005)}]{Balents05}%
	\BibitemOpen
	\bibfield  {author} {\bibinfo {author} {\bibfnamefont {Leon}\ \bibnamefont
			{Balents}}\ and\ \bibinfo {author} {\bibfnamefont {Pierre~Le}\ \bibnamefont
			{Doussal}},\ }\bibfield  {title} {\enquote {\bibinfo {title} {{Thermal
					fluctuations in pinned elastic systems: field theory of rare events and
					droplets}},}\ }\href {\doibase https://doi.org/10.1016/j.aop.2004.10.001}
	{\bibfield  {journal} {\bibinfo  {journal} {Ann. Phys.}\ }\textbf {\bibinfo
			{volume} {315}},\ \bibinfo {pages} {213} (\bibinfo {year} {2005})},\ \bibinfo
	{note} {special Issue}\BibitemShut {NoStop}%
	\bibitem [{\citenamefont {Fisher}\ and\ \citenamefont
		{Huse}(1986)}]{Fisher86a}%
	\BibitemOpen
	\bibfield  {author} {\bibinfo {author} {\bibfnamefont {Daniel~S.}\
			\bibnamefont {Fisher}}\ and\ \bibinfo {author} {\bibfnamefont {David~A.}\
			\bibnamefont {Huse}},\ }\bibfield  {title} {\enquote {\bibinfo {title}
			{Ordered phase of short-range Ising spin-glasses},}\ }\href {\doibase
		10.1103/PhysRevLett.56.1601} {\bibfield  {journal} {\bibinfo  {journal}
			{Phys. Rev. Lett.}\ }\textbf {\bibinfo {volume} {56}},\ \bibinfo {pages}
		{1601} (\bibinfo {year} {1986})}\BibitemShut {NoStop}%
	\bibitem [{not({\natexlab{f}})}]{not230}%
	\BibitemOpen
	\href@noop {} {}  \bibinfo {note} {Note that this
			reasoning should be based on the exact quantum states rather than the
			droplets (which are classical field configurations). Here we assume that the
			energy distribution of the low-energy quantum states is given by
			$P_L(\w)$.}\BibitemShut {Stop}%
	\bibitem [{not({\natexlab{g}})}]{not130}%
	\BibitemOpen
	\href@noop {} {} \bibinfo {note} {Note that the droplet
		picture differs from the description of the
		Bose-glass phase as a Griffiths phase, as often advocated in particular near
		the transition to the Mott insulating
		phase.\cite{Fisher89,Pollet09,Niederle13,Hegg13} In the Griffiths phase the
		superfluid regions are exponentially rare while they are power-law rare in
		the droplet scenario.}\BibitemShut {Stop}%
	\bibitem [{not({\natexlab{h}})}]{not200}%
	\BibitemOpen
	\href@noop {} {}  \bibinfo {note} {See Appendix D in
		Ref.~\onlinecite{Giamarchi01}. Note that the case relevant to our discussion
		corresponds to $\mu=m=0$.}\BibitemShut {Stop}%
	\bibitem [{not({\natexlab{i}})}]{not210}%
	\BibitemOpen
	\href@noop {} {} \bibinfo {note} {We have verified
		numerically that in the PFRG approach, the fixed-point solution $\delta^*(u)$
		is reached at a nonzero momentum scale $k$ where the dynamical critical
		exponent $z_k$ diverges.}\BibitemShut {Stop}%
	\bibitem [{\citenamefont {Delamotte}\ \emph {et~al.}(2016)\citenamefont
		{Delamotte}, \citenamefont {Tissier},\ and\ \citenamefont
		{Wschebor}}]{Delamotte16a}%
	\BibitemOpen
	\bibfield  {author} {\bibinfo {author} {\bibfnamefont {Bertrand}\
			\bibnamefont {Delamotte}}, \bibinfo {author} {\bibfnamefont {Matthieu}\
			\bibnamefont {Tissier}}, \ and\ \bibinfo {author} {\bibfnamefont {Nicol\'as}\
			\bibnamefont {Wschebor}},\ }\bibfield  {title} {\enquote {\bibinfo {title}
			{Scale invariance implies conformal invariance for the three-dimensional
				Ising model},}\ }\href {\doibase 10.1103/PhysRevE.93.012144} {\bibfield
		{journal} {\bibinfo  {journal} {Phys. Rev. E}\ }\textbf {\bibinfo {volume}
			{93}},\ \bibinfo {pages} {012144} (\bibinfo {year} {2016})}\BibitemShut
	{NoStop}%
	\bibitem [{\citenamefont {Papenbrock}\ and\ \citenamefont
		{Wetterich}(1995)}]{Papenbrock95}%
	\BibitemOpen
	\bibfield  {author} {\bibinfo {author} {\bibfnamefont {T.}~\bibnamefont
			{Papenbrock}}\ and\ \bibinfo {author} {\bibfnamefont {C.}~\bibnamefont
			{Wetterich}},\ }\bibfield  {title} {\enquote {\bibinfo {title} {Two-loop
				results from improved one loop computations},}\ }\href {\doibase
		10.1007/BF01556140} {\bibfield  {journal} {\bibinfo  {journal} {Z. Phys. C}\
		}\textbf {\bibinfo {volume} {65}},\ \bibinfo {pages} {519} (\bibinfo {year}
		{1995})}\BibitemShut {NoStop}%
	\bibitem [{\citenamefont {Reuter}\ and\ \citenamefont
		{Wetterich}(1994)}]{Reuter94}%
	\BibitemOpen
	\bibfield  {author} {\bibinfo {author} {\bibfnamefont {M.}~\bibnamefont
			{Reuter}}\ and\ \bibinfo {author} {\bibfnamefont {C.}~\bibnamefont
			{Wetterich}},\ }\bibfield  {title} {\enquote {\bibinfo {title} {Effective
				average action for gauge theories and exact evolution equations},}\ }\href
	{\doibase https://doi.org/10.1016/0550-3213(94)90543-6} {\bibfield  {journal}
		{\bibinfo  {journal} {Nucl. Phys. B}\ }\textbf {\bibinfo {volume} {417}},\
		\bibinfo {pages} {181} (\bibinfo {year} {1994})}\BibitemShut {NoStop}%
	\bibitem [{\citenamefont {Kownacki}\ and\ \citenamefont
		{Mouhanna}(2009)}]{Kownacki09}%
	\BibitemOpen
	\bibfield  {author} {\bibinfo {author} {\bibfnamefont {J.-P.}\ \bibnamefont
			{Kownacki}}\ and\ \bibinfo {author} {\bibfnamefont {D.}~\bibnamefont
			{Mouhanna}},\ }\bibfield  {title} {\enquote {\bibinfo {title} {Crumpling
				transition and flat phase of polymerized phantom membranes},}\ }\href
	{\doibase 10.1103/PhysRevE.79.040101} {\bibfield  {journal} {\bibinfo
			{journal} {Phys. Rev. E}\ }\textbf {\bibinfo {volume} {79}},\ \bibinfo
		{pages} {040101(R)} (\bibinfo {year} {2009})}\BibitemShut {NoStop}%
	\bibitem [{\citenamefont {Essafi}\ \emph {et~al.}(2014)\citenamefont {Essafi},
		\citenamefont {Kownacki},\ and\ \citenamefont {Mouhanna}}]{Essafi14}%
	\BibitemOpen
	\bibfield  {author} {\bibinfo {author} {\bibfnamefont {K.}~\bibnamefont
			{Essafi}}, \bibinfo {author} {\bibfnamefont {J.-P.}\ \bibnamefont
			{Kownacki}}, \ and\ \bibinfo {author} {\bibfnamefont {D.}~\bibnamefont
			{Mouhanna}},\ }\bibfield  {title} {\enquote {\bibinfo {title} {First-order
				phase transitions in polymerized phantom membranes},}\ }\href {\doibase
		10.1103/PhysRevE.89.042101} {\bibfield  {journal} {\bibinfo  {journal} {Phys.
				Rev. E}\ }\textbf {\bibinfo {volume} {89}},\ \bibinfo {pages} {042101}
		(\bibinfo {year} {2014})}\BibitemShut {NoStop}%
	\bibitem [{\citenamefont {Chou}\ \emph {et~al.}(2018)\citenamefont {Chou},
		\citenamefont {Nandkishore},\ and\ \citenamefont {Radzihovsky}}]{Chou18}%
	\BibitemOpen
	\bibfield  {author} {\bibinfo {author} {\bibfnamefont {Yang-Zhi}\
			\bibnamefont {Chou}}, \bibinfo {author} {\bibfnamefont {Rahul~M.}\
			\bibnamefont {Nandkishore}}, \ and\ \bibinfo {author} {\bibfnamefont {Leo}\
			\bibnamefont {Radzihovsky}},\ }\bibfield  {title} {\enquote {\bibinfo {title}
			{{Mott glass from localization and confinement}},}\ }\href {\doibase
		10.1103/PhysRevB.97.184205} {\bibfield  {journal} {\bibinfo  {journal} {Phys.
				Rev. B}\ }\textbf {\bibinfo {volume} {97}},\ \bibinfo {pages} {184205}
		(\bibinfo {year} {2018})}\BibitemShut {NoStop}%
	\bibitem [{\citenamefont {{Dupuis}}(2020)}]{Dupuis20a}%
	\BibitemOpen
	\bibfield  {author} {\bibinfo {author} {\bibfnamefont {Nicolas}\ \bibnamefont
			{{Dupuis}}},\ }\href@noop {} {\enquote {\bibinfo {title} {{Is there a
					Mott-glass phase in a one-dimensional disordered quantum fluid with linearly
					confining interactions?}}}\ } (\bibinfo {year} {2020}),\ \Eprint
	{http://arxiv.org/abs/2001.05682} {arXiv:2001.05682 [cond-mat.quant-gas]}
	\BibitemShut {NoStop}%
	\bibitem [{\citenamefont {Orignac}\ \emph {et~al.}(1999)\citenamefont
		{Orignac}, \citenamefont {Giamarchi},\ and\ \citenamefont
		{Le~Doussal}}]{Orignac99}%
	\BibitemOpen
	\bibfield  {author} {\bibinfo {author} {\bibfnamefont {E.}~\bibnamefont
			{Orignac}}, \bibinfo {author} {\bibfnamefont {T.}~\bibnamefont {Giamarchi}},
		\ and\ \bibinfo {author} {\bibfnamefont {P.}~\bibnamefont {Le~Doussal}},\
	}\bibfield  {title} {\enquote {\bibinfo {title} {{Possible New Phase of
					Commensurate Insulators with Disorder: The Mott Glass}},}\ }\href {\doibase
		10.1103/PhysRevLett.83.2378} {\bibfield  {journal} {\bibinfo  {journal}
			{Phys. Rev. Lett.}\ }\textbf {\bibinfo {volume} {83}},\ \bibinfo {pages}
		{2378} (\bibinfo {year} {1999})}\BibitemShut {NoStop}%
	\bibitem [{\citenamefont {Giamarchi}\ \emph {et~al.}(2001)\citenamefont
		{Giamarchi}, \citenamefont {Le~Doussal},\ and\ \citenamefont
		{Orignac}}]{Giamarchi01}%
	\BibitemOpen
	\bibfield  {author} {\bibinfo {author} {\bibfnamefont {T.}~\bibnamefont
			{Giamarchi}}, \bibinfo {author} {\bibfnamefont {P.}~\bibnamefont
			{Le~Doussal}}, \ and\ \bibinfo {author} {\bibfnamefont {E.}~\bibnamefont
			{Orignac}},\ }\bibfield  {title} {\enquote {\bibinfo {title} {{Competition of
					random and periodic potentials in interacting fermionic systems and classical
					equivalents: The Mott glass}},}\ }\href {\doibase 10.1103/PhysRevB.64.245119}
	{\bibfield  {journal} {\bibinfo  {journal} {Phys. Rev. B}\ }\textbf {\bibinfo
			{volume} {64}},\ \bibinfo {pages} {245119} (\bibinfo {year}
		{2001})}\BibitemShut {NoStop}%
	\bibitem [{\citenamefont {Zinn-Justin}(2002)}]{Zinn_book}%
	\BibitemOpen
	\bibfield  {author} {\bibinfo {author} {\bibfnamefont {J.}~\bibnamefont
			{Zinn-Justin}},\ }\href@noop {} {\emph {\bibinfo {title} {{Quantum Field
					Theory and Critical Phenomena}}}}\ (\bibinfo  {publisher} {Fourth Edition,
		Clarendon Press, Oxford},\ \bibinfo {year} {2002})\BibitemShut {NoStop}%
	\bibitem [{\citenamefont {Hasselmann}(2012)}]{Hasselmann12}%
	\BibitemOpen
	\bibfield  {author} {\bibinfo {author} {\bibfnamefont {N.}~\bibnamefont
			{Hasselmann}},\ }\bibfield  {title} {\enquote {\bibinfo {title}
			{{Effective-average-action-based approach to correlation functions at finite
					momenta}},}\ }\href {\doibase 10.1103/PhysRevE.86.041118} {\bibfield
		{journal} {\bibinfo  {journal} {Phys. Rev. E}\ }\textbf {\bibinfo {volume}
			{86}},\ \bibinfo {pages} {041118} (\bibinfo {year} {2012})}\BibitemShut
	{NoStop}%
	\bibitem [{\citenamefont {Pollet}\ \emph {et~al.}(2009)\citenamefont {Pollet},
		\citenamefont {Prokof'ev}, \citenamefont {Svistunov},\ and\ \citenamefont
		{Troyer}}]{Pollet09}%
	\BibitemOpen
	\bibfield  {author} {\bibinfo {author} {\bibfnamefont {L.}~\bibnamefont
			{Pollet}}, \bibinfo {author} {\bibfnamefont {N.~V.}\ \bibnamefont
			{Prokof'ev}}, \bibinfo {author} {\bibfnamefont {B.~V.}\ \bibnamefont
			{Svistunov}}, \ and\ \bibinfo {author} {\bibfnamefont {M.}~\bibnamefont
			{Troyer}},\ }\bibfield  {title} {\enquote {\bibinfo {title} {{Absence of a
					Direct Superfluid to Mott Insulator Transition in Disordered Bose
					Systems}},}\ }\href {\doibase 10.1103/PhysRevLett.103.140402} {\bibfield
		{journal} {\bibinfo  {journal} {Phys. Rev. Lett.}\ }\textbf {\bibinfo
			{volume} {103}},\ \bibinfo {pages} {140402} (\bibinfo {year}
		{2009})}\BibitemShut {NoStop}%
	\bibitem [{\citenamefont {Niederle}\ and\ \citenamefont
		{Rieger}(2013)}]{Niederle13}%
	\BibitemOpen
	\bibfield  {author} {\bibinfo {author} {\bibfnamefont {A.~E.}\ \bibnamefont
			{Niederle}}\ and\ \bibinfo {author} {\bibfnamefont {H.}~\bibnamefont
			{Rieger}},\ }\bibfield  {title} {\enquote {\bibinfo {title} {{Superfluid
					clusters, percolation and phase transitions in the disordered,
					two-dimensional Bose-Hubbard model}},}\ }\href {\doibase
		10.1088/1367-2630/15/7/075029} {\bibfield  {journal} {\bibinfo  {journal}
			{New J. Phys.}\ }\textbf {\bibinfo {volume} {15}},\ \bibinfo {pages} {075029}
		(\bibinfo {year} {2013})}\BibitemShut {NoStop}%
	\bibitem [{\citenamefont {Hegg}\ \emph {et~al.}(2013)\citenamefont {Hegg},
		\citenamefont {Kr\"uger},\ and\ \citenamefont {Phillips}}]{Hegg13}%
	\BibitemOpen
	\bibfield  {author} {\bibinfo {author} {\bibfnamefont {A.}~\bibnamefont
			{Hegg}}, \bibinfo {author} {\bibfnamefont {F.}~\bibnamefont {Kr\"uger}}, \
		and\ \bibinfo {author} {\bibfnamefont {P.~W.}\ \bibnamefont {Phillips}},\
	}\bibfield  {title} {\enquote {\bibinfo {title} {{Breakdown of self-averaging
					in the Bose glass}},}\ }\href {\doibase 10.1103/PhysRevB.88.134206}
	{\bibfield  {journal} {\bibinfo  {journal} {Phys. Rev. B}\ }\textbf {\bibinfo
			{volume} {88}},\ \bibinfo {pages} {134206} (\bibinfo {year}
		{2013})}\BibitemShut {NoStop}%
\end{thebibliography}
